\documentclass[preprint,10pt]{aastex}
\pdfoutput=1 
\renewcommand{\clearpage}{\relax}

\slugcomment{Submitted to Ap. J.}

\shorttitle{Tracking Vector Magnetograms with the Magnetic Induction Equation} 
\shortauthors{P. W. Schuck}

\usepackage{color}
\newcommand{\vv}{\mbox{\boldmath{$v$}}}

\newcommand{\Slopes}{S}
\newcommand{\RO}{\rho}
\newcommand{\PCC}{C}

\newcommand{\CCC}{\mathcal{C}}

\newcommand{\xhat}{\widehat{\vectorfont{x}}}
\newcommand{\yhat}{\widehat{\vectorfont{y}}}
\newcommand{\E}{\vectorfont{E}}
\newcommand{\B}{\vectorfont{B}}
\newcommand{\A}{\vectorfont{A}}
\newcommand{\f}{\vectorfont{f}}
\newcommand{\gA}{g_A}
\newcommand{\GA}{G_A}

\newcommand{\grad}{\vectorfont{\nabla}}
\newcommand{\gradh}{\vectorfont{\nabla}_{h}}
\newcommand{\cross}{\vectorfont{\times}}

\newcommand{\zhat}{\widehat{\vectorfont{z}}}
\newcommand{\x}{\vectorfont{x}}

\newcommand{\X}{\vectorfont{\chi}}

\newcommand{\T}{\tau}

\newcommand{\evec}{\vectorfont{\eta}}

\renewcommand{\v}{\vectorfont{v}}
\newcommand{\V}{\vectorfont{V}}
\newcommand{\Ftv}{\vectorfont{F}}
\newcommand{\ftv}{\vectorfont{f}}
\newcommand{\vt}{\vectorfont{v}_{h}}
\newcommand{\Vt}{\vectorfont{V}_{h}}
\renewcommand{\u}{\vectorfont{\vartheta}}
\newcommand{\tensorfont}[1]{\mbox{\boldmath{$\mathsf{#1}$}}}
\newcommand{\sss}{\mathsf{s}}
\newcommand{\GGG}{\mathcal{G}}

\newcommand{\AAaa}{\tensorfont{A}}

\newcommand{\bb}{\vectorfont{b}}
\newcommand{\bbb}{\left\langle\vectorfont{b}\right\rangle}

\newcommand{\SING}{\vectorfont{\Sigma}}
\newcommand{\SSS}{\left\langle\SS\right\rangle}

\renewcommand{\SS}{\tensorfont{S}}

\newcommand{\UA}{\tensorfont{U}}
\newcommand{\VA}{\tensorfont{V}}

\newcommand{\vectorfont}[1]{\mbox{\boldmath{$#1$}}}

\newcommand{\MM}{\vectorfont{P}}

\newcommand{\AAA}{\left\langle\AAaa\right\rangle}

\newcommand{\PreserveBackslash}[1]{\let\temp=\\#1\left\\=\temp}

\newcommand{\uF}{\vectorfont{u}}
\newcommand{\UF}{\vectorfont{U}}
\newcommand{\Ccs}{C_{\mathrm{CS}}}

\tablewidth{0pt}
\definecolor{Mygrey}{gray}{0.75}
\begin{document} 


\title{Tracking Vector Magnetograms with the Magnetic Induction Equation}


\author{P. W. Schuck\altaffilmark{\dag}}
\affil{Plasma Physics Division, United States Naval Research Laboratory}
\affil{4555 Overlook Ave., SW, Washington, DC 20375-5346}
\altaffiltext{\dag}{schuck@ppdmail.nrl.navy.mil}


\begin{abstract}
The differential affine velocity estimator (DAVE) developed in
\cite{Schuck2006a} for estimating velocities from line-of-sight magnetograms
is modified to directly incorporate horizontal magnetic fields to produce a
differential affine velocity estimator for vector magnetograms (DAVE4VM).  The
DAVE4VM's performance is demonstrated on the synthetic data from the anelastic
pseudospectral ANMHD simulations that were used in the recent comparison of
velocity inversion techniques by \cite{Welsch2007}. The DAVE4VM predicts
roughly 95\% of the helicity rate and 75\% of the power transmitted through
the simulation slice. Inter-comparison between DAVE4VM and DAVE and further
analysis of the DAVE method demonstrates that line-of-sight tracking methods
capture the shearing motion of magnetic footpoints but are insensitive to flux
emergence \---- the velocities determined from line-of-sight methods are more
consistent with horizontal plasma velocities than with flux transport
velocities. These results suggest that previous studies that rely on
velocities determined from line-of-sight methods such as the DAVE or local
correlation tracking may substantially misrepresent the total helicity rates
and power through the photosphere.
\end{abstract}




\section{Introduction}
Coronal mass ejections (CMEs) are now recognized as the primary solar
driver of geomagnetic storms \cite[]{Gosling1993}. Several theoretical
mechanisms have been proposed as drivers of CMEs, including large
scale coronal reconnection
\cite[]{Sweet1958,Parker1957,Antiochos1999}, emerging flux
cancellation of the overlying coronal field \cite[]{Linker2001}, flux
injection \cite[]{Chen1989a,Chen1996}, the kink instability of
filaments \cite[]{Rust1996,Torok2004,Kliem2004}, and photospheric
footpoint shearing
\cite[]{Amari2000,Amari2003a,Amari2003b,Schrijver2005a}. All of these
CME mechanisms are driven by magnetic forces. The main differences
depend on whether the magnetic helicity and energy are first stored in
the corona and later released by reconnection and instability or
whether the helicity and Poynting fluxes are roughly concomitant with
the eruption. The timing and magnitude of the transport of magnetic
helicity and energy through the photosphere provides an important
discriminator between the mechanisms.  In addition, eruption
precursors in the photospheric magnetic field might provide reliable
forecasting for space weather events. However, reliable, repeatable
photospheric precursors of CMEs have so far eluded detection
\cite[]{Leka2003a,Leka2003b,Leka2007}.  \par
The magnetic helicity and Poynting flux may be estimated from photospheric
velocities inferred from a sequence of magnetograms
\cite[]{Berger2000,Demoulin2003}. However, accurately estimating velocities
from a sequence of images is extremely challenging because image motion is
ambiguous.  The ``aperture problem'' occurs when different velocities produce
image dynamics that are indistinguishable
\cite[]{Stumpf1911,Marr1981,Hildreth1983,Hildreth1984}.  Optical flow methods
solve these under-determined or ill-posed problems that have no unique
velocity field solution by applying additional assumptions about flow
structure or flow properties.  Both \cite{Schuck2006a} and \cite{Welsch2007}
provide an overview of optical flow methods for recovering estimates of
photospheric velocities from a sequence of magnetograms
\cite[]{Kusano2002a,Kusano2004a,Welsch2004,Longcope2004,Schuck2005a,Schuck2006a,Georgoulis2006}.
Currently, most methods for estimating photospheric velocities implement some
form of the normal component of the induction equation
\begin{equation}
\partial_t B_z+\gradh\cdot\left(B_z\,\Vt-V_z\,\B_h\right)=0,\label{eqn:induction}
\end{equation}
where the plasma velocity $\V$ and the magnetic fields $\B$ are
decomposed into a local right-handed Cartesian coordinate system with
vertical direction along the $z$-axis and the horizontal plane,
denoted generically by the subscript ``h,'' containing the $x$- and
$y$-axes. \par
\cite{Demoulin2003} observed that the geometry of magnetic fields
embedded in the photosphere implied that
\begin{mathletters}
\label{eqn:LOS}
\begin{equation}
 \Ftv=\UF\,B_z\equiv{B}_z\,\Vt-V_z\,\B_h=\zhat\cross\left(\V\cross\B\right)=\zhat\cross\left(\V_\perp\cross\B\right),\label{eqn:flux_transport}
\end{equation}
where $\Ftv$ denotes the flux transport vector, $\UF$ is the horizontal
footpoint velocity or flux transport velocity $\left(\UF\cdot\zhat=0\right)$
and $\V_\perp$ is the plasma velocity perpendicular to the magnetic field
$\V_\perp\cdot\B=0$. The flux transport vectors are composed of two terms
${B}_z\,\Vt$ and $V_z\,\B_h$ representing shearing due to 
horizontal motion and flux emergence due to vertical motion respectively.
Equation~(\ref{eqn:flux_transport}) may be used to transform
(\ref{eqn:induction}) into a continuity equation for the vertical magnetic
field
\begin{equation}
\partial_t B_z+\gradh\cdot\left(\UF\,B_z\right)=0,\label{eqn:continuity}
\end{equation}
\end{mathletters}
where plasma velocity may be written generally in terms of the flux
transport velocity as
\begin{mathletters}
\label{eqn:ADC}
\begin{eqnarray}
\V&=&\UF-\frac{\left(\UF\cdot\B_h\right)\,\B}{|\B|^2}+V_\parallel\,\frac{\B}{\left|\B\right|},\\
\V_{\perp{h}}&=&\UF-\frac{\left(\UF\cdot\B_h\right)\,\B_h}{|\B|^2},\label{eqn:vph}\\
\V_{\perp{z}}&=&-\frac{\left(\UF\cdot\B_h\right)\,B_z}{|\B|^2},\label{eqn:vpz}
\end{eqnarray}
\end{mathletters}
and the subscripts ``$\parallel$'' and ``$\perp$'' denote plasma
velocities parallel and perpendicular to the magnetic field
respectively. Equations~(\ref{eqn:ADC}a\--c) are the algebraic
decomposition \cite[]{Welsch2004} generalized for arbitrary parallel
velocity $V_\parallel$, but the value of $V_\parallel$ does not affect
the perpendicular plasma velocity~(\ref{eqn:vph})\--(\ref{eqn:vpz}) or
the perpendicular electric field
\begin{equation}
c\,\E_\perp=-\V\cross\B=-\overbrace{\UF\cross\zhat\,B_z}^{\mbox{$\E_{\perp{h}}$}}-\overbrace{\UF\cross\B_h}^{\mbox{$E_{\perp{z}}$}},
\end{equation}
which both depend only on the flux transport velocity $\UF$.\par
Equations~(\ref{eqn:induction})\--(\ref{eqn:ADC}) should be formally
distinguished from the \textit{inverse problem} for determining an
estimate of the plasma velocity $\vv$ from vector magnetograms using
the normal component of the magnetic induction equation
\begin{mathletters} 
\label{eqn:DAVE4VM:all}
\begin{equation}
\partial_t B_z+\gradh\cdot\left(B_z\,\vt-v_z\,\B_h\right)=0,\label{eqn:DAVE4VM}
\end{equation}
where
\begin{equation}
\ftv=\uF\,B_z=B_z\,\vt-v_z\,\B_h=\zhat\cross\left(\v\cross\B\right)=\zhat\cross\left(\v_\perp\cross\B\right),\label{eqn:u}
\end{equation}
\end{mathletters}
and the \textit{inverse problem} for determining flux transport
velocity $\uF$ from the evolution of the vertical magnetic field or
line-of-sight component
\begin{equation}
\partial_t B_z+\gradh\cdot\left(\u\,B_z\right)=0.\label{eqn:DAVE}
\end{equation}
The notation $\u$, denoting an optical flow estimate, emphasizes that $\u$
determined from~(\ref{eqn:DAVE}) is not necessarily immediately identified
with the flux transport velocity $\uF$.
Equations~(\ref{eqn:DAVE4VM:all})\--(\ref{eqn:DAVE}) are ill-posed inverse
problems because of two ambiguities:
\begin{enumerate}
\item The Helmholtz decomposition
  of the flux transport vectors \cite[]{Welsch2004,Longcope2004}
\begin{equation}
\ftv=\uF\,B_z=B_z\,\vt-v_z\,\B_h=-\left(\gradh\phi+\gradh\psi\cross\zhat\right),\label{eqn:Helmholtz}
\end{equation}
where $\phi$ is the inductive potential and $\psi$ is the
electrostatic potential manifestly demonstrates that only inductive
potential $\psi$ may be unambiguously determined from the local
evolution of $B_z$ in~(\ref{eqn:DAVE4VM:all}).  The electrostatic potential
$\phi$ must be constrained by additional assumptions.  By
analogy,~(\ref{eqn:DAVE}) is also ill-posed for the same reason;
$\gradh\cross\left(\u\,B_z\right)$ is not constrained by the local
evolution of $\partial_t B_z$.
\item For~(\ref{eqn:DAVE4VM}) and~(\ref{eqn:u}) $V_\parallel$ is not
constrained by the local evolution of $\partial_t B_z$.
For~(\ref{eqn:DAVE}), there is no \textit{a priori} relationship
between $\u$ and $\uF$ or $\u$ and $\v$ for the inverse
problem. However, if $\u$ is identified with the flux transport
velocity $\uF$ then $\uF$ and $\v$ will satisfy the same relationships
as $\UF$ and $\V$ in~(\ref{eqn:ADC}).
\end{enumerate}\par
The first ambiguity may be resolved for~(\ref{eqn:DAVE4VM}) by the
induction method (IM) \cite[]{Kusano2002a,Kusano2004a}, minimum energy
fit (MEF) \cite[]{Longcope2004}, or the differential affine velocity
estimator for vector magnetograms (DAVE4VM) presented in
\S~\ref{sec:model}.  These methods produce \textit{a unique
solution} for $\v$, but not necessarily \textit{the unique solution}
that corresponds to $\V$.  The first ambiguity may be resolved
for~(\ref{eqn:DAVE}) by local optical flow methods such as the
differential affine velocity estimator (DAVE) \cite[]{Schuck2006a},
its nonlinear generalization \cite[]{Schuck2005a}, global methods
\cite[]{Wildes2000}, the minimum structure reconstruction (MSR)
\cite[]{Georgoulis2006} which imposes $\v_{\perp{z}}=0$ as an
assumption, or hybrid local\--global methods such as inductive local
correlation tracking (ILCT) \cite[]{Welsch2004}. These methods produce
\textit{a unique solution} for $\u$, but not necessarily \textit{the
unique solution} that corresponds to $\uF$.\par
Several assumptions have been used either explicitly or implicitly to
resolve the second ambiguity. \cite{Chae2001a} conjecture that local
correlation tracking (LCT) \cite[]{Leese1970,Leese1971,November1988}
provides a direct estimate of the horizontal photospheric plasma
velocity: $\u^{\left(\mathrm{LCT}\right)}=\Vt$. \cite{Demoulin2003}
conjecture that line-of-sight tracking methods, and in particular LCT,
estimate the total flux transport velocity
$\u^{\left(\mathrm{LCT}\right)}=\UF$.  \cite{Schuck2005a} formally
demonstrated that LCT is consistent with the advection equation
\begin{equation}
\partial_t B_z+\u^{\left(\mathrm{LCT}\right)}\cdot\gradh{B_z}=0,\label{eqn:advection}
\end{equation}
not the continuity equation in~(\ref{eqn:DAVE}), but that LCT could be
modified to be consistent with~(\ref{eqn:DAVE}) by direct integration
along Lagrangian trajectories in an affine velocity
profile. Nonetheless, both conjectures may be considered in the context
of~(\ref{eqn:DAVE}).  Under \citeauthor{Chae2001a}'s
(\citeyear{Chae2001a}) assumption, the flux transport velocity would
be derived from line-of-sight optical flow methods via
\begin{equation}
\uF\,B_z\equiv\u\,B_z-v_z\,\B_h,\label{eqn:chae}
\end{equation}
where in principle, $v_z$ might be approximately determined from Doppler
velocities near disk center.  Under \citeauthor{Demoulin2003}'s
(\citeyear{Demoulin2003}) assumption, the total flux transport velocity would
be derived from line-of-sight optical flow methods via
\begin{equation}
\uF\equiv\u\mbox{ for }B_z\neq0.\label{eqn:demoulin}
\end{equation}
The \textit{Ansatz} $\uF=\u$ has important implications for solar
observations. This conjecture implies that the total helicity and
Poynting flux may be estimated by tracking the vertical magnetic field
or by tracking the line-of-sight component near disk center as a proxy
for the vertical magnetic field. \citeauthor{Demoulin2003}'s
(\citeyear{Demoulin2003}) \textit{Ansatz} has largely been accepted by
the solar community
\cite[]{Welsch2004,Welsch2007,Kusano2004a,Schuck2005a,Schuck2006a,Labonte2007,Santos2007,Tain2008,Zhang2008a,Wang2008}. However,
equivalence between $\u$ and $\uF$ for line-of-sight methods has never
been practically established. These two different
hypotheses~(\ref{eqn:chae}) and~(\ref{eqn:demoulin}) for the
interpretation of $\u$ inferred by DAVE will be considered in
\S~\ref{sec:conclusions}.\par
The second ambiguity usually is not resolved using only information about the
magnetic fields. The velocity field inferred by the IM
\cite[]{Kusano2002a,Kusano2004a} does produce a component of the plasma velocity
along the magnetic field, but this was simply subtracted off in
\cite{Welsch2007}. In the absence of a reference flow, possibly derived from
Doppler measurements or LCT, the MEF imposes $v_\parallel=0$
\cite[]{Longcope2004}. ILCT and the original algebraic decomposition both
assume $v_\parallel=0$ \cite[]{Welsch2004}. \cite{Georgoulis2006} describe a
method for inferring $v_\parallel$ from Doppler measurements for MSR. For
DAVE4VM the second ambiguity is resolved simultaneously with the first. The
DAVE4VM method estimates a \textit{field aligned plasma velocity} from only
magnetic field observations! \par
Using established computer vision techniques
\cite[]{Lucas1981,Lucas1984,Baker2004}, \cite{Schuck2006a} developed the DAVE
from a short time-expansion of the modified LCT method discussed in
\cite{Schuck2005a} for estimating velocities from line-of-sight
magnetograms. The DAVE locally minimizes the square of the continuity
equation~(\ref{eqn:continuity}) subject to an affine velocity profile. Using
``moving paint'' experiments, \cite{Schuck2006a} demonstrated that this
technique was faster and more accurate than existing LCT algorithms for data
satisfying~(\ref{eqn:continuity}).  The DAVE method has been used to study the
apparent motion of active regions \cite[]{Schuck2006a}, flux pile up in the
photosphere \cite[]{Litvinenko2007}, and helicity flux in the photosphere
\cite[]{Chae2007a}. However, nagging questions remain about its
performance. \par
\cite{Welsch2007} set an important new standard for evaluating scientific
optical flow methods used for studying the Sun. For the first time many
existing methods for estimating photospheric velocities from magnetograms were
tested on a reasonable approximation to synthetic photospheric data from
anelastic pseudospectral ANMHD simulations
\cite[]{Fan1999,Abbett2000,Abbett2004}. The methods tested were Lockheed
Martin's Solar and Astrophysical Laboratory's (LMSAL) LCT code
\cite[]{DeRosa2001}, Fourier LCT (FLCT) \cite[]{Welsch2004}, the DAVE
\cite[]{Schuck2006a}, the IM \cite[]{Kusano2002a,Kusano2004a}, ILCT
\cite[]{Welsch2004}, the MEF \cite[]{Longcope2004}, and MSR
\cite[]{Georgoulis2006}. Unfortunately the results were not entirely
encouraging.  \cite{Welsch2007} treated the velocities estimated from
line-of-sight methods as the flux transport velocities consistent with the
hypothesis of \cite{Demoulin2003} in~(\ref{eqn:demoulin}). Evaluation of the
DAVE's performance on the ANMHD data under this assumption revealed that the
DAVE method did not estimate the helicity flux or Poynting flux reliably. In
fact none of the pure line-of-sight methods: LMSAL's LCT, FLCT, or the DAVE\----estimated these fluxes reliably, reproducing (at best) respectively 11\%, 9\%,
and 23\% of the helicity rate, and reproducing respectively 6\%, 11\%, and
22\% of the power injected through the surface.\par
Of course the ANMHD data have limitations. The simulation models the rise
of a buoyant magnetic flux rope in the convection zone and represents
the magnetic structure of granulation or super-granulation rather than
the dynamics of an active region \cite[See \S~2 in][for a
complete discussion]{Welsch2007}. In addition, \cite{Welsch2007}
noted that tracking methods performed better on real magnetograms than
on the synthetic ANMHD data using ``moving paint'' experiments where
images were simply shifted relative to one another. These results
provoked them to comment ``that the ANMHD data set either lacks some
characteristic present in real solar magnetograms or contains
artifacts not present in solar data.'' Consequently, the poor
performance of tracking methods on ANMHD data might be attributed to
the de-aliasing method for nonlinear terms in ANMHD (truncating the
spatial Fourier spectrum effectively smoothes small-scale structures)
or perhaps to the Fourier ringing near strong fields in the ANMHD data
set. While these issues are important to resolve, \textit{they fail to
fully explain the poor performance of the tracking methods to
accurately reproduce the quantity they were designed to estimate,
namely the helicity flux!}\par
This paper has two primary goals:
\begin{enumerate}
\item Develop a modified DAVE \cite[]{Schuck2006a} that incorporates
horizontal magnetic fields, termed the ``differential affine velocity
estimator for vector magnetograms'' (DAVE4VM), and demonstrate its
performance on the ANMHD simulation data. DAVE4VM performs much better
than the original DAVE technique and roughly on par with the minimum
energy fit (MEF) method developed by \cite{Longcope2004} which was
deemed to have performed the best overall in \citeauthor{Welsch2007}'s
(\citeyear{Welsch2007}) comparison of velocity-inversion techniques.
\item Identify the reasons for the poor performance of DAVE in
\cite{Welsch2007}.
\end{enumerate}
The paper attempts to follow, as closely as possible, the presentation
of the DAVE in \cite{Schuck2006a} and the analysis of velocity
inversion techniques by \cite{Welsch2007}. For the remainder of this
paper, lower case variables are used to represent the flux transport
vector, flux transport velocity, plasma velocity, electric field,
Poynting flux, and helicity flux estimates from the DAVE4VM and DAVE:
$\ftv$, $\uF$, $\v_\perp$, $\vectorfont{e}_\perp$, $s_z$ and $h$ and
the corresponding uppercase variables are used to represent the
``ground truth'' from ANMHD: $\Ftv$, $\UF$, $\V_\perp$,
$\vectorfont{E}_\perp$, $S_z$, and $H$. The one deviation from this
notation involves $\u$ which denotes an optical flow estimate based
on~(\ref{eqn:DAVE}).  Section~(\ref{sec:model}) describes the DAVE4VM
model and \S~\ref{sec:ANMHD} describes its application to the
ANMHD data.  For the most part, the plots and quantitative analysis
presented in \cite{Welsch2007} are produced for the DAVE4VM and DAVE
to facilitate inter-comparison and comparison to the other methods
considered in \cite{Welsch2007}. For the DAVE this analysis involves
the explicit assumption that $\u=\uF$. In
\S~\ref{sec:conclusions} the assumption $\u=\uF$ for the DAVE is
relaxed and compared with an alternative hypotheses that $\u=\v_h$
\---- that the DAVE produces a biased estimate of the total horizontal
plasma velocity.
\section{The DAVE4VM Model\label{sec:model}}
The extension of the DAVE for horizontal magnetic fields is
straight-forward. The plasma velocity is modeled with a
three-dimensional affine velocity profile:
\begin{equation}
\v\left(\MM;\x\right)=\left(\begin{array}{c}\widehat{u}_0\\ \widehat{v}_0\\ \widehat{w}_0\end{array}\right)+\left(\begin{array}{cc}\widehat{u}_x&\widehat{u}_y\\ \widehat{v}_x&\widehat{v}_y\\ \widehat{w}_x&\widehat{w}_y
\end{array}\right)\,\left(\begin{array}{c}x\\ y\end{array}\right),\label{eqn:affine}
\end{equation}
where the hatted variables model the local plasma velocity
profile. The coordinate system for the affine velocity profile is
\textit{not} aligned with the magnetic field. Therefore, the
velocities are \textit{not} guaranteed to be orthogonal to $\B$.
However, the parallel $\left(\parallel\right)$ and perpendicular
$\left(\perp\right)$ components of the plasma velocity may be
determined from
\begin{mathletters}
\label{eqn:v}
\begin{eqnarray}
v_\parallel&=&\frac{\left(\v\cdot\B\right)\,\B}{B^2},\\
\v_\perp&=&\v-\frac{\left(\v\cdot\B\right)\,\B}{B^2},\\
\v_{\perp{h}}&=&\v_h-\frac{\left(\v\cdot\B\right)\,\B_h}{B^2},\\
v_{\perp{z}}&=&v_z-\frac{\left(\v\cdot\B\right)\,B_z}{B^2}.
\end{eqnarray}
\end{mathletters}\par
The error metric
\begin{mathletters}
%
\begin{eqnarray}
\CCC_{\mbox{SSD}}&=&\int{dt}{dx^2}\,w\left(\x-\X,t-\T\right)\left\lbrace\partial_t B_{z}\left(\x,t\right)+\gradh\cdot\left[B_{z}\left(\x,t\right)\,\v_{h}\left(\MM,\x-\X\right)\right.\right.,\label{eqn:error}\\
&&\qquad\qquad\qquad\qquad\qquad\qquad\qquad\qquad\qquad\qquad\qquad\qquad\left.\left.-v_{z}\left(\MM,\x-\X\right)\,\B_{h}\left(\x,t\right)\right]\right\rbrace^2,\nonumber\\
&=&\evec\cdot\SSS\cdot\evec
\end{eqnarray}
characterizes how well the local velocity profile satisfies the
magnetic induction equation over a subregion of the magnetogram sequence
defined by the window function $w\left(\x-\X,t-\T\right)$ where
$\MM=\left(\widehat{u}_0,\widehat{v}_0,\widehat{u}_x,\widehat{v}_y,\widehat{u}_y,\widehat{v}_x,\widehat{w}_0,\widehat{w}_x,\widehat{w}_y\right)$
is a vector of parameters and $\evec\equiv\left(\MM,1\right)$. The plasma
velocity $\v\left(\MM,\x-\X\right)$ in~(\ref{eqn:error}) is referenced
from the center of the window at $\x=\X$ so that $\widehat{u}_0$,
$\widehat{v}_0$, and $\widehat{w}_0$ represent the plasma velocities
at the center of the window and the subscripted parameters represent
the best fit local shears in the plasma flows,
i.e. $\widehat{u}_x=\partial_x\,\left(\xhat\cdot\v\right)$.  The matrix
elements of $\SSS$ are defined by
\begin{equation}
\SSS=\int{dt}\,{dx^2}\,w\left(\x-\X,t-\T\right)\,\SS\left(\X;x,t\right),
\end{equation}
\end{mathletters}
where
\begin{equation}
\SS\left(\X;x,t\right)\equiv\left[\begin{array}{cc}\AAaa&\bb\\ \bb&\GGG_{99}\end{array}\right]=\left[
\begin{array}{cccccccccc}
\GGG_{00}&\cdot&\cdot&\cdot&\cdot&\cdot&\cdot&\cdot&\cdot&\cdot\\
\GGG_{10}&\GGG_{11}&\cdot&\cdot&\cdot&\cdot&\cdot&\cdot&\cdot&\cdot\\
\GGG_{20}&\GGG_{21}&\GGG_{22}&\cdot&\cdot&\cdot&\cdot&\cdot&\cdot&\cdot\\
\GGG_{30}&\GGG_{31}&\GGG_{32}&\GGG_{33}&\cdot&\cdot&\cdot&\cdot&\cdot&\cdot\\
\GGG_{40}&\GGG_{41}&\GGG_{42}&\GGG_{43}&\GGG_{44}&\cdot&\cdot&\cdot&\cdot&\cdot\\
\GGG_{50}&\GGG_{51}&\GGG_{52}&\GGG_{53}&\GGG_{54}&\GGG_{55}&\cdot&\cdot&\cdot&\cdot\\
\sss_{60}&\sss_{61}&\sss_{62}&\sss_{63}&\sss_{64}&\sss_{65}&\sss_{66}&\cdot&\cdot&\cdot\\
\sss_{70}&\sss_{71}&\sss_{72}&\sss_{73}&\sss_{74}&\sss_{75}&\sss_{76}&\sss_{77}&\cdot&\cdot\\
\sss_{80}&\sss_{81}&\sss_{82}&\sss_{83}&\sss_{84}&\sss_{85}&\sss_{86}&\sss_{87}&\sss_{88}&\cdot\\
\GGG_{90}&\GGG_{91}&\GGG_{92}&\GGG_{93}&\GGG_{94}&\GGG_{95}&\sss_{96}&\sss_{97}&\sss_{98}&\GGG_{99}\\
\end{array}\right],\label{eqn:structure}
\end{equation}
is a real symmetric $\SS=\SS^*$ positive semidefinite structure tensor
where a superscript ``*'' indicates the matrix transpose.  The matrix
elements of $\SS$ are provided in Appendix~\ref{app:matrix}. The
elements $\GGG_{ij}$ correspond to the original DAVE method
\cite[]{Schuck2006a} and the remainder $\sss_{ij}$ represent
corrections due to the horizontal components of the magnetic field and
flows normal to the surface. The least-squares solution is
\begin{equation}
\MM=-\AAA^{-1}\cdot\bbb,\label{eqn:least}
\end{equation}
when the aperture problem is completely resolved
$\det\left(\AAA\right)\neq0$ and the velocity field is
unambiguous. However, there are important new terms in the structure
tensor $\SSS$ involving $\B_h$. Situations where
$\det\left(\AAA\right)=0$ because $\B_h=0$ or $B_z=0$ over the region
contained within the window must be considered.  In general, the
Moore\--Penrose pseudo-inverse $\AAA^\dag$ provides a numerically
stable estimate of the optical flow parameters even when
$\det\left(\AAA\right)=0$
\begin{mathletters}
\begin{equation}
\MM=-\AAA^{\dag}\cdot\bbb,
\end{equation}
where
\begin{equation}
\AAA^\dag\equiv\VA\,\SING^\dag\,\UA^*,
\end{equation}
is defined in terms of the singular value decomposition \cite[]{Golub1980}
\begin{equation}
\AAA=\UA\,\SING\,\VA^*.
\end{equation}
\end{mathletters}
Here $\UA$ and $\VA$ are orthonormal matrices corresponding to the
nine principle directions, $\SING$ is a diagonal matrix containing the
nine singular values, and $\SING^\dag$ is computed by replacing
\textit{every nonzero element of $\SING$} by its reciprocal. If
$\B_h=0$, the singular values along the vertical direction are zero
and $\AAA$ is rank deficient. In this case, the method implemented
produces the minimum norm least squares solution resulting in no vertical
flows: $w_0=w_x=w_y=0$.
\section{Application to ANMHD Simulations\label{sec:ANMHD}}
This paper considers the pair of vector magnetograms $\B$ separated by
the \textit{shortest} time interval $\Delta t\approx250$~s between
data dumps of the ANMHD simulation slice archived by \cite{Welsch2007}. The
``ground truth'' data are derived from the time-averaged velocity and
magnetic fields from ANMHD over the \textit{shortest} time interval.
The region of interest in the ANMHD simulations corresponds to a
$101\times101$ pixel region centered on a convection cell.
\cite{Welsch2007} thresholded on the vertical magnetic field and
considered only pixels with $|B_z|>370$~G for all plots and quantities
except for the total helicity where a different masking of results was
used \cite[]{Welsch2008a}.  In a departure from the original
presentation of \cite{Welsch2007} this paper considers pixels with
$B=\sqrt{B_x^2+B_y^2+B_z^2}>370$~G. This corresponds to 7013 pixels or
roughly 70\% of the region of interest. This difference in
thresholding is important because most of the flux emergence and
helicity flux in the simulation occurs along the neutral line in
regions of weak vertical field that are missed with vertical field
thresholding used in the original study. Note that the modified
thresholding mask contains weak vertical field regions and contains
substantially more points than the roughly 3800 used in
\cite{Welsch2007}. The difference between the helicity flux in this
comparison region and the total simulation is 0.023\%.  \par
Since the DAVE4VM is a local optical flow method that determines the
plasma velocities within a windowed subregion by constraining the
local velocity profile, the choice of window size is a crucial issue
for estimating velocities accurately.  The window must be large enough
to contain enough structure to uniquely determine the coefficients of
the flow profile and resolve the aperture problem, but not so large as
to violate the affine velocity profile~(\ref{eqn:affine}) which is
only valid locally \cite[See][for discussion of the aperture problem
in the context of the DAVE]{Schuck2006a}.  In \cite{Welsch2007}, the
optimal window size was chosen for the DAVE by examining the Pearson
correlation and slope between $\grad_h\cdot\left(\u\,B_z\right)$ and
$\Delta B_z/\Delta t$ from ANMHD.  If the method satisfies the
induction equation exactly everywhere, the Pearson correlation and
slope would both be equal to $-1$. However both the DAVE and DAVE4VM
were conceived with the recognition that real magnetograms contain
noise and should not satisfy the magnetic induction equation exactly;
these methods satisfy the induction equation statistically within the
window by minimizing the mean squared deviations from the ideal
induction equations. Consequently, how well these methods satisfy the
magnetic induction equation globally or over a subset of pixels can be
used to assess overall performance.  In \cite{Welsch2007}, the DAVE
was ``optimized'' over a subset of pixels with
$\left|B_z\right|>370$~G that did not include the weak vertical field
regions discussed in this study. Using the Pearson correlation and
slope, an asymmetric window of $21\times39$ performed the best on the
ANMHD data.  For the present study, the DAVE was ``re-optimized'' over
the new criteria $\left|\B\right|>370$~G to provide an
``apples-to-apples'' comparison. However, I emphasize that generally
this optimization cannot be carried out for the DAVE because this
method was proposed for deriving flux-transport velocities from
line-of-sight magnetograms. In this situation, the threshold mask can
only be applied on the line-of-sight component \textit{as a proxy} for
the vertical magnetic field. Nonetheless, as a practical matter,
understanding the accuracy limitation of the line-of-sight method in
comparison to the vector method when synthetic vector magnetograms are
available will reveal the relative reliability of helicity flux
studies that used optical flow methods that rely exclusively on the
line-of-sight magnetic field under the ``best case scenario'' for the
tracking methods: the true vertical magnetic field is tracked and
regions that contain interesting physics are known
\textit{a-priori}. By using the ``ground truth'' vertical magnetic
field, the evaluation is biased to \textit{favor} the performance of
the DAVE method over what would be possible under realistic conditions
where only the line-of-sight magnetic field is available.\par
\newlength{\size}
\setlength{\size}{2.9in}
\newcommand{\hspc}{\hskip0.1in}
\begin{figure}
\centerline{\includegraphics[width=\size]{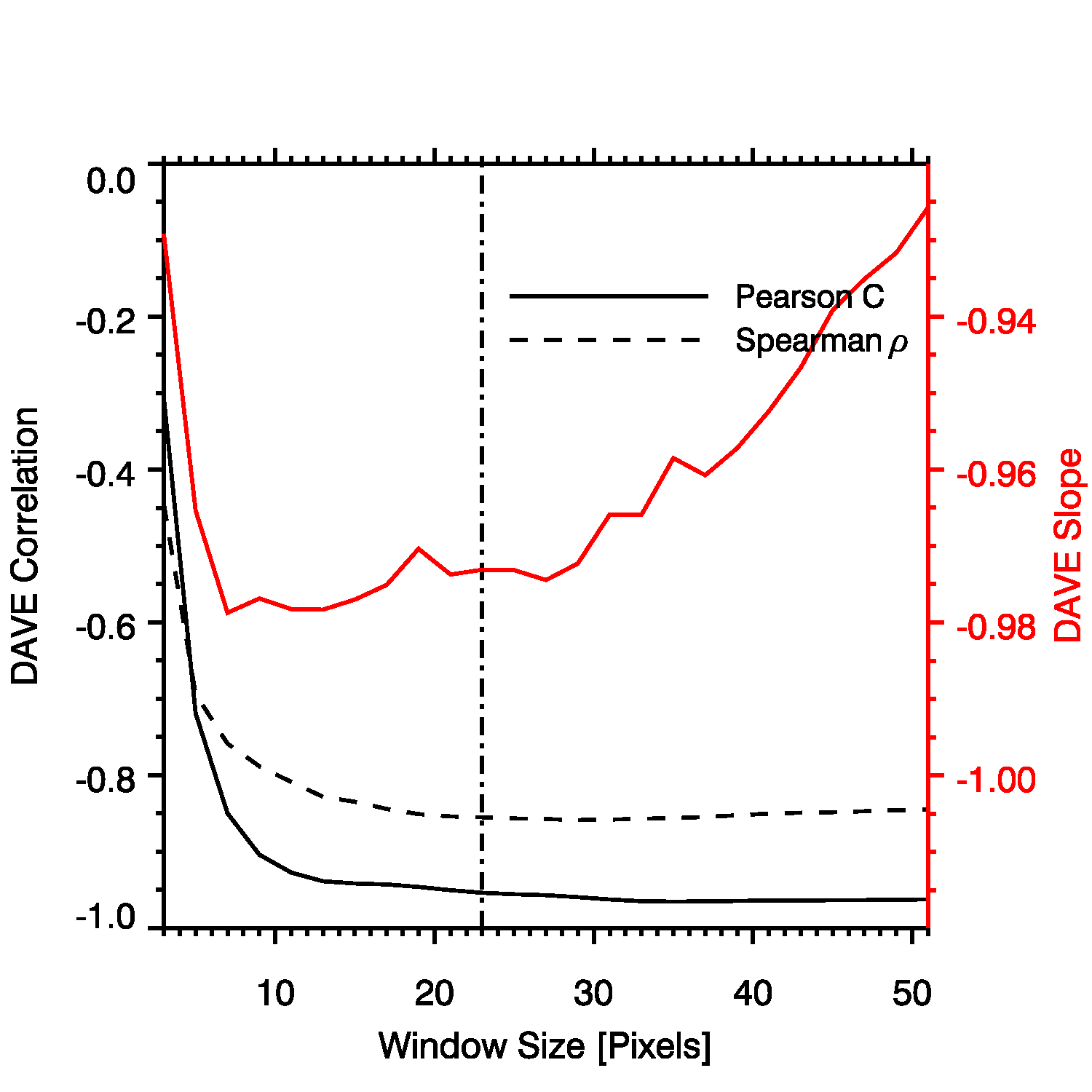}\hspc\includegraphics[width=\size]{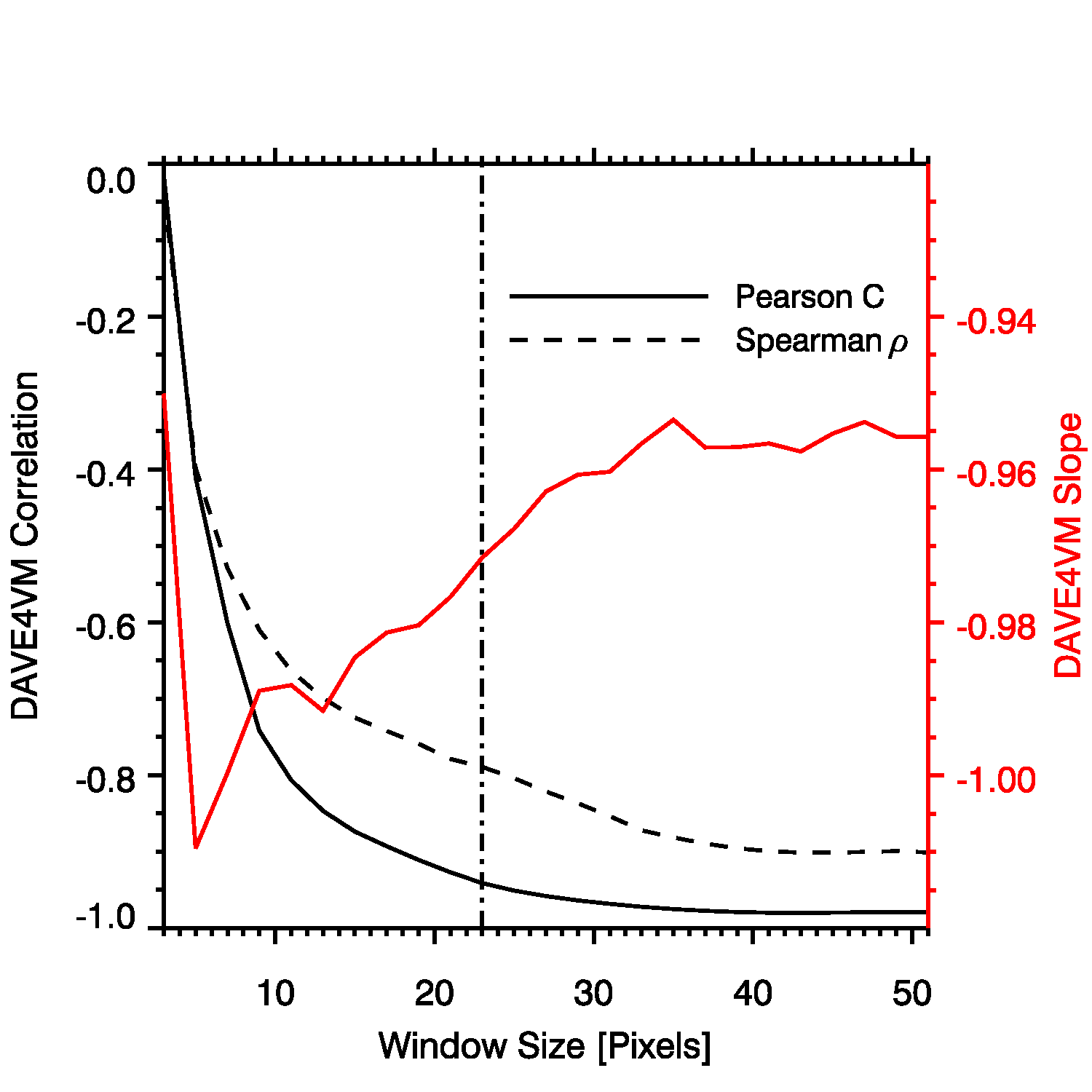}}
\centerline{\includegraphics[width=\size]{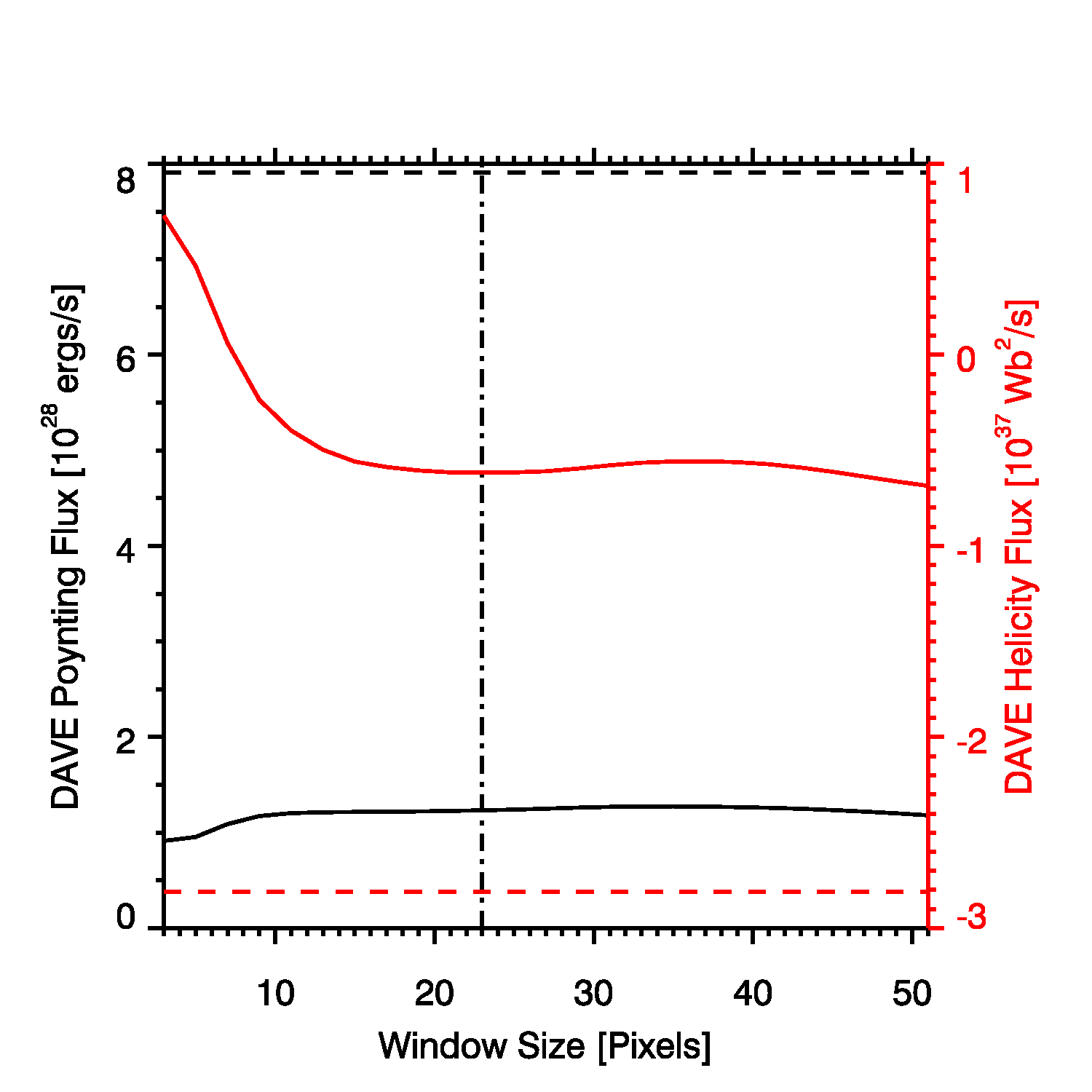}\hspc\includegraphics[width=\size]{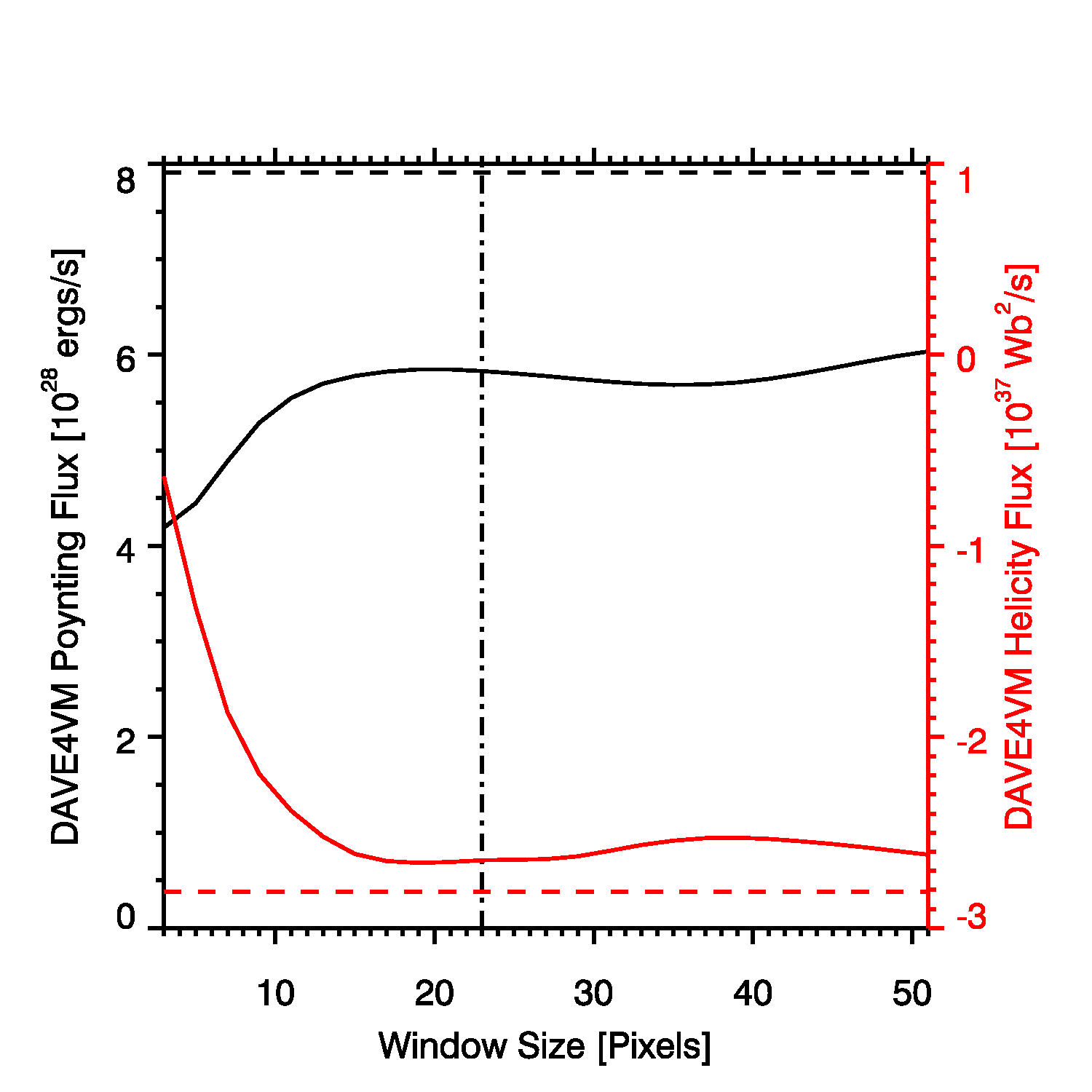}}
\caption{Optimization curves for the DAVE (\textit{left}) and DAVE4VM
  (\textit{right}). (top) The Spearman rank order (\textit{dashed black curve}) and Pearson (\textit{solid black
    curve}) correlations and slope (\textit{red curve}) between
  $\grad_h\cdot\left(\u\,B_z\right)$ and $\Delta B_z/\Delta t$ for
  DAVE and $\grad_h\cdot\left(\uF\,B_z\right)$ and $\Delta B_z/\Delta
  t$ for DAVE4VM as a function of window size. (bottom) The power
  (black) and helicity (red) as a function of window size; this
  assumes $\u=\uF$ for DAVE. The horizontal black and red dashed lines
  correspond to the ground truth power and helicity rate through the
  simulation slice from ANMHD. The vertical dot-dashed lines indicate
  the ``optimal'' window size of 23 pixels chosen for both DAVE and
  DAVE4VM.\label{fig:cal}}
\end{figure}
\begin{deluxetable}{cccccccc}
\tablecaption{Comparison between the DAVE and
DAVE4VM over the 7013 pixels that satisfy $\left|\B\right|>370$~G in
Figure~\ref{fig:ftv}. This mask contain regions of weak vertical field
not considered in \cite{Welsch2007}.\label{tab:weak}}
\tablehead{&&\multicolumn{3}{c}{DAVE (Assuming $\u=\uF$)}&\multicolumn{3}{c}{DAVE4VM}\\
\multicolumn{2}{c}{Quantities}&\colhead{Spearman}&\colhead{Pearson}&\colhead{Slope}&\colhead{Spearman}&\colhead{Pearson}&\colhead{Slope}}
\startdata
                                             $u_x\,B_z$&                                             $U_x\,B_z$& 0.34& 0.57& 0.15& 0.88& 0.89& 0.80\\
                                             $u_y\,B_z$&                                             $U_y\,B_z$& 0.70& 0.76& 0.71& 0.94& 0.90& 0.89\\
                                         $v_{\perp{x}}$&                                         $V_{\perp{x}}$& 0.87& 0.85& 0.91& 0.89& 0.88& 0.94\\
                                         $v_{\perp{y}}$&                                         $V_{\perp{y}}$& 0.93& 0.92& 1.20& 0.94& 0.94& 1.00\\
                                         $v_{\perp{z}}$&                                         $V_{\perp{z}}$& 0.17& 0.28& 0.07& 0.80& 0.80& 0.79\\
                    $\grad_h\cdot\left(\uF\,B_z\right)$&                                  $\Delta B_z/\Delta t$&-0.85&-0.95&-0.97&-0.79&-0.94&-0.97\\
                                         $e_{\perp{x}}$&                                         $E_{\perp{x}}$& 0.70& 0.76& 0.71& 0.94& 0.90& 0.89\\
                                         $e_{\perp{y}}$&                                         $E_{\perp{y}}$& 0.34& 0.57& 0.15& 0.88& 0.89& 0.80\\
                                         $e_{\perp{z}}$&                                         $E_{\perp{z}}$& 0.96& 0.96& 1.20& 0.94& 0.97& 1.00\\
                                                  $s_z$&                                                  $S_z$& 0.20& 0.12& 0.04& 0.88& 0.83& 0.71\\

\enddata
\end{deluxetable}
Five different criteria were used to optimize window selection.  Only
symmetric windows were considered and some improvement in the results
can be achieved by implementing asymmetric windows as in
\cite{Welsch2007}. Figure~\ref{fig:cal} shows the optimization curves
for the DAVE (\textit{left}) and DAVE4VM (\textit{right}). The top
plots show the Spearman rank order ($\RO$; \textit{dashed black}) and
Pearson ($\PCC$; \textit{solid black curve}) correlations and slope
($S$; \textit{red curve}) between $\grad_h\cdot\left(\u\,B_z\right)$
and $\Delta B_z/\Delta t$ for DAVE and
$\grad_h\cdot\left(\uF\,B_z\right)$ and $\Delta B_z/\Delta t$ for
DAVE4VM using pixels that satisfy $\left|\B\right|>370$~G. The
gradients were computed with 5-point least-squares optimized
derivatives \cite[]{Jahne2004}. Window sizes between 15 and 30 pixels
provide the best balance for achieving performance approaching
$\PCC=\RO=\Slopes=-1$.  Increasing the window size beyond 30 pixels
continues to improve the Spearman and Pearson correlations but with
diminishing returns while the slopes degrade significantly. The bottom
plots show the power (\textit{black curve}) and helicity rate
(\textit{red curve}) as a function of window size for the DAVE and
DAVE4VM. For DAVE these calculations require the assumption that
$\u=\uF$.  The horizontal black and red dashed lines correspond to the
ground truth Poynting and helicity flux from ANMHD.  The magnitude of
these fluxes are maximum near 20~pixels with roughly uniform
performance between 15 and 30 pixels. A window size\footnote{Only
  windows with odd numbers of pixels on each side of the window are
  possible in these implementations of the DAVE and DAVE4VM.} of 23
pixels was chosen for both the DAVE and DAVE4VM as indicated by the
vertical dot-dashed lines in Figure~\ref{fig:cal}. These objective
metrics for evaluation of global performance can be implemented
without knowledge of ground truth. Only future tests with more
realizations of synthetic data can reveal whether they are robust
metrics for optimizing window choice. Table~\ref{tab:weak} presents a
summary of the correlation coefficients on the mask
$\left|\B\right|>370$~G characterizing the accuracy of the DAVE and
DAVE4VM for the quantities discussed in this section.\par
\begin{deluxetable}{lcccccccccc}
\tablecaption{Comparison of accuracy of the velocity estimates between
the DAVE and DAVE4VM over the 7013 pixels that satisfy
$\left|\B\right|>370$~G in Figure~\ref{fig:ftv}. This mask contains
regions of weak vertical field not considered in
\cite{Welsch2007}. Here
$\left\langle\left|\delta\smash{\widetilde{\f}}\right|\right\rangle$ is the
average fractional error,
$\left\langle\delta\left|\smash{\widetilde{\f}}\right|\right\rangle$ is the
average error in magnitude, $C_{\mathrm{vec}}$ is the vector
correlation, $C_{\mathrm{CS}}$ is the direction
correlation, and $\left\langle\cos\theta\right\rangle_W$ is weighted direction cosine.\label{tab:vweak}}
\tablehead{&\multicolumn{5}{c}{DAVE (Assuming $\u=\uF$)}&\multicolumn{5}{c}{DAVE4VM}\\
\colhead{\f}&\colhead{$\left\langle\left|\delta\smash{\widetilde{\f}}\right|\right\rangle$}&\colhead{$\left\langle\delta\left|\smash{\widetilde{\f}}\right|\right\rangle$}&\colhead{$C_{\mathrm{vec}}$}&\colhead{$C_{\mathrm{CS}}$}&\colhead{$\left\langle\cos\theta\right\rangle_W$}&
\colhead{$\left\langle\left|\delta\smash{\widetilde{\f}}\right|\right\rangle$}&\colhead{$\left\langle\delta\left|\smash{\widetilde{\f}}\right|\right\rangle$}&\colhead{$C_{\mathrm{vec}}$}&\colhead{$C_{\mathrm{CS}}$}&\colhead{$\left\langle\cos\theta\right\rangle_W$}}
\startdata
  $\uF\,B_z$& 0.83$\pm$0.25&-0.47$\pm$0.34&0.61&0.52&0.71& 0.38$\pm$0.26&-0.09$\pm$0.21&0.91&0.92&0.95\\
  $\v_\perp$& 0.72$\pm$0.30& 0.09$\pm$0.27&0.81&0.77&0.87& 0.40$\pm$0.25& 0.01$\pm$0.14&0.93&0.90&0.92\\

\enddata
\end{deluxetable}
\subsection{Flux Transport Vectors and Plasma Velocities}
\begin{figure}
\centerline{\includegraphics[width=\size]{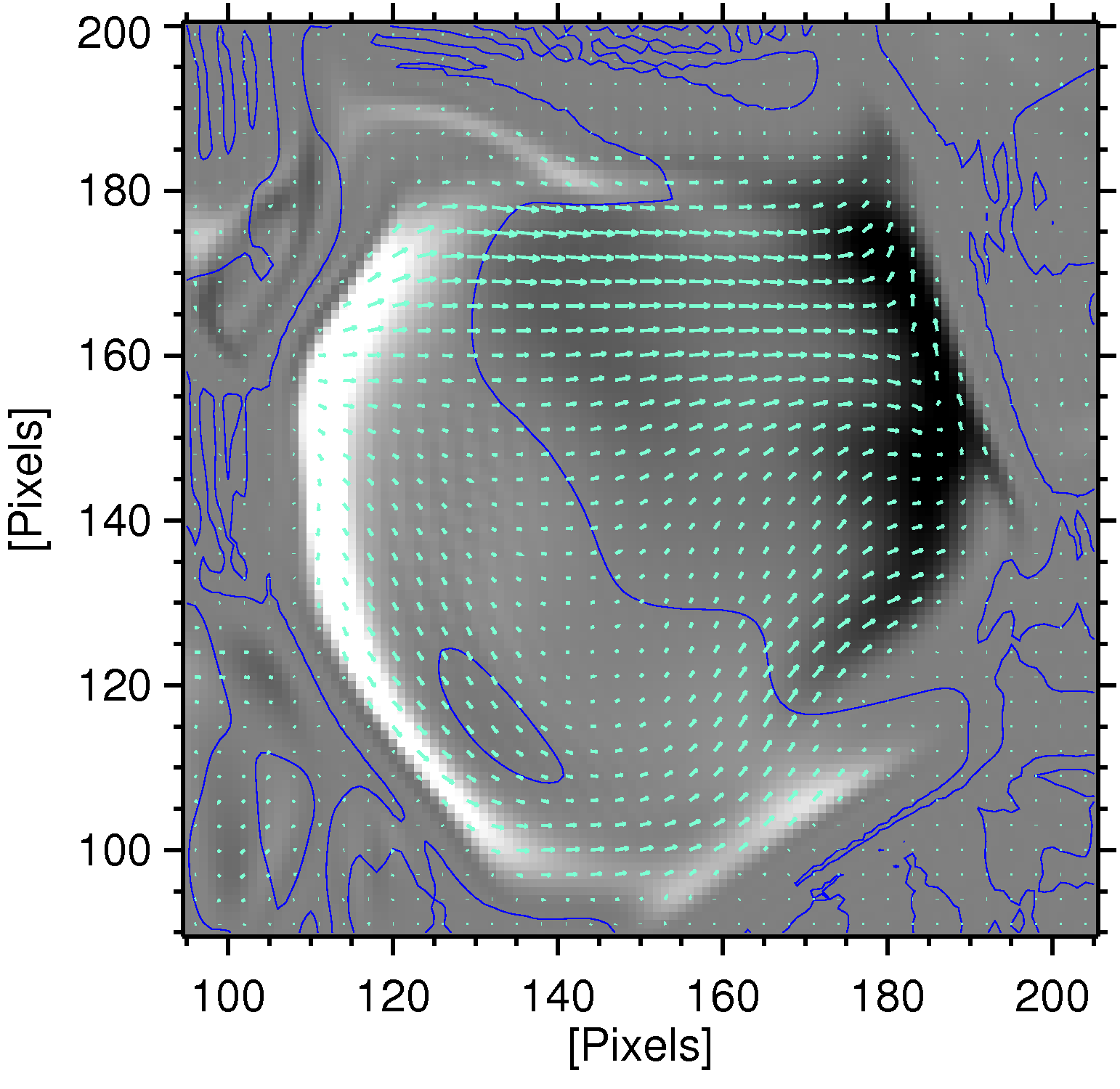}\hspc\includegraphics[width=\size]{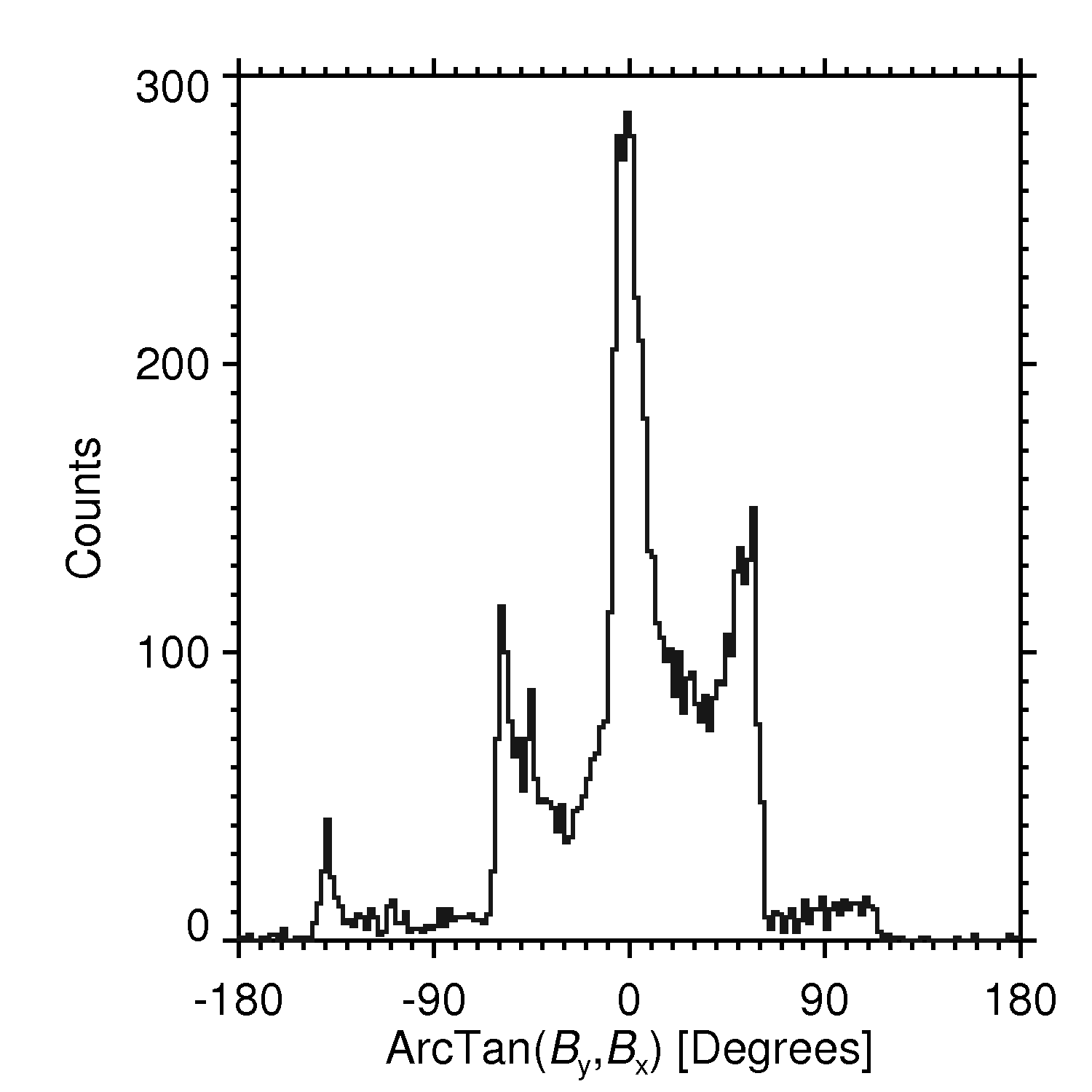}}
\centerline{\includegraphics[width=\size]{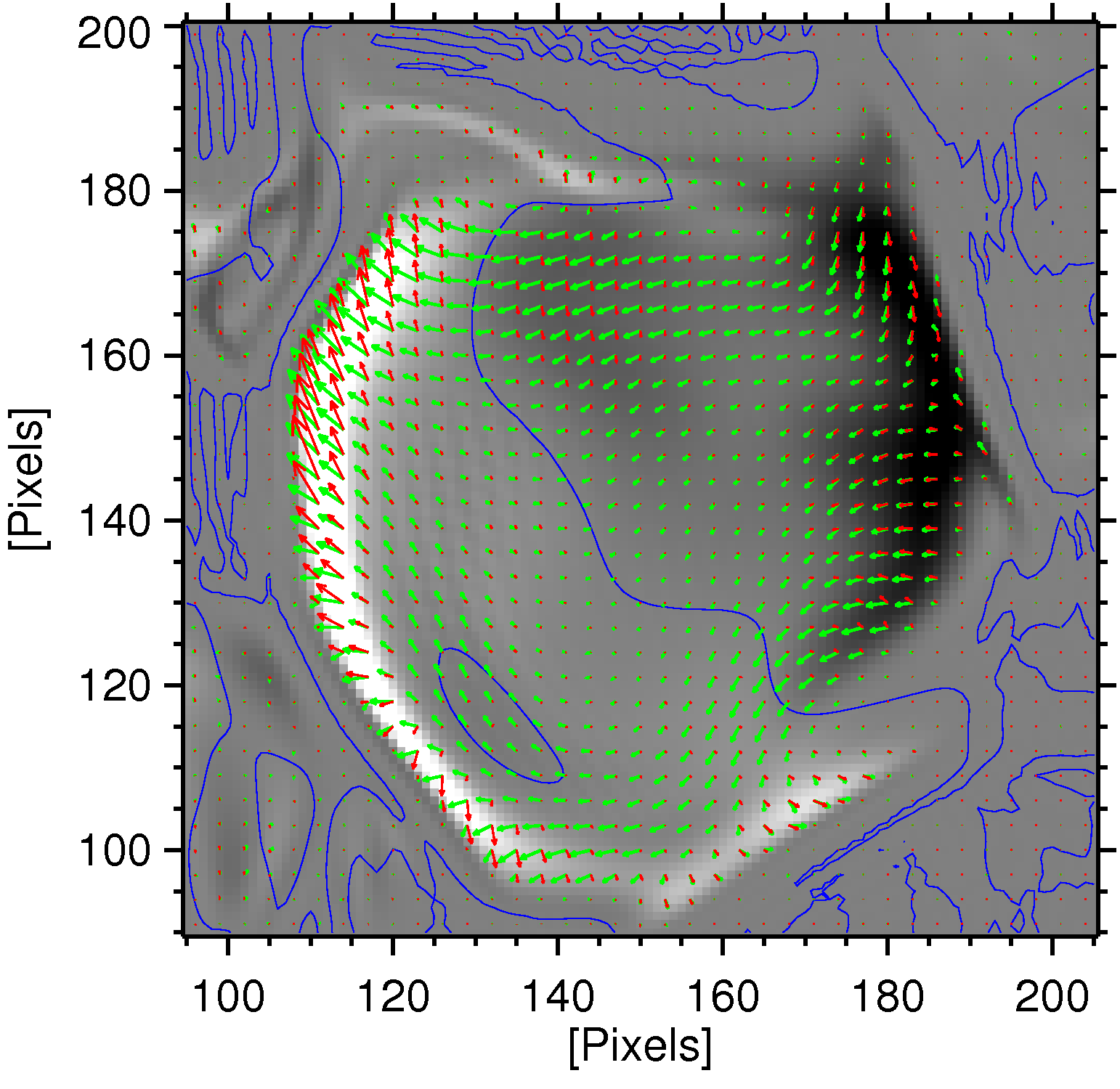}\hspc\includegraphics[width=\size]{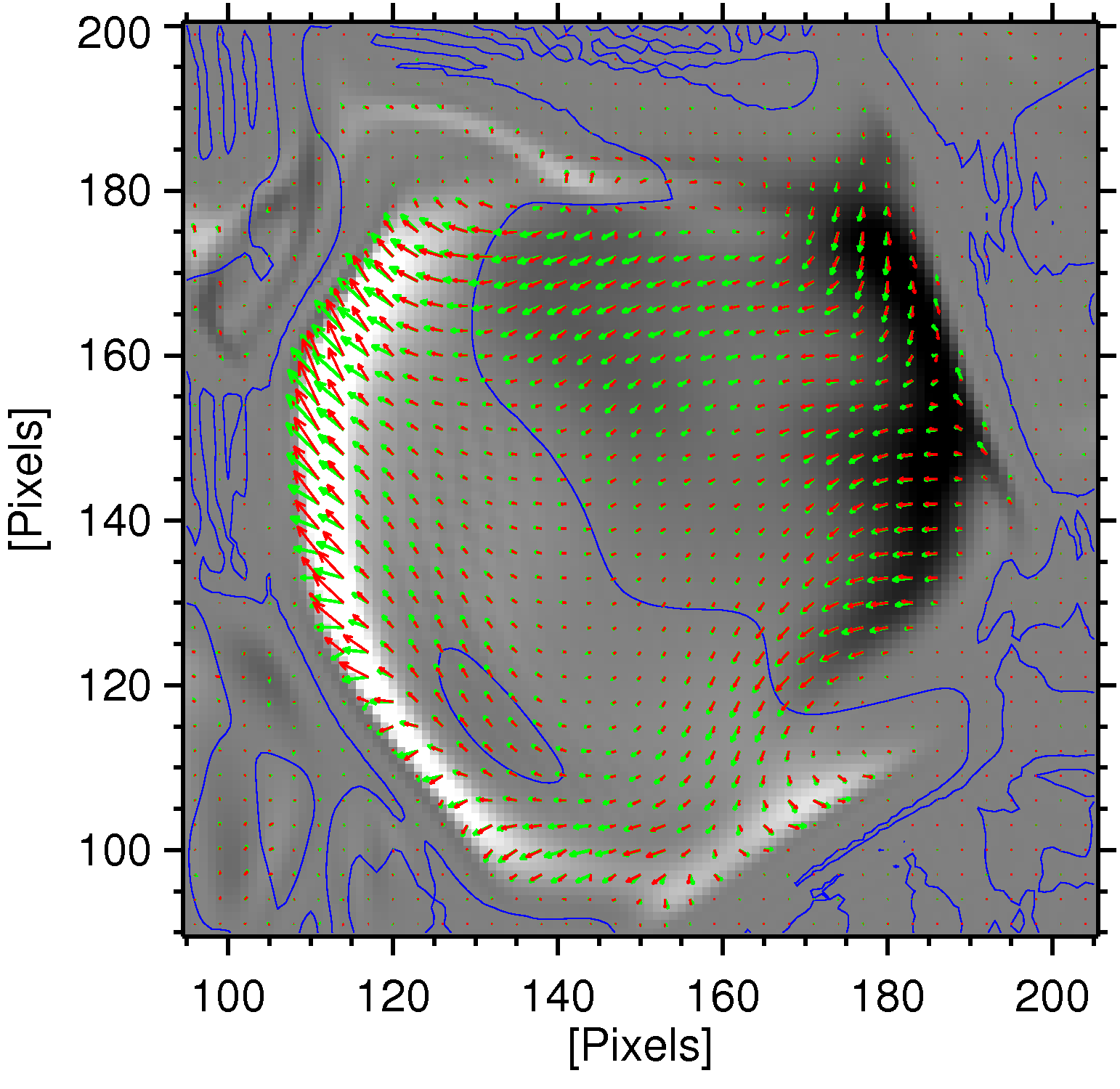}}
\centerline{\includegraphics[width=\size]{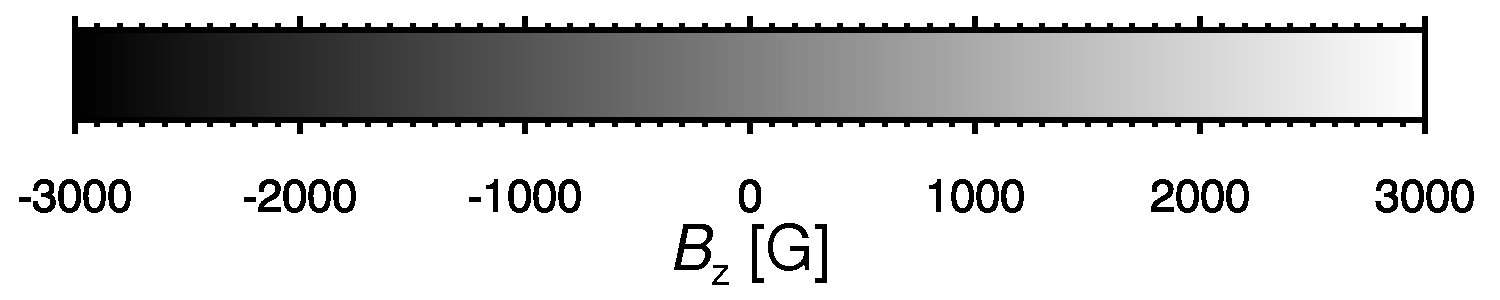}}
\caption{Top left: Grey scale image of the vertical magnetic field
$B_z$ overlaid with the horizontal magnetic field vectors $\B_h$ in
aqua.  Blue contours indicate smoothed neutral lines.  Top right:
Distribution of angles for $\B_h$ in the horizontal plane.  Bottom:
(red arrows) $\u\,B_z$ for the DAVE (\textit{left}) and $\uF\,B_z$ for the
DAVE4VM (\textit{right}) plotted over ANMHD's flux transport vectors $\UF\,B_z$
(green arrows). Vectors are shown only for pixels in which
$|B|>370$~G, and for clarity, only every third vector is displayed.
\label{fig:ftv}}
\end{figure}
Determining the flux transport vectors $\f=\uF\,B_z$ is equivalent to
determining the perpendicular plasma velocities $\v_\perp$. The
accuracy of the flux transport vectors is critical for estimating
other MHD quantities: perpendicular plasma velocities, electric field,
helicity flux, Poynting flux, etc, since all of these
quantities may be derived directly from flux transport
vectors. 
\subsubsection{Flux Transport Velocities and Perpendicular Plasma Velocities}
The top left of Figure~\ref{fig:ftv} shows the region of interest from
the ANMHD simulations with grey scale image of vertical magnetic field
overlaid with the horizontal magnetic field vectors $\B_h$ in
aqua. The blue contours indicate smoothed neutral lines. The top right
shows the distribution of angles for $\B_h$ in the horizontal
plane. The horizontal magnetic field is largely aligned with the
$\xhat$\--axis as indicated by the aqua vectors in the left panel and
the strong peak in the histogram near
$\arctan\left(B_y,B_x\right)\approx0^\circ$ in the right
panel.\footnote{$\arctan\left(y,x\right)\equiv\arctan\left(y/x\right)$.}
There is also significant alignment of the magnetic field with
$\pm60^\circ$ and alignment of weak fields with $-140^\circ$.  The
bottom panels show $\u\,B_z$ from DAVE (\textit{left}) and $\uF\,B_z$ from
DAVE4VM in red (\textit{right}) and the flux transport vectors $\Ftv=\UF\,B_z$
from ANMHD in green. The improvement between the DAVE and DAVE4VM is
manifest \---- finding a region where the DAVE4VM performs
qualitatively worse than the DAVE is difficult. The DAVE4VM performs
the worst in the region $140\--160\times150\--170$ where the flux
transport vectors run roughly anti-parallel to the horizontal magnetic
field and there is little structure in the vertical component.\par
\begin{figure}
\centerline{\includegraphics[width=\size]{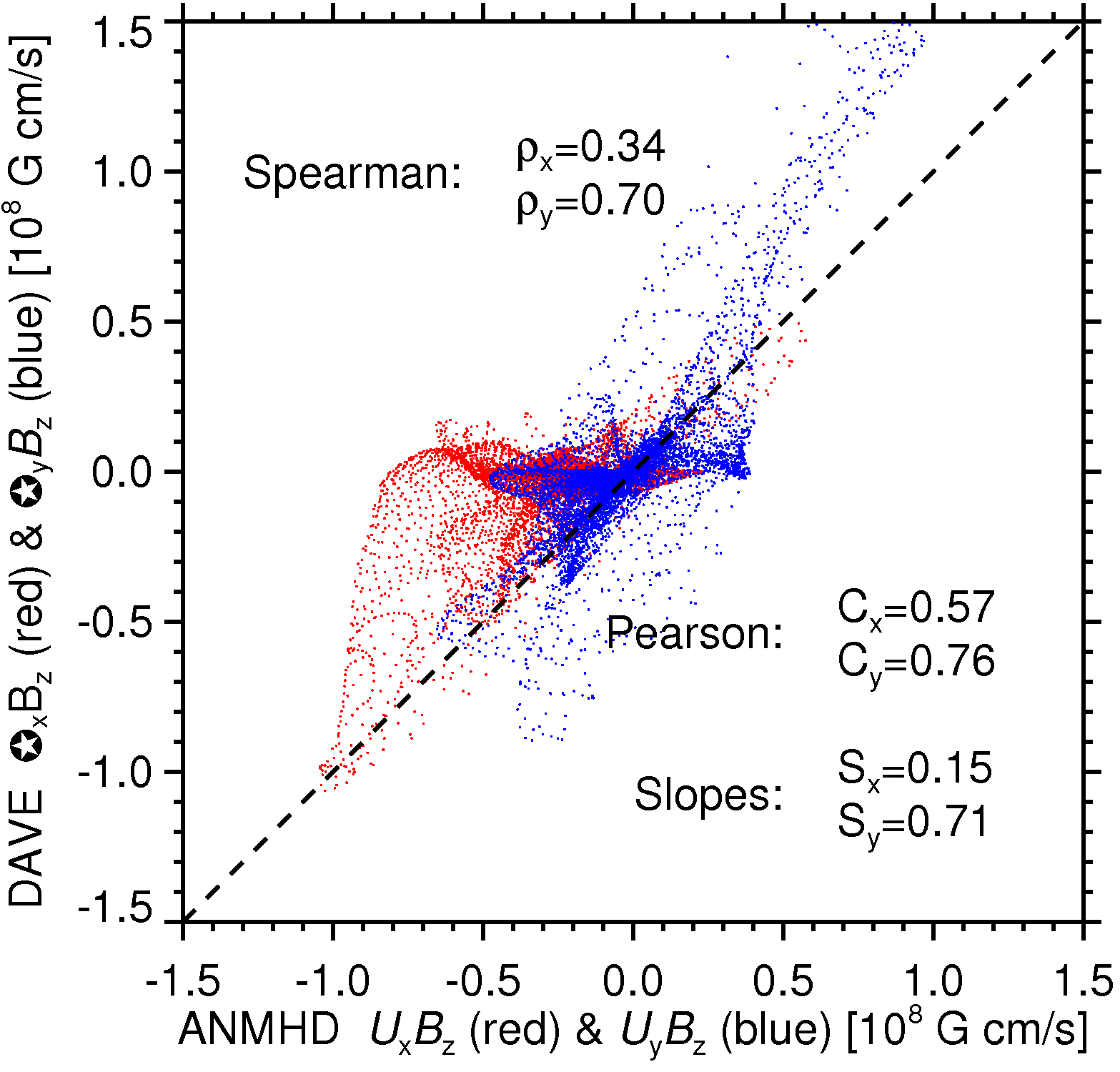}\hspc\includegraphics[width=\size]{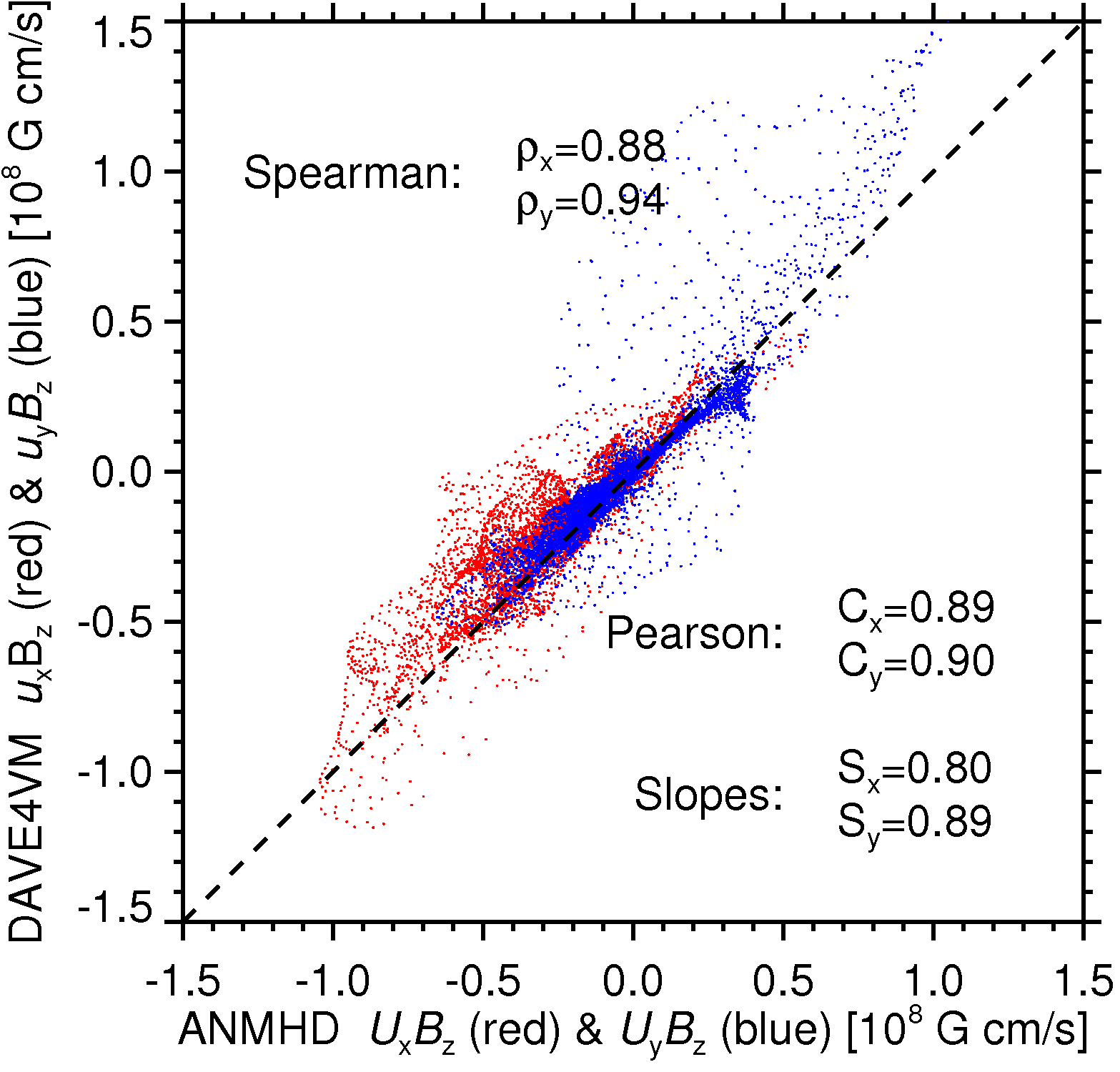}}
\caption{Scatter plots of $\u\,B_z$ for the DAVE (\textit{left}) and $\uF\,B_z$
for the DAVE4VM (\textit{right}) versus ANMHD's flux transport vectors,
$\Ftv=\UF\,B_z$. Red and blue are used to distinguish $x$- and
$y$-components, respectively.  The nonparametric Spearman rank-order
correlation coefficients ($\RO$), Pearson correlation coefficients
($\PCC$), and slopes ($\Slopes$) estimated by the least absolute
deviation method are shown for both components of the flux transport
vectors.\label{fig:ftv_corr}}
\end{figure}
Figure~\ref{fig:ftv_corr} shows scatter plots of the $\u\,B_z$ from
DAVE (\textit{left}) and $\uF\,B_z$ from DAVE4VM (\textit{right}) versus the flux
transport vectors $\UF\,B_z$ from ANMHD. Red and blue are used to
distinguish $x$- and $y$-components, respectively.  The nonparametric
Spearman rank-order correlation coefficients ($\RO$), Pearson
correlation coefficients ($\PCC$), and slopes ($\Slopes$) estimated by
the least absolute deviation method are shown for both components of
the flux transport vectors. Both visually and quantitatively the
DAVE4VM's correlation with ANMHD is much higher than the DAVE's. The
correlation coefficients even match or exceed the correlation
coefficients for the flux transport vectors from the DAVE and MEF
reported for the restricted mask $\left|B_z\right|>370$~G in
\cite{Welsch2007}. In particular, DAVE does not accurately estimate
the flux transport vectors in the $\xhat$-direction. The correlation
coefficients for this $\xhat$-component of the flux transport vectors
are $\RO_x=0.34$ and $\PCC_x=0.57$ with a slope of $\Slopes_x=0.15$.
Since that the $\xhat$-direction is the predominant direction of the
horizontal magnetic for the ANMHD data, the low correlation
coefficients suggest that DAVE is insensitive to flux emergence which
is proportional to $v_z\,\B_h$. This will be discussed further in
\S~\ref{sec:conclusions}. \par
\begin{figure}
\centerline{\includegraphics[width=\size]{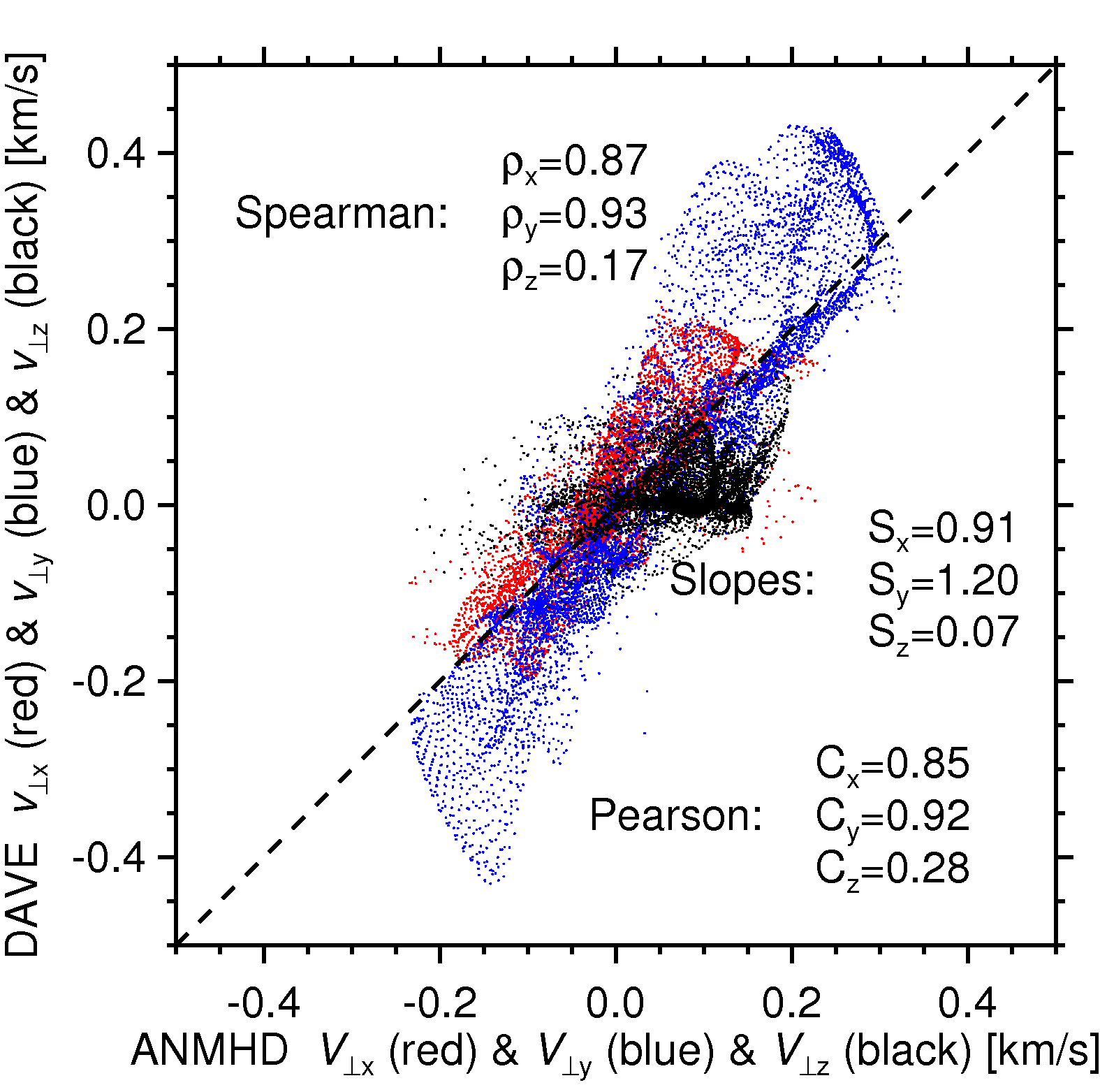}\hspc\includegraphics[width=\size]{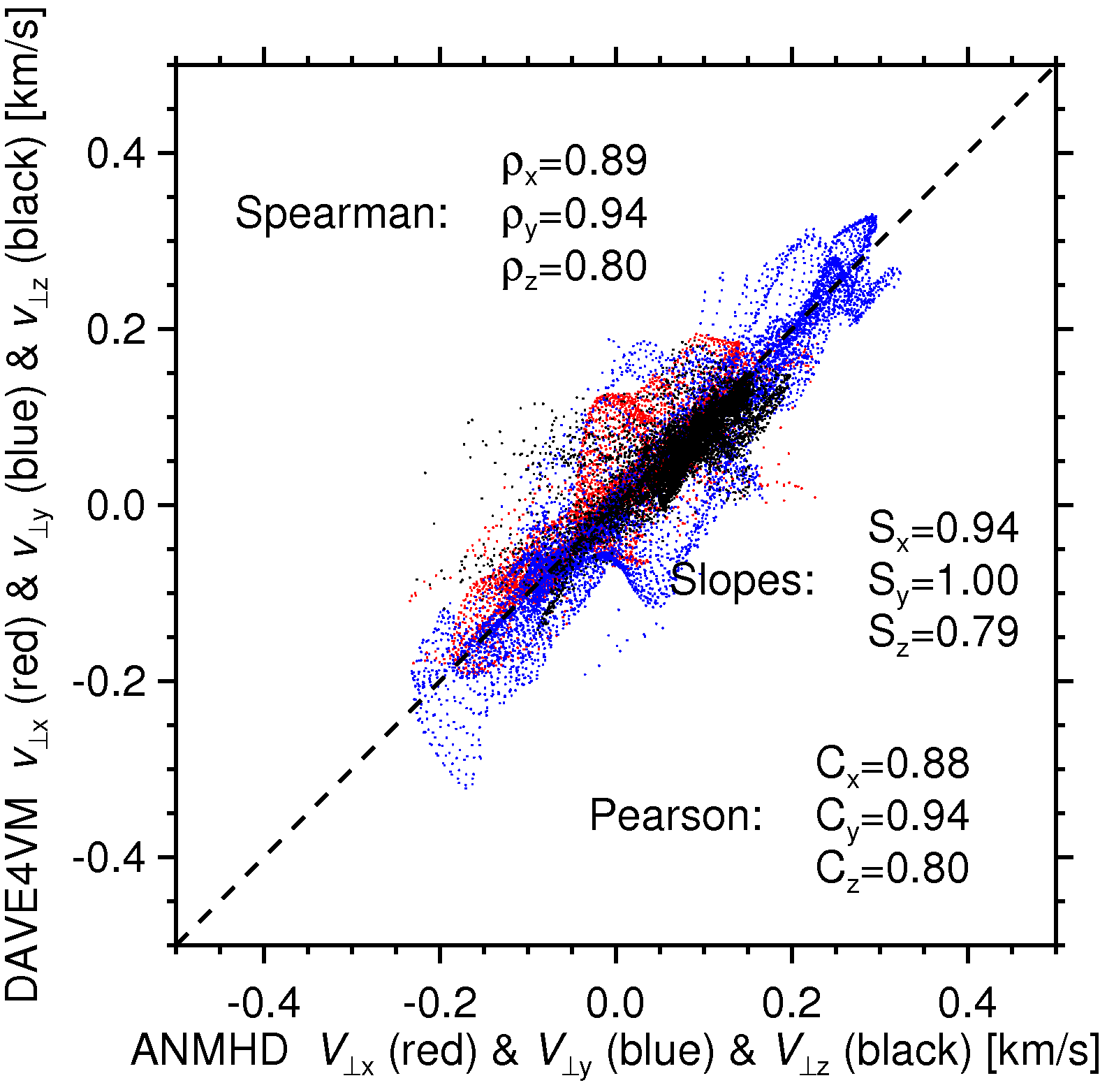}}
\caption{Scatter plots of the estimated perpendicular plasma
  velocities $\v_\perp$ from DAVE assuming $\u=\uF$ (\textit{left}) and DAVE4VM
  (\textit{right}) versus the perpendicular plasma velocities $\V_\perp$ from
  ANMHD. Red, blue, and black correspond to the $x$-, $y$- and
  $z$-components respectively.  The nonparametric Spearman rank-order
  correlation coefficients ($\RO$), Pearson correlation coefficients
  ($\PCC$), and slopes ($\Slopes$) are shown for each component of the
  perpendicular plasma velocities.\label{fig:vp_corr}}
\end{figure}
Figure~\ref{fig:vp_corr} shows scatter plots of the estimated
perpendicular plasma velocities $\v_\perp$ from the DAVE assuming
$\u=\uF$ (\textit{left}) and DAVE4VM (\textit{right}) versus ANMHD's perpendicular
plasma velocities $\V_\perp$. Red, blue, and black correspond to the
$x$-, $y$-, and $z$-components respectively. The nonparametric
Spearman rank-order correlation coefficients ($\RO$), Pearson
correlation coefficients ($\PCC$), and slopes ($\Slopes$) estimated by
the least absolute deviation method are shown for each component of
the perpendicular plasma velocities. The DAVE4VM's correlation
coefficients match or exceed the correlation coefficients for the
perpendicular plasma velocities from the DAVE. Particularly striking
is the DAVE4VM's relatively higher correlation for the perpendicular
vertical plasma velocity $\v_{\perp{z}}$ which \textit{exceeds} the
correlation for the DAVE by roughly $0.5\--0.6$. The improvement in the
DAVE4VM's estimate is due to the \textit{explicit} inclusion of
horizontal magnetic fields and vertical flows.
The flux transport and perpendicular plasma velocity estimates are
further quantified by considering the metrics used by
\cite{Schrijver2006}, \cite{Welsch2007}, and \cite{Metcalf2008}. The
fractional error between the estimated vector $\f$ and the true vector
$\Ftv$ at the $i$th pixel is
\begin{mathletters}
\label{eqn:vector:metrics}
\begin{equation}
\left|\delta\smash{\widetilde{\f}}_i\right|\equiv\frac{\left|\f_i-\Ftv_i\right|}{\left|\Ftv_i\right|},
\end{equation}
whereas the fractional error in magnitude is
\begin{equation}
\delta\left|\smash{\widetilde{\f}}_i\right|\equiv\frac{\left|\f_i\right|-\left|\Ftv_i\right|}{\left|\Ftv_i\right|}.
\end{equation}
\end{mathletters}
The moments of these error metrics or any quantity $q$ may be
accumulated over the $N$ pixels within the masks (either
$\left|\B\right|>370$~G or $\left|B_z\right|>370$~G) producing the
average
\begin{mathletters}
\begin{equation}
\left\langle{q}\right\rangle\equiv\frac{1}{N}\sum_{i=1}^N q_i,
\end{equation}
and the variance
\begin{equation}
\sigma^2_q\equiv\frac{1}{N-1}\sum_{i=1}^N 
\left(q_i-\left\langle q\right\rangle\right)^2.
\end{equation}
\end{mathletters}
For perfect agreement between the estimates and the ``ground truth''
from ANMHD,
$\left\langle\left|\delta\smash{\widetilde{\f}}\right|\right\rangle$,
$\left\langle\delta\left|\smash{\widetilde{\f}}\right|\right\rangle$,
and their associated variances would be zero. Two measures of
directional error are considered, the vector correlation
\begin{mathletters}
\label{eqn:correlation:metrics}
\begin{equation}
C_{\mathrm{vec}}=\frac{\left\langle\f\cdot\Ftv\right\rangle}{\sqrt{\left\langle\f^2\right\rangle\,\left\langle\Ftv^2\right\rangle}},
\end{equation}
and the direction correlation
\begin{equation}
C_{\mathrm{CS}}=\left\langle\frac{\f\cdot\Ftv}{\sqrt{\f^2\,\Ftv^2}}\right\rangle\equiv\left\langle\cos\theta\right\rangle.
\end{equation}
\end{mathletters}
Both metrics range from $-1$ for antiparallel vector fields, to $0$ for
orthogonal vector fields, and to $1$ for parallel vector fields (perfect
agreement).
Table~\ref{tab:vweak} shows these metrics for the DAVE and DAVE4VM
over the mask $\left|\B\right|>370$~G. The DAVE4VM has fractional
errors less than or equal to $0.4$ whereas the fractional errors for
the DAVE exceed $0.7$ for both the flux transport vectors and the
perpendicular plasma velocities. The average bias error in the
magnitude is improved for the DAVE4VM over the DAVE. For the flux
transport vectors the bias error in magnitude is $-0.09$ and $-0.47$ for
DAVE4VM and DAVE respectively which corresponds to a factor of $5$
improvement. For the plasma velocity, the bias error in magnitude is
0.01 and 0.09 for DAVE4VM and DAVE respectively which corresponds to
a factor of $9$ improvement. The vector correlation is larger for
DAVE4VM than for DAVE. For DAVE4VM $C_{\mathrm{vec}}\gtrsim0.9$ for
both the flux transport velocity and the perpendicular plasma
velocities. In contrast, for the DAVE there is a substantial
difference in the accuracy of the flux transport vectors with
$C_{\mathrm{vec}}=0.61$ and the perpendicular plasma velocities with
$C_{\mathrm{vec}}=0.81$.  Finally the directional errors are smaller
for the DAVE4VM than for the DAVE. For the DAVE4VM $\Ccs\gtrsim0.9$
for both the flux transport vectors and the perpendicular plasma
velocities. Again, for the DAVE there is a substantial difference in
the accuracy of the flux transport vectors with $\Ccs=0.52$ and the
perpendicular plasma velocities with $\Ccs=0.77$.\par
\begin{figure}
\centerline{\includegraphics[width=\size]{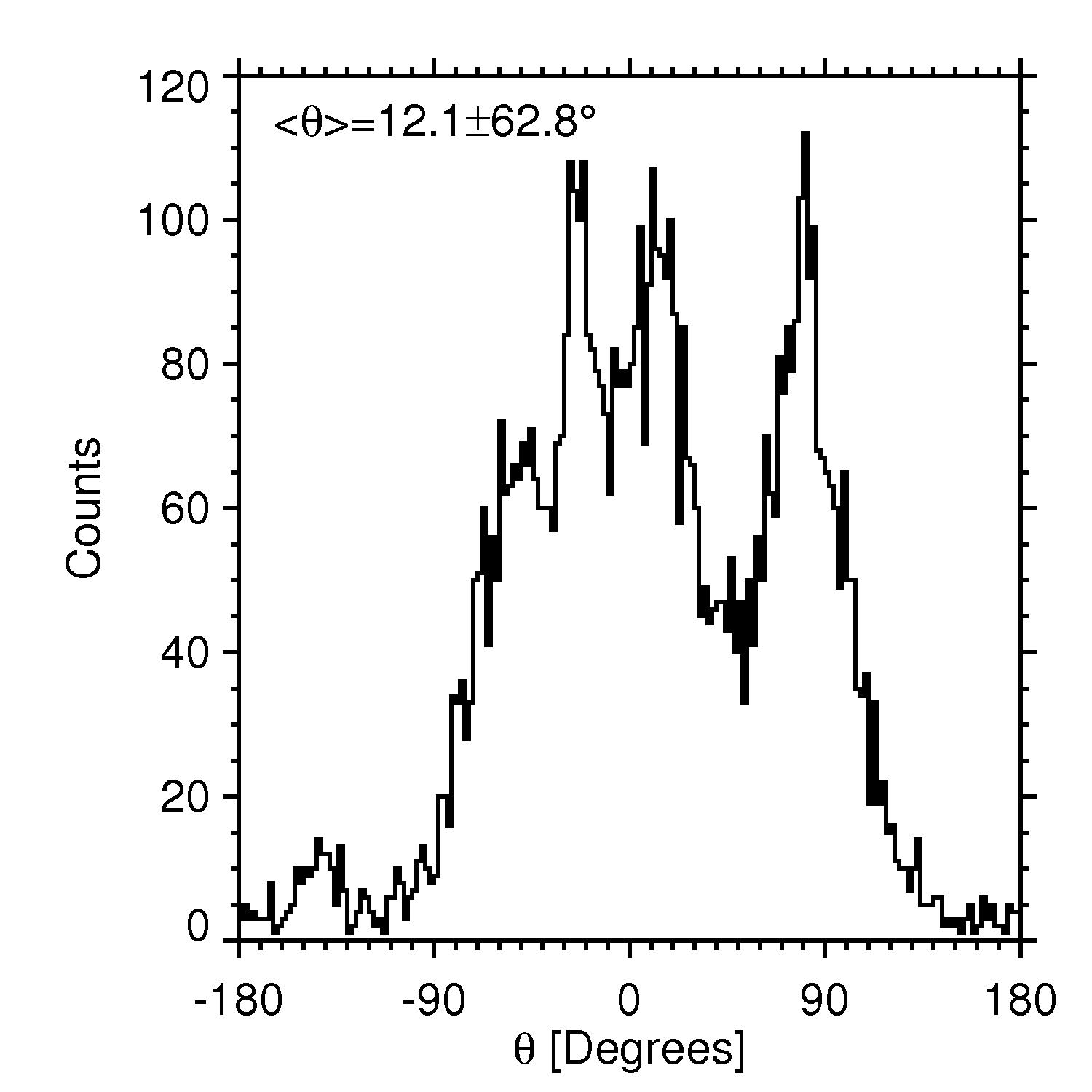}\hspc\includegraphics[width=\size]{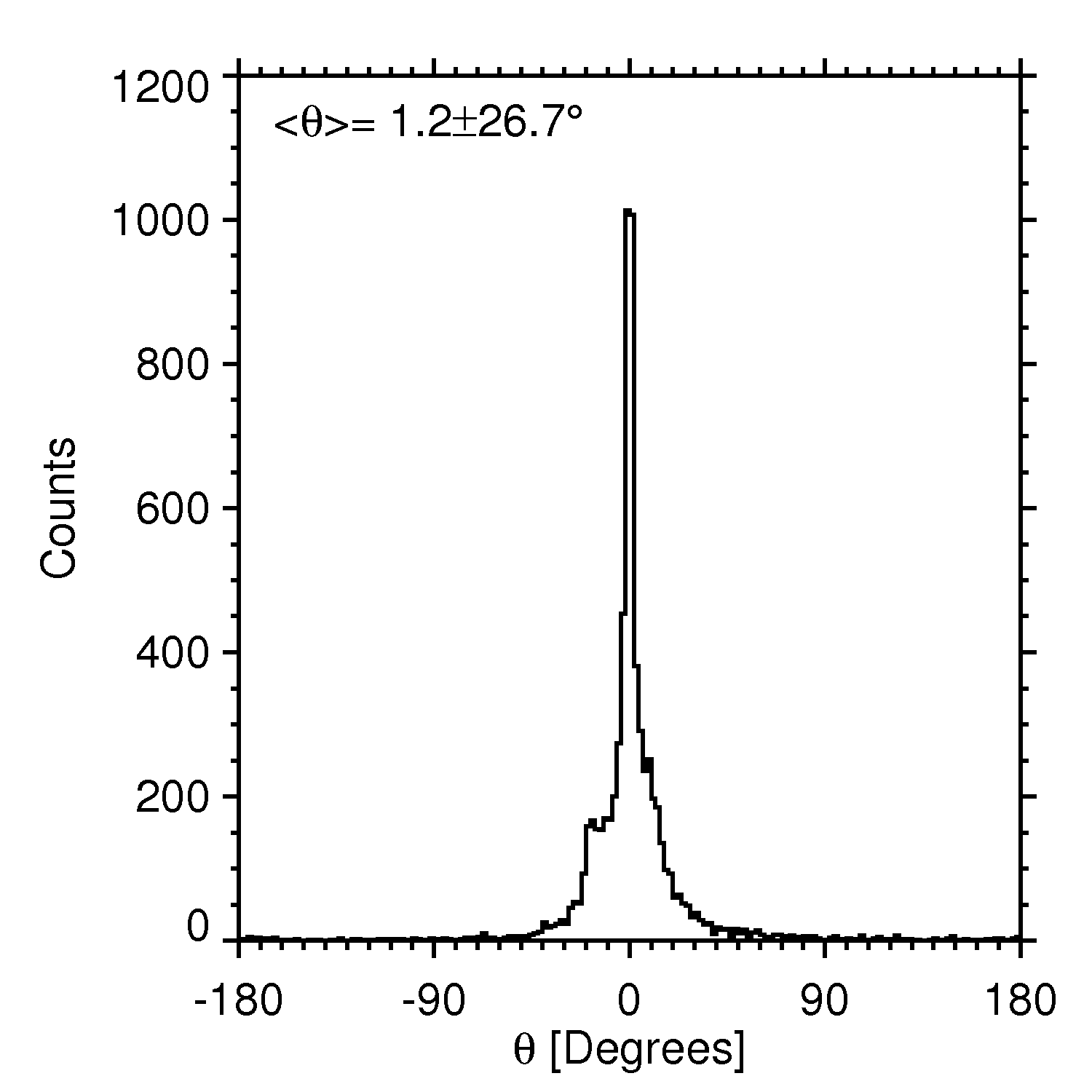}}
\caption{Histograms of the angles between $\u\,B_z$ and $\UF\,B_z$
for the DAVE (\textit{left}) and between $\uF\,B_z$ and $\UF\,B_z$ for the DAVE4VM (\textit{right}). The mean (bias) and standard
deviation are reported in the upper left-hand
corners.\label{fig:angle}}
\end{figure}
The direction correlation $\Ccs$ is difficult to translate into
average angular error because it is a nonlinear function of $\theta$
and does not indicate whether the estimated vectors ``lead'' or
``lag'' the ``ground truth'' on average.  For the 2D flux transport
vectors the moments of the distribution of angular errors
\begin{equation}
\theta=\arctan\left[\left(\uF\cross\UF\right)_z,\uF\cdot\UF\right],
\end{equation}
can be more informative. Figure~\ref{fig:angle} shows histograms of
the angles between $\uF\,B_z$ and $\UF\,B_z$ for the DAVE (\textit{left}) and
DAVE4VM (\textit{right}).  This is a quantitative estimate of the errors in
directions of the flux transport vectors. The DAVE4VM represents a
dramatic improvement over the DAVE. The DAVE4VM produces a nearly
unimodal distribution peaked near $0^\circ$, whereas the DAVE produces
a multi-peaked distribution with the largest peak at $80^\circ$ and a
variance that is more than twice as large as DAVE4VM. \par
Metrics such as~(\ref{eqn:vector:metrics})
and~(\ref{eqn:correlation:metrics}) weight all estimates equally. To
address this \cite{Metcalf2008} suggested weighting the errors. For
example, the weighted direction cosine between an inferred vector and
the ground truth vector may be defined as
\begin{equation}
\left\langle\cos\theta\right\rangle_W\equiv\frac{\left\langle{W}\,\cos\theta\right\rangle}{\left\langle{W}\right\rangle}
\end{equation}
where $W$ represents weights.  For the flux transport
velocity, the errors in the orientation of $\uF$ are more important where
$\left|\UF\,B_z\right|$ is large and less important where
$\left|\UF\,B_z\right|$ is small which suggests a weighting factor
$W_i=\left|\left(\UF\,B_z\right)\right|$. For perfect agreement
$\left\langle\cos\theta\right\rangle_W=1$. The weighted direction
cosines for the flux transport vectors and the plasma velocities are
reported in Table~\ref{tab:vweak}. Comparing the values of
$C_{\mathrm{CS}}$ and $\left\langle\cos\theta\right\rangle_W$
demonstrates that weighting the direction cosine improves the apparent
performance of DAVE but the results for DAVE4VM are essentially
unchanged. This suggests that DAVE4VM estimates velocities better than
DAVE in regions of weak flux transport.\par 
\subsubsection{Parallel Velocity\label{sec:vll}}
Under ideal conditions, the magnetic field is only affected by
$\grad\cross\left(\v\cross\B\right)$. Consequently, only the inductive
potential $\phi$ in~(\ref{eqn:Helmholtz}) may be uniquely determined
from the evolution of $\B$ alone. The electrostatic potential $\psi$
must \textit{and can} be estimated with additional judicious
assumptions.  These additional assumptions correspond to the minimum
photospheric velocity consistent with~(\ref{eqn:Helmholtz}) for the
global method MEF and to the prescribed affine form of the local
plasma velocity for the local method DAVE4VM. The constraint of the
affine velocity profile permits DAVE4VM to determine the electrostatic
potential $\psi$ from the \textit{nonlocal} structure of the inductive
potential $\phi$. DAVE4VM uses unambiguous ``pieces'' of the plasma
velocity within the window aperture to reconstruct the total plasma
velocity at the center of the aperture. Within the notation of the
Helmholtz decomposition, DAVE4VM estimates the local electrostatic
field from the structure of the nonlocal inductive field within the
aperture window by imposing a smoothness constraint on the velocity
(the affine velocity profile).  The accuracy of this estimate for the
electrostatic field depends on the validity of the local affine
velocity profile and the amount of structure in the local magnetic
field; local methods cannot detect motion in regions of uniform
magnetic field.\par
\begin{figure}
\centerline{\includegraphics[width=5in]{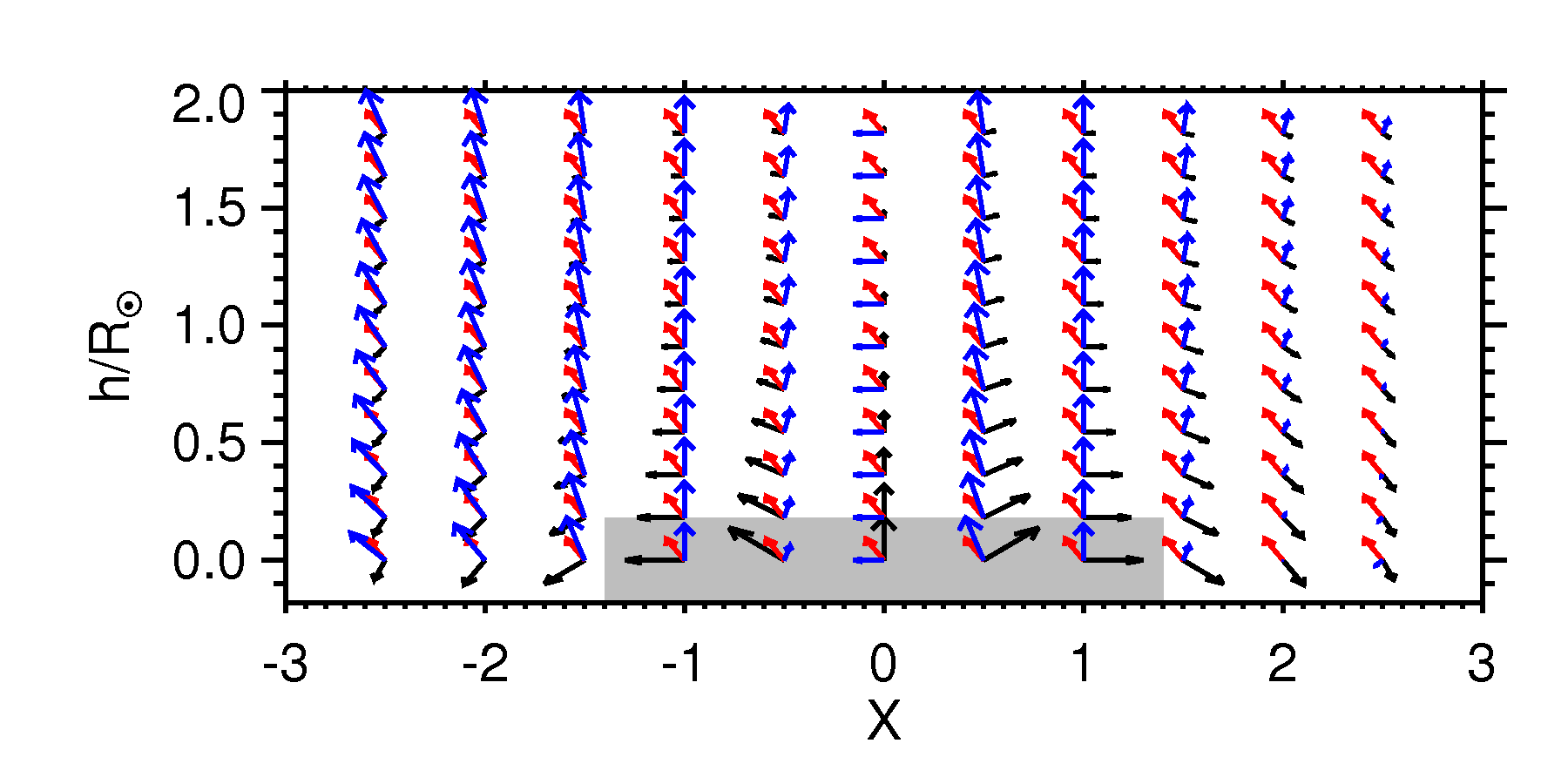}}
\caption{Schematic diagram of a uniform plasma flow across a diverging
  magnetic field above the photosphere at $h=0$. The black arrows
  indicate the strength and direction of the magnetic field, the red
  arrows indicate the direction of the spatially uniform total plasma
  velocity $\v$, and the blue arrows indicate the magnitude and
  direction of the perpendicular plasma velocity $\v_\perp$. The
  aperture in the photosphere is indicated by the gray
  box.\label{sec:parallel}}
\end{figure}
While researchers have widely recognized that estimating the
electrostatic potential $\psi$ from~(\ref{eqn:DAVE4VM}) requires
additional assumptions, they have not generally recognized that the
parallel velocity $v_\parallel$ may be estimated by the analogous
arguments. In the absences of a reference flow, the MEF constrains the
velocity to be perpendicular to the magnetic field:
$v_\parallel=0$. In contrast to the velocity estimated by DAVE4VM is
not constrained to be perpendicular to the local magnetic
field. Instead, DAVE4VM fits an affine velocity model to the magnetic
induction equation in an aperture window. This affine velocity model
couples the dynamics across pixels within the window aperture.  If
there is sufficient structuring in the direction of the magnetic field
within the aperture, i.e., the perpendicular plasma velocity points in
different directions at different pixels within the aperture, then
DAVE4VM can resolve the ambiguity in the field-aligned component of
the plasma velocity at the center of the aperture.\par
Consider the simplified two-dimensional situation illustrated by the
schematic diagram in Figure~\ref{sec:parallel} of spatially uniform
plasma flow across a diverging magnetic field above the photosphere at
$h=0$. The black arrows indicate the magnitude and direction of the
magnetic field, the red arrows indicate the direction of the spatially
uniform total plasma velocity $\v$, and the blue arrows indicate the
magnitude and direction of the perpendicular plasma velocity
$\v_\perp$. The aperture in the photosphere is indicated by the gray
box. Within the aperture, the perpendicular plasma velocity captures a
different component of the total plasma velocity at different
locations; this is a consequence of the structuring of the magnetic
field. Under the smoothness assumption of a uniform velocity profile,
the velocity along the magnetic field in the $\zhat$-direction {at
  $x=0$} may be determined from the components of the perpendicular
plasma velocity in the $\zhat$ direction \textit{at other locations
  within the aperture}.  Using the $N$ pixels in the window aperture
results in an overdetermined system for the \textit{total plasma
  velocity}:
\begin{mathletters}
\begin{equation}
\underbrace{\left[\begin{array}{cc}
\frac{B_z^2\left(\x_1\right)}{B^2\left(\x_1\right)}&-\frac{B_x\left(\x_1\right)\,B_z\left(\x_1\right)}{B^2\left(\x_1\right)}\\
-\frac{B_x\left(\x_1\right)\,B_z\left(\x_1\right)}{B^2\left(\x_1\right)}&\frac{B_x^2\left(\x_1\right)}{B^2\left(\x_1\right)}\\
\vdots&\vdots\\
\frac{B_z^2\left(\x_N\right)}{B^2\left(\x_N\right)}&-\frac{B_x\left(\x_N\right)\,B_z\left(\x_N\right)}{B^2\left(\x_N\right)}\\
-\frac{B_x\left(\x_N\right)\,B_z\left(\x_N\right)}{B^2\left(\x_N\right)}&\frac{B_x^2\left(\x_N\right)}{B^2\left(\x_N\right)}
\end{array}
\right]}_{\tensorfont{D}}\,\left(\begin{array}{cc}
v_x\\ v_z\end{array}\right)=\underbrace{\left[\begin{array}{cc}
v_{\perp{x}}\left(\x_1\right)\\ v_{\perp{z}}\left(\x_1\right)\\
\vdots\\
v_{\perp{x}}\left(\x_N\right)\\ v_{\perp{z}}\left(\x_N\right)
\end{array}\right]}_{{\tensorfont{d}}},
\end{equation}
which has the solution \cite[]{Golub1980}
\begin{equation}
\left(\begin{array}{cc}
\widehat{v}_x\\ \widehat{v}_z\end{array}\right)=\left({\tensorfont{D}}^*\,{\tensorfont{D}}\right)^{-1}\,{\tensorfont{D}}^*{\tensorfont{d}}.\label{eqn:vt}
\end{equation}
\end{mathletters}
Note that ${\tensorfont{D}}^*\,{\tensorfont{D}}$ is analogous to $\AAA$ and ${\tensorfont{D}}^*{\tensorfont{d}}$ is analogous to $\bbb$ in~(\ref{eqn:least}).\par
This pedagogical example illustrates how DAVE4VM may analogously
estimate the field-aligned plasma velocity for the more general case
of a spatially variable plasma flow in an inhomogeneous magnetic field
for~(\ref{eqn:least}). The accuracy of the estimate of the parallel
velocity will be limited by the structuring in direction of the
magnetic field within the aperture; if the magnetic field has a
uniform orientation in the aperture window, no useful estimate of the
field-aligned plasma velocity can be made from the magnetic
measurements alone. The quality of the estimate may be assessed with
the conditioning of ${\tensorfont{D}}^*\,{\tensorfont{D}}$
in~(\ref{eqn:vt}) for the pedagogical example or $\AAA$
in~(\ref{eqn:least}) for the full system.\par
\begin{figure}
\centerline{\includegraphics[width=\size]{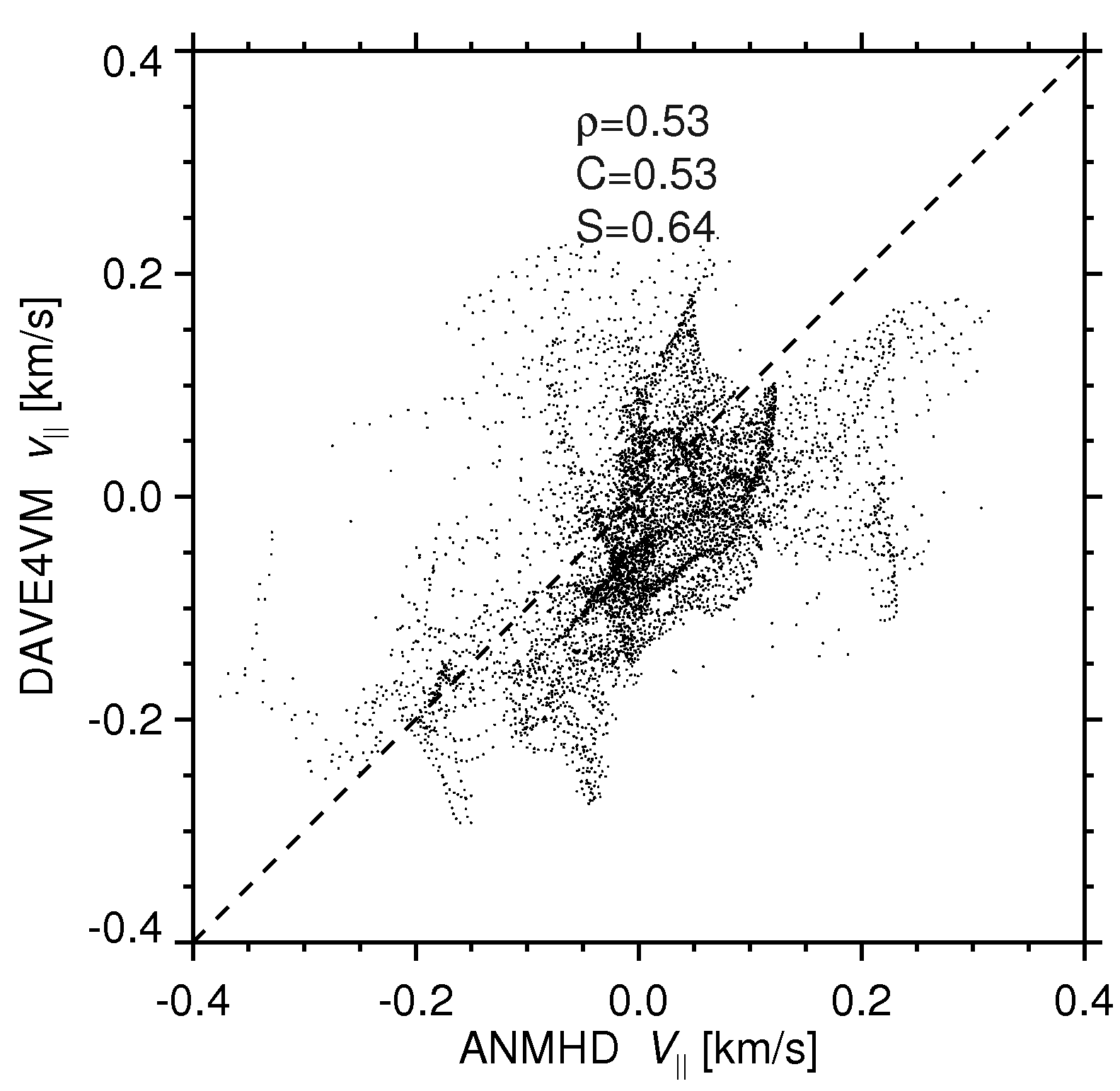}\hspc\includegraphics[width=\size]{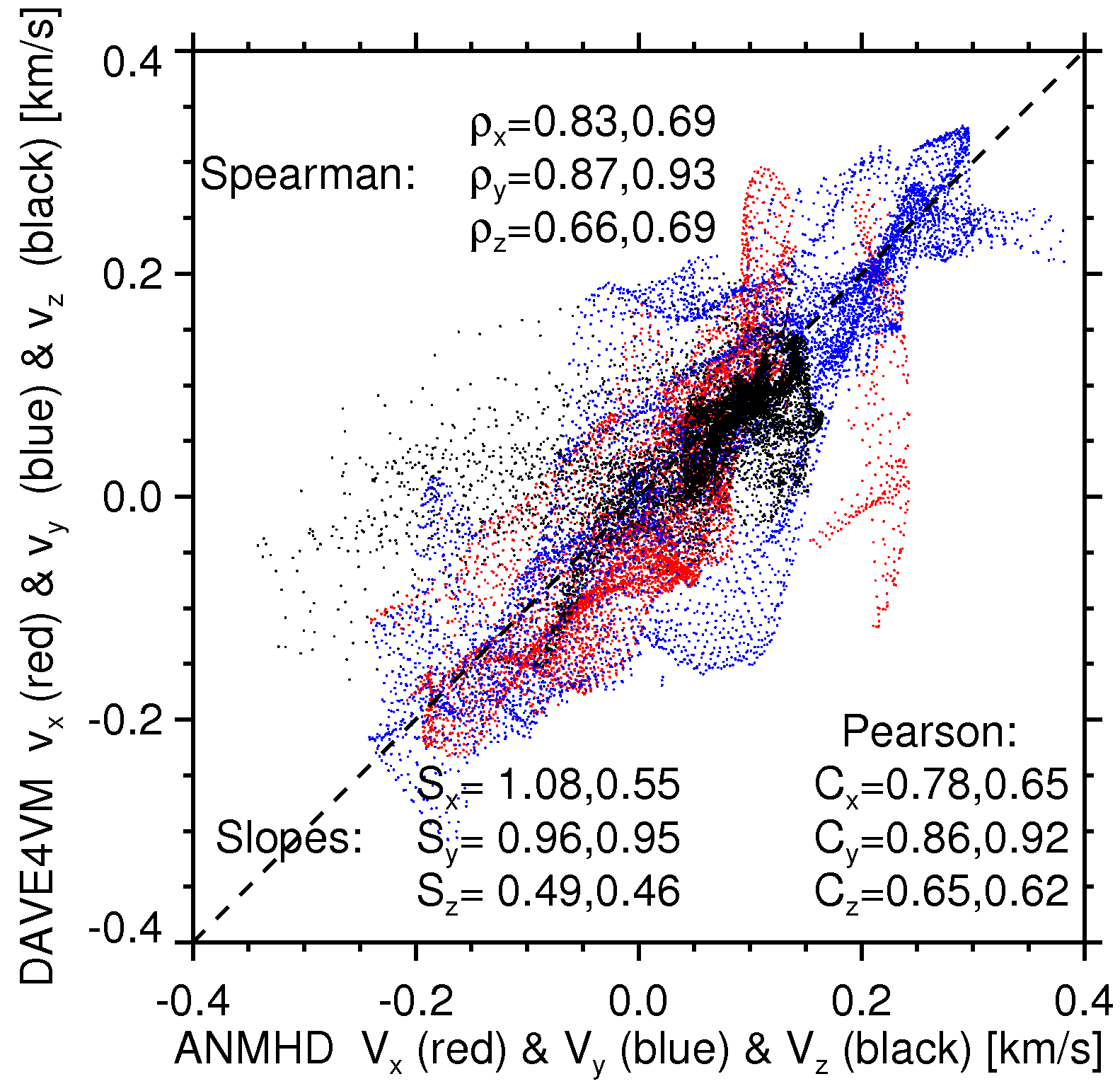}}
\caption{Scatter plots of (\textit{left}) the estimated parallel plasma
  velocities $v_\parallel$ from DAVE4VM versus the parallel plasma
  velocities $V_\parallel$ from ANMHD and (\textit{right}) the estimated
  \textit{total} plasma velocity $\v$ from DAVE4VM versus the
  \textit{total} plasma velocities $\V$ from ANMHD. The nonparametric
  Spearman rank-order correlation coefficients ($\RO$), Pearson
  correlation coefficients ($\PCC$), and slopes ($\Slopes$) estimated
  by the least absolute deviation method are shown. The pairs of
  numbers represent correlations between $\v$ and $\V$ and $\v_\perp$
  and $\V$. \label{fig:vll}}
\end{figure}
Figure~(\ref{fig:vll}) shows scatter plots of (\textit{left}) the estimated
parallel plasma velocities $v_\parallel$ from DAVE4VM versus the
parallel plasma velocities $V_\parallel$ from ANMHD and (\textit{right}) the
estimated \textit{total} plasma velocity $\v$ from DAVE4VM versus the
\textit{total} plasma velocities $\V$ from ANMHD.  The nonparametric
Spearman rank-order correlation coefficients ($\RO$), Pearson
correlation coefficients ($\PCC$), and slopes ($\Slopes$) estimated by
the least absolute deviation method are shown. The comma-separated
pairs of numbers in the right plot, corresponding to correlations
between $\v$ and $\V$ and $\v_\perp$ and $\V$ respectively, represent
the relative improvement in total velocity estimate over the simple
null hypothesis $H_0:\v=\v_\perp$ that the total plasma velocity is
the perpendicular plasma velocity. The correlation of the
$\xhat$-component of total velocity is significantly improved over the
null hypothesis $H_0$. This improvement id interesting since the
horizontal magnetic field is predominantly aligned with the $\x$-axis
(See Figure~\ref{fig:ftv}).  The correlation of the $\yhat$-component
of total velocity is slightly worse than the null hypothesis.
Finally, the correlation of the $\zhat$-component of total velocity is
mixed with the Spearman correlation $\RO$ slightly worse than the null
hypothesis and the Pearson correlation $\PCC$ slightly better than the
null hypothesis. However, the slopes of all three components are
improved over the null hypothesis.\par
The significance of the correlations in the left plot may be tested
against the null hypothesis $H_0:\RO=0$ by the Fisher permutation
test. \cite{Fieller1957} have demonstrated with analysis backed
Monte-Carlo simulation that Fisher's $z$\--transform of the
correlation coefficient
\begin{equation}
z_{\mathrm{S}}\left(\RO\right)=\frac{1}{2}\,\log\left|\frac{1+\RO}{1-\RO}\right|,
\end{equation}
produces approximately normally distributed values. For example,
permuting the values of $v_\parallel$ and $V_\parallel$ 10,000 times
generates the null hypothesis distribution with $\left\langle
{z_{\mathrm{S}}}\right\rangle=0.00\pm0.01$. The Spearman correlation
coefficient $\RO=0.53$ has a $z$-transform of
$z_{\mathrm{S}}\left(0.53\right)=0.60$ which is roughly 50 standard deviations
from the mean of the null distribution indicating that the parallel
velocity correlation is statistically significant and not due to
sampling error. However, the correlation $\RO=0.53$ is small and the
parallel velocity estimates may not be scientifically significant for
accurately predicting the parallel velocity. The plasma velocities may
be further constrained by introducing Doppler velocities, but this is
beyond the scope of the present discussion.
\subsubsection{Are $\v_h$ and $v_z$ Redundant?}
\begin{figure}
\centerline{\includegraphics[width=\size]{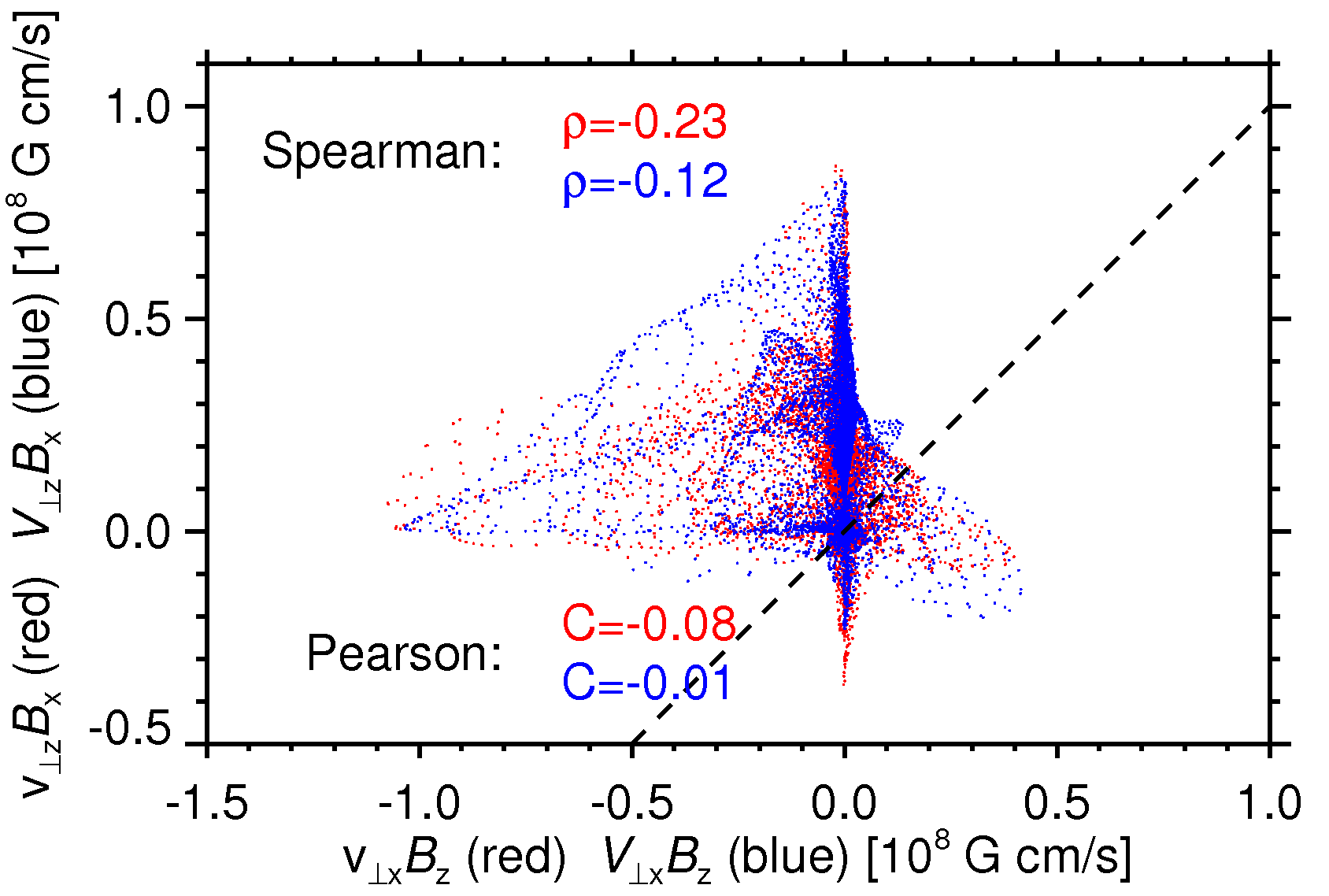}\hspc\hspc\hspc\includegraphics[width=\size]{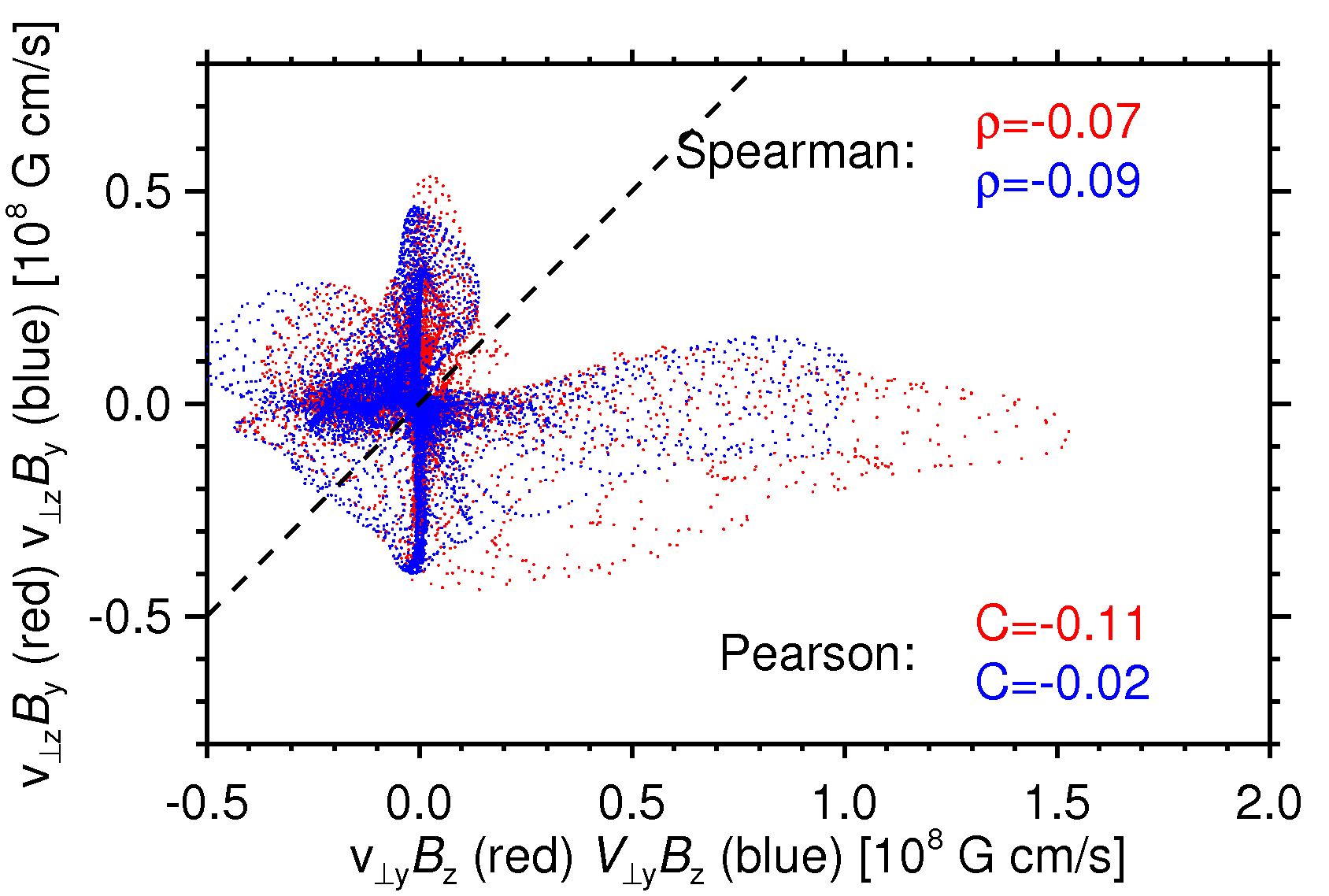}}
\caption{ Scatterplots for the $\xhat$ (\textit{left}) and $\yhat$ (\textit{right})
components of~(\ref{eqn:flux_transport}). The scatterplots indicate
the \textit{lack} of correlation between the terms describing
shearing motion ${B}_z\,\Vt$ and emergence $V_z\,\B_h$.  Red
points indicate the results for DAVE4VM and blue points indicate the
results for ANMHD. The Spearman rank order ($\RO$) and Pearson
($\PCC$) correlations between the two terms are very low for both
components of the flux transport velocities.\label{fig:redundent}}
\end{figure}
DAVE4VM has incorporated an additional component of the velocity over
DAVE by introducing three additional variables $\widehat{w}_0$,
$\widehat{w}_x$ and $\widehat{w}_y$. Consequently, one may reasonably
wonder ``are the terms $\v_h\,B_z$ and $v_z\,\B_h$ redundant for
DAVE4VM?''  The answer is a clear ``No'' for the ANMHD
data. Equation~(\ref{eqn:flux_transport}) is composed of two terms
${B}_z\,\Vt$ and $V_z\,\B_h$ describing shearing motion and emergence
respectively. Figure~\ref{fig:redundent} shows scatterplots of the two
terms for the $\xhat$ (\textit{left}) and $\yhat$ (\textit{right}) components
of~(\ref{eqn:flux_transport}). The scatterplots indicate the
\textit{lack} of correlation between the terms describing shearing
motions ${B}_z\,\Vt$ and emergence $V_z\,\B_h$.  Red points indicate
the results for DAVE4VM and blue points indicate the results for
ANMHD. The Spearman rank order ($\RO$) and Pearson ($\PCC$)
correlations between the two terms, summarized in
Table~\ref{tab:redundent}, are very low for both components of the
flux transport velocities from DAVE4VM or ANMHD. These terms describe
different physics, that are uncorrelated, and which require
independent variables to describe. \par
\begin{deluxetable}{cccccc}
\tablecaption{The Spearman rank order ($\RO$) and Pearson ($\PCC$)
  correlations between the terms in the flux transport velocity describing
  shearing motion and flux emergence.\label{tab:redundent}}
\tablehead{\multicolumn{2}{c}{Correlates}&\multicolumn{2}{c}{DAVE4VM}&\multicolumn{2}{c}{ANMHD}\\
&&\colhead{Spearman}&\colhead{Pearson}&\colhead{Spearman}&\colhead{Pearson}}
\startdata
$v_{\perp x}\,B_z$&$v_{\perp z}\,B_x$&-0.23&-0.08&-0.12&-0.01\\
$v_{\perp y}\,B_z$&$v_{\perp z}\,B_y$&-0.07&-0.11&-0.09&-0.02\\
$v_{\perp x}$&$v_{\perp z}$&-0.09&-0.10&-0.15&-0.15\\
$v_{\perp y}$&$v_{\perp z}$&-0.34&-0.33&-0.30&-0.20\\

\enddata
\end{deluxetable}
\subsection{Induction Equation and Electric Fields}
\begin{figure}
\centerline{\includegraphics[width=\size]{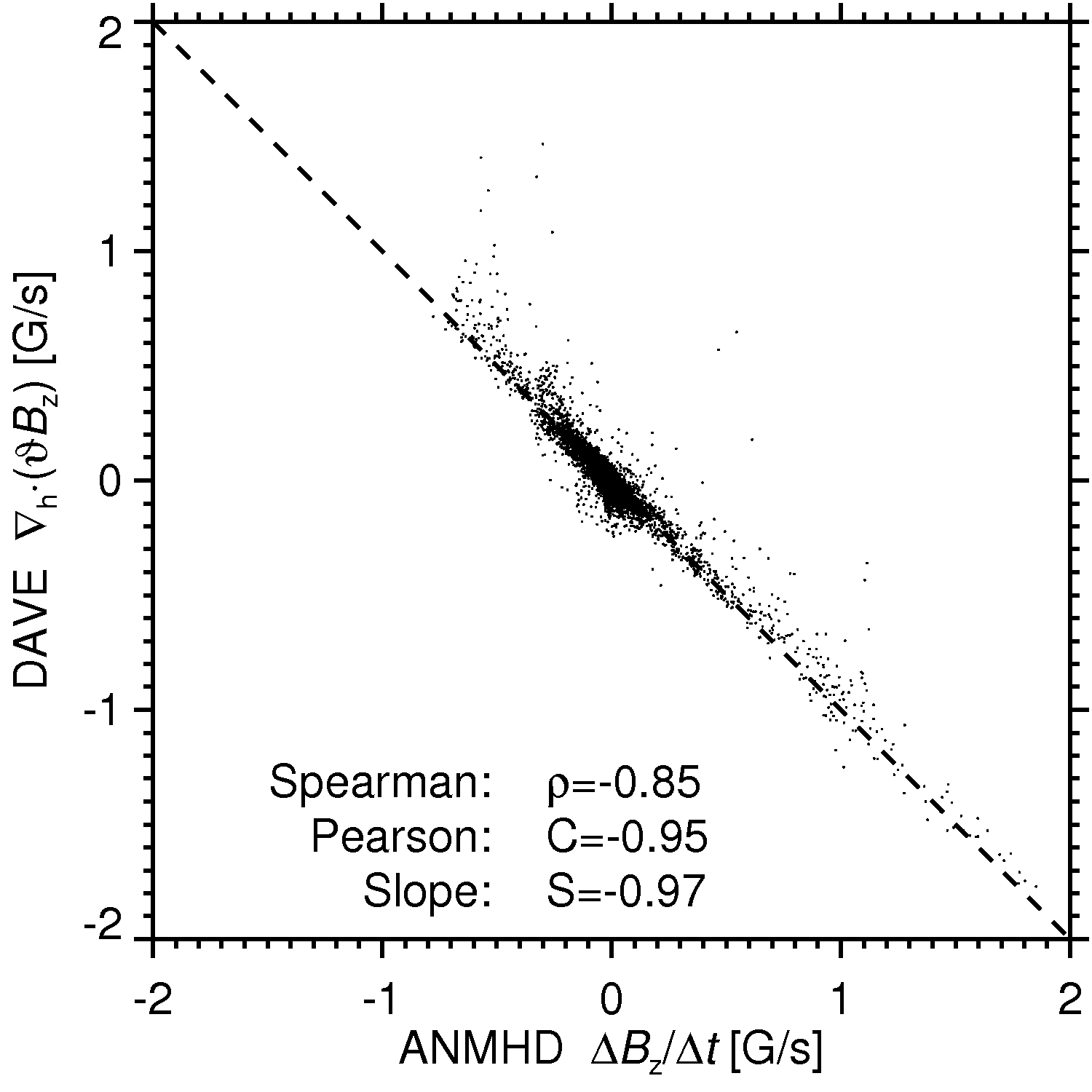}\hspc\includegraphics[width=\size]{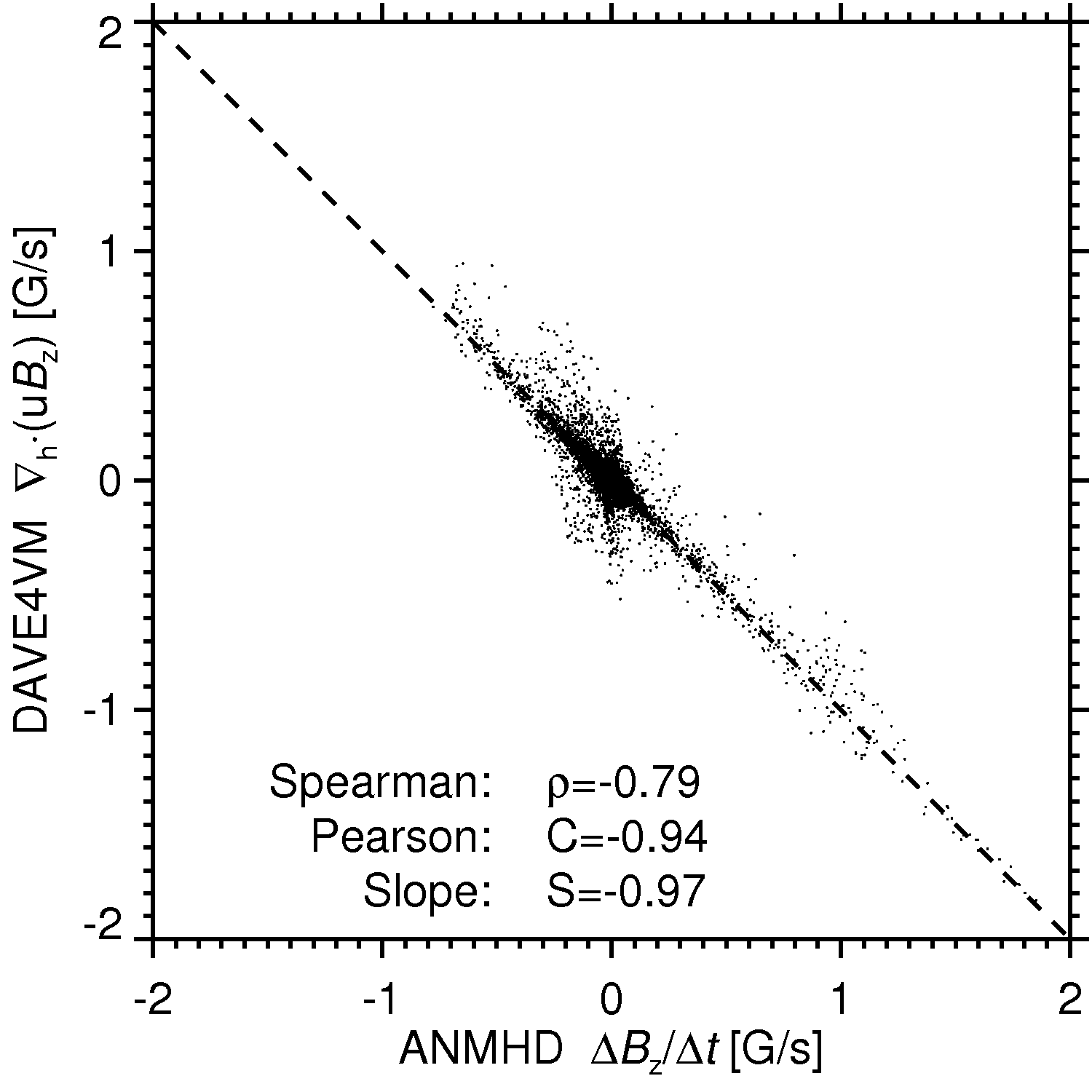}}
\caption{Scatter plots of $\grad_h\cdot\left(\u\,B_z\right)$ from the
DAVE (\textit{left}) and $\grad_h\cdot\left(\uF\,B_z\right)$ from the DAVE4VM
(\textit{right}) versus $\Delta B_z/\Delta t$ from ANMHD.  The nonparametric
Spearman rank-order correlation coefficients ($\RO$), Pearson
correlation coefficients ($\PCC$), and slopes ($\Slopes$) estimated by
the least absolute deviation method are shown.\label{fig:induction}}
\end{figure}
Figure~\ref{fig:induction} shows $\grad_h\cdot\left(\u\,B_z\right)$
from the DAVE (\textit{left}) and $\grad_h\cdot\left(\uF\,B_z\right)$
from DAVE4VM (\textit{right}) versus $\Delta B_z/\Delta t$ from
ANMHD. The derivatives for these plots were estimated from 5-point
optimized least squares. These plots indicate how well the two methods
satisfy the MHD induction equation globally. The DAVE has higher
correlations than the DAVE4VM but the slopes are equivalent. For the
DAVE4VM, the most significant deviations from the MHD induction
equation occur near $\Delta B_z/\Delta t\approx0$.  Neither the DAVE
nor DAVE4VM satisfy the induction equation exactly. This is by design,
because real magnetogram data are likely to contain significant noise
which will contaminate velocity estimates if the induction equation is
satisfied exactly. Furthermore, how well a method satisfies the
induction equation will generally depend on the differencing
template. Consequently, if the velocity estimates are to be used as
boundary values for ideal MHD coronal field models, then the
velocities of any method will have to be adjusted to satisfy the
induction equation on the differencing template implemented by the
simulation.  Using the Helmholtz decomposition~(\ref{eqn:Helmholtz}),
the inductive potential may be computed for the simulation directly
from the magnetogram sequence (on the simulation differencing
template)
\begin{mathletters}
\begin{equation}
\partial_t{B_z}=\nabla_h^2\phi,
\end{equation}
and the electrostatic potential may be derived from the flux transport
vectors determined by the optical flow method \cite[]{Welsch2004}
\begin{equation}
\nabla_h^2\psi=\zhat\cdot\left[\grad\cross\left(\uF\,B_z\right)\right].\label{eqn:psi}
\end{equation}
\end{mathletters}
Incorporating photospheric velocity estimate into boundary
conditions for a coronal MHD simulation, in a minimally consistent way
with the normal component of the magnetic induction equation, requires
solving two Poisson equations on the photospheric boundary using the
differencing template of the MHD code.\par
\begin{figure}
\centerline{\includegraphics[width=\size]{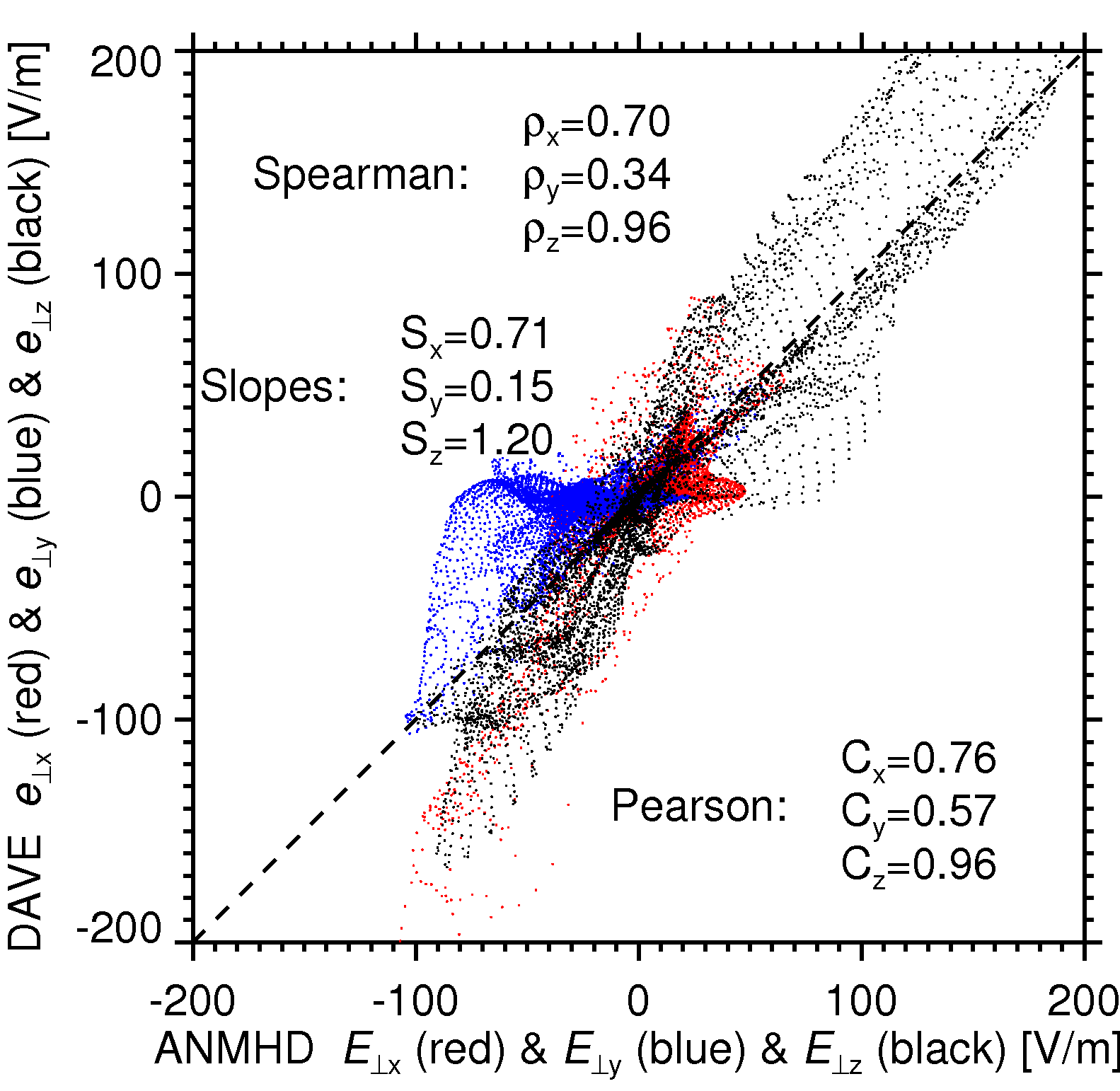}\hspc\hspc\hspc\includegraphics[width=\size]{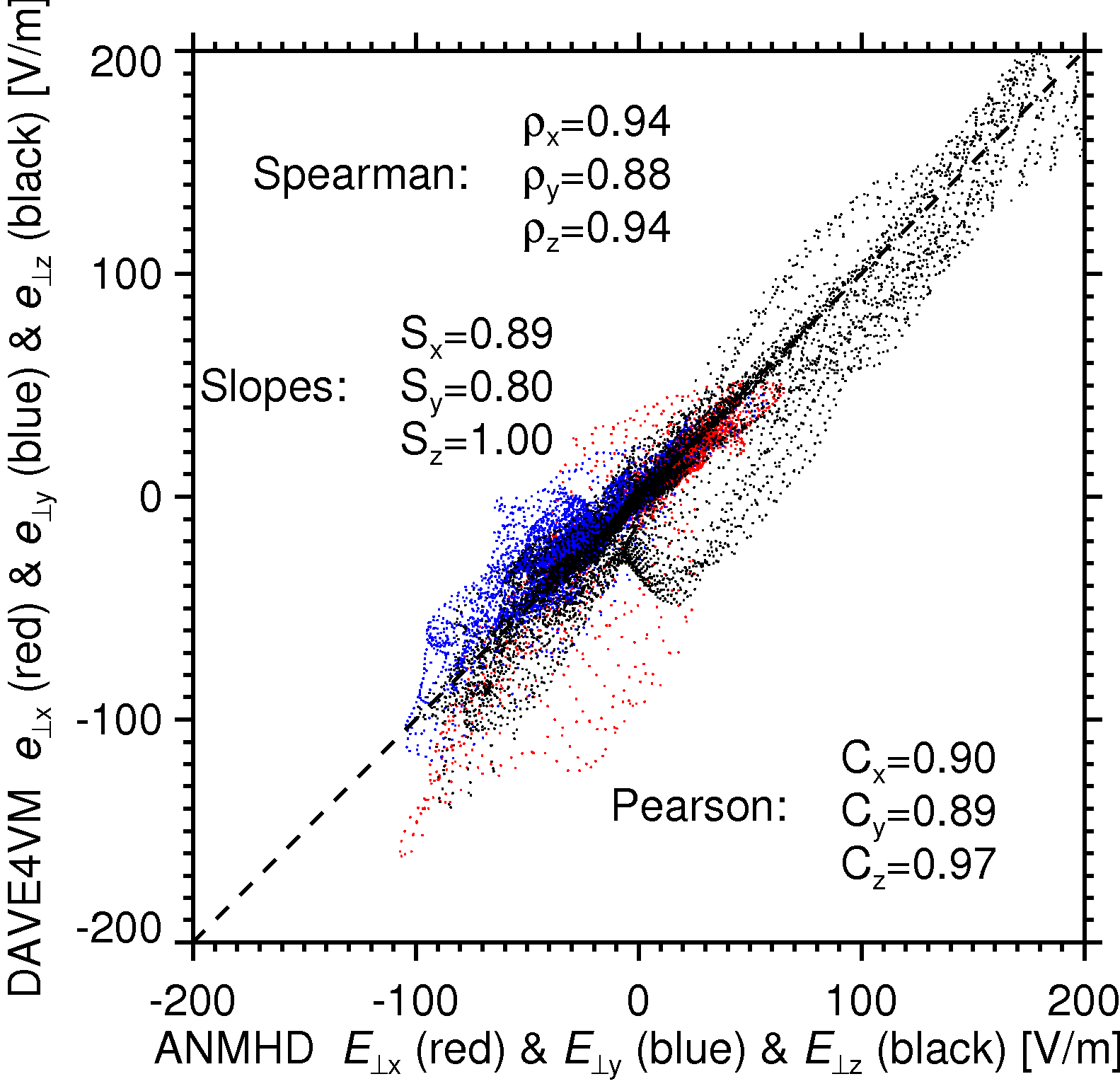}}
\caption{Scatter plots of the estimated perpendicular electric field
  $\vectorfont{e}_\perp$ from DAVE assuming $\u=\uF$ (\textit{left}) and from DAVE4VM
  (\textit{right}) versus the electric field $\E_\perp$ from ANMHD. Red, blue, and
  black correspond to the $x$-, $y$-, and $z$-components respectively. The
  nonparametric Spearman rank-order correlation coefficients ($\RO$), Pearson
  correlation coefficients ($\PCC$), and slopes ($\Slopes$) estimated by the
  least absolute deviation method are shown for each component of the electric
  field.\label{fig:E_corr}}
\end{figure}
Figure~\ref{fig:E_corr} shows scatter plots of the estimated
perpendicular electric fields $\vectorfont{e}_\perp$ from DAVE
assuming $\u=\uF$ (\textit{left}) and from DAVE4VM (\textit{right}) versus the electric
fields $\E_\perp$ from ANMHD. Red, blue, and black correspond to the
$x$-, $y$-, and $z$-components, respectively. The nonparametric Spearman
rank-order correlation coefficients ($\RO$) and Pearson correlation
coefficients ($\PCC$) are shown for each component of the electric
field. On the present mask the DAVE4VM estimates improve or
essentially match the correlation and slopes of the DAVE's estimates
for all three components of the electric field. Particularly dramatic
is the improvement in the $\yhat$ component of the electric field
which the DAVE does not estimate accurately either on the present mask
$\left|\B\right|>370$~G or the restricted mask
$\left|B_z\right|>370$~G \cite[]{Welsch2007}. \par
\subsection{Poynting and Helicity Fluxes}
\begin{figure}
\centerline{\includegraphics[width=\size]{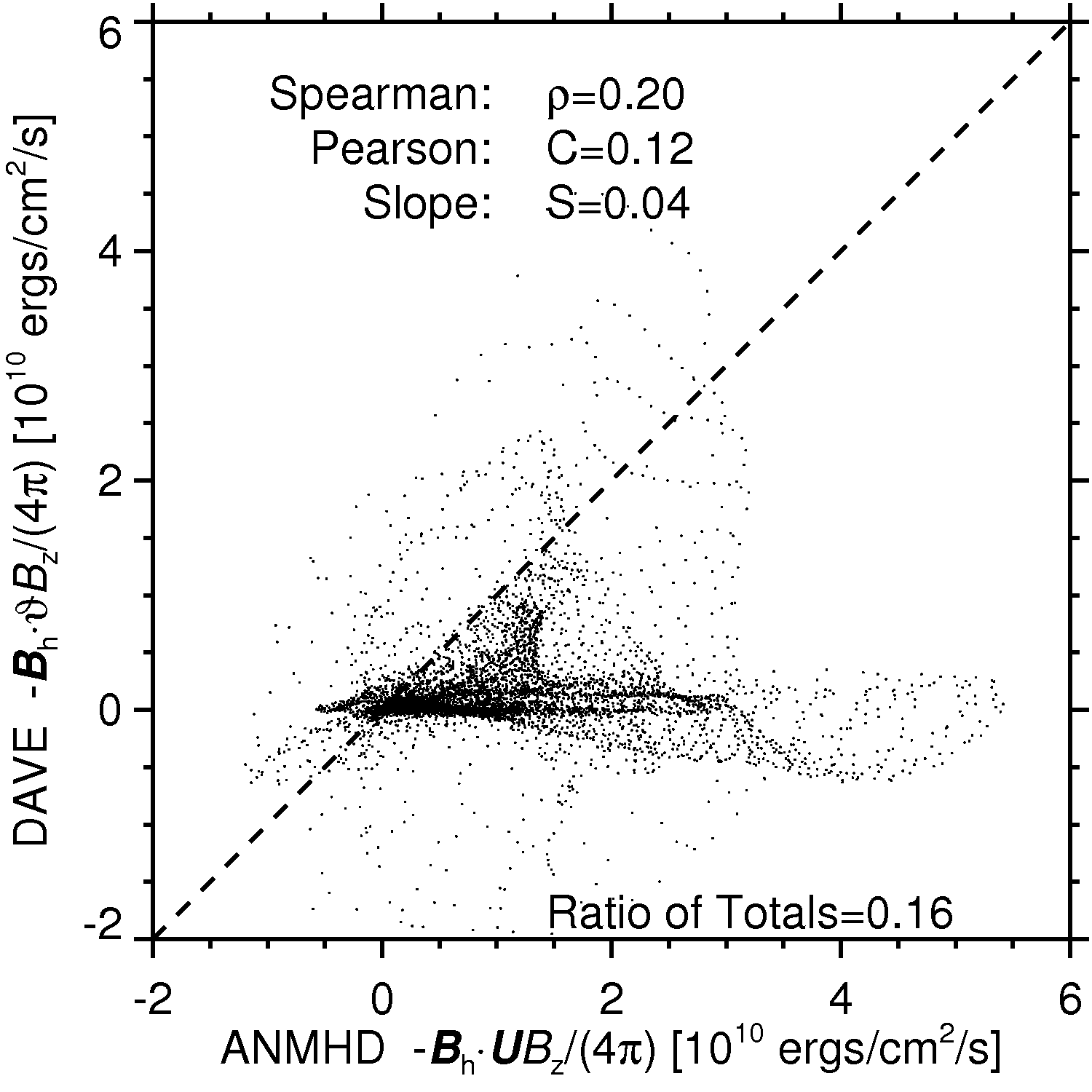}\hspc\includegraphics[width=\size]{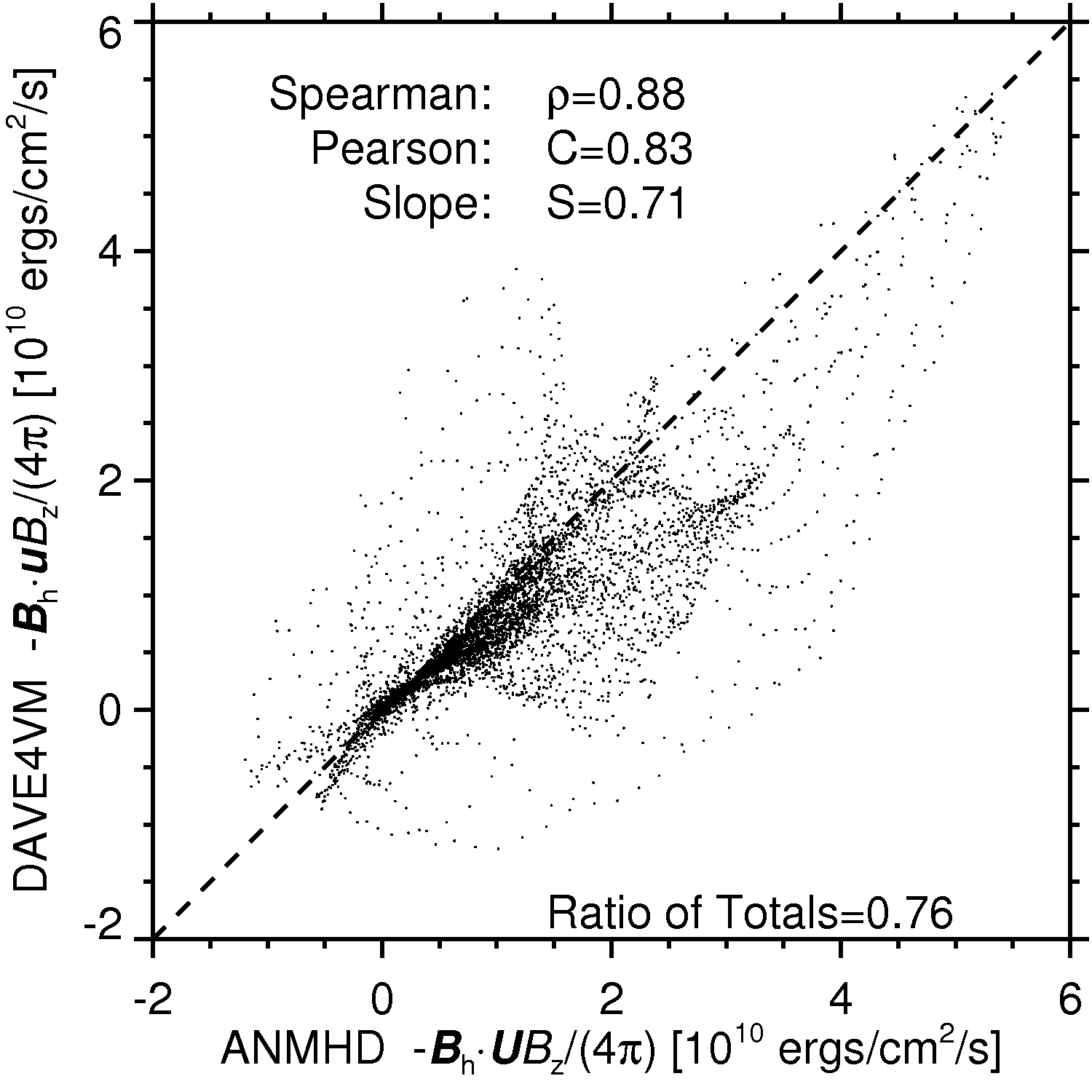}}
\caption{Scatter plots of the estimated Poynting flux from DAVE
 assuming $\u=\uF$ (\textit{left}) and DAVE4VM (\textit{right}) versus the Poynting flux
 from ANMHD. The nonparametric Spearman rank-order correlation
 coefficients ($\RO$), Pearson correlation coefficients ($\PCC$), and
 slopes ($\Slopes$) estimated by the least absolute deviation method
 are shown, as is the ratio of the integrated estimated Poynting flux
 to the integrated ANMHD Poynting flux.\label{fig:poynting}}
\end{figure}
\cite{Demoulin2003} show that the Poynting flux can be expressed
concisely in terms of the flux transport vectors $\uF\,B_z$
\begin{equation}
s_z\left(\x\right)=-\frac{1}{4\,\pi}\,\B_h\cdot\left(B_z\,\v_h-v_z\,\B_h\right)=-\frac{\B_h\cdot\left(\uF\,\,B_z\right)}{4\,\pi}.\label{eqn:poynting}
\end{equation}
Figure~\ref{fig:poynting} shows scatterplots of the estimated Poynting
flux $s_z$ from the DAVE assuming $\u=\uF$ (\textit{left}) and DAVE4VM (\textit{right})
versus ANMHD's Poynting flux $S_z$. The correspondence for DAVE4VM, or
lack there of for DAVE, indicates the accuracy of the velocity
estimates in the direction of the horizontal magnetic field
$\B_h$. The nonparametric Spearman rank-order correlation coefficients
($\RO$), Pearson correlation coefficients ($\PCC$), and slopes
($\Slopes$) estimated by the least absolute deviation method are
shown, as is the ratio of the integrated estimated Poynting flux to
the integrated ANMHD Poynting flux
$\mathcal{R}_{{s}_z}=\sum{s}_z/\sum{S}_z$. The DAVE4VM's estimate of
Poynting flux is a significant improvement over the DAVE's.  The
correlations have improved by roughly a factor of $4\--6$, the slope
has improved by nearly a factor of 18, and the ratio of the totals has
improved by nearly a factor of 5. Again, DAVE does not reliably
estimate the flux transport velocity in the direction of the
horizontal magnetic field $\B_h$ suggesting that DAVE is insensitive
to flux emergence which is proportional to $v_z\,\B_h$. The ``ground
truth'' total power through the mask is
$dP/dt=\sum{S_z}=7.7\times10^{28}\,\mathrm{ergs/s}$. \par
\begin{figure}
\centerline{\includegraphics[width=\size]{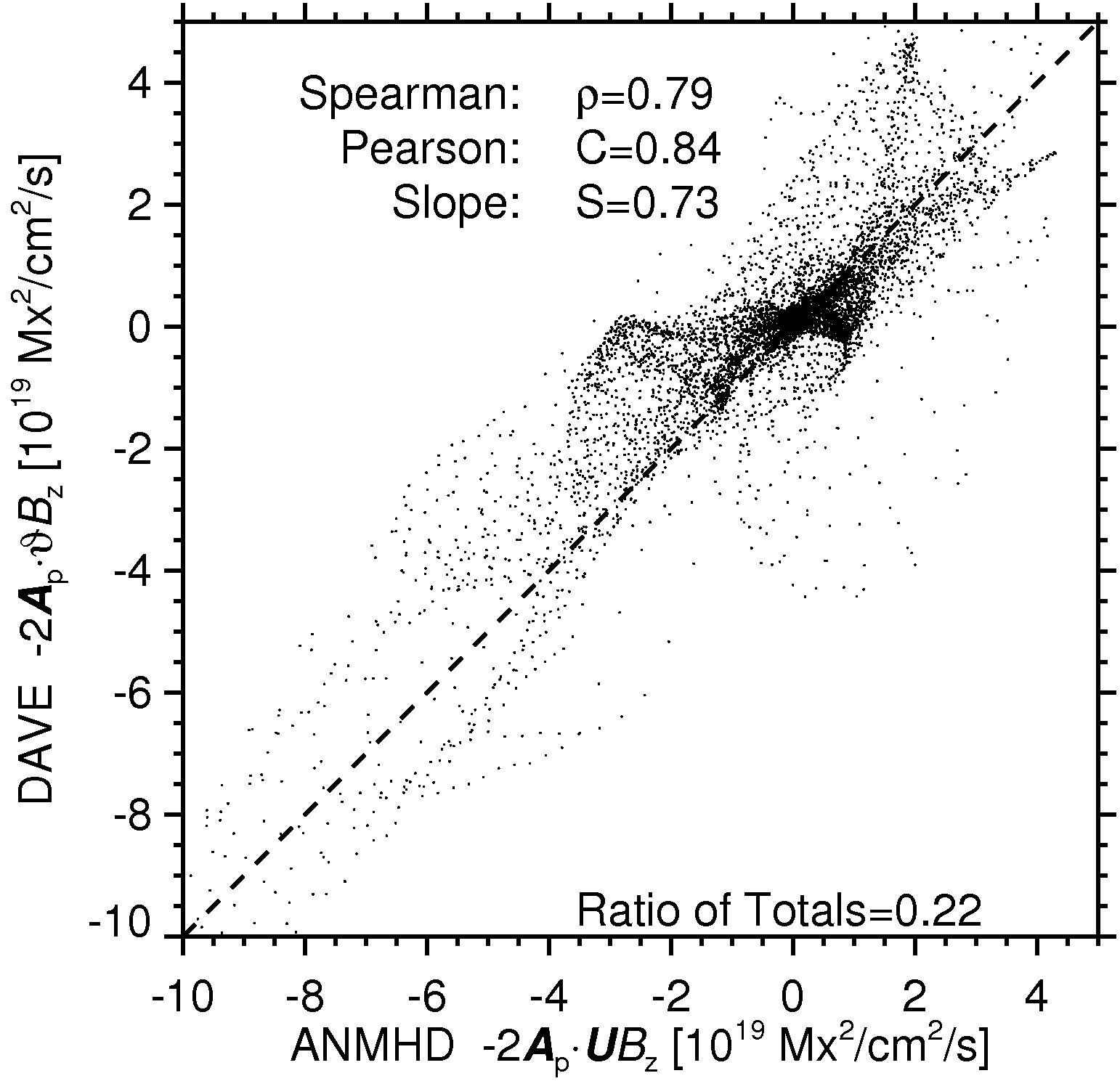}\hspc\includegraphics[width=\size]{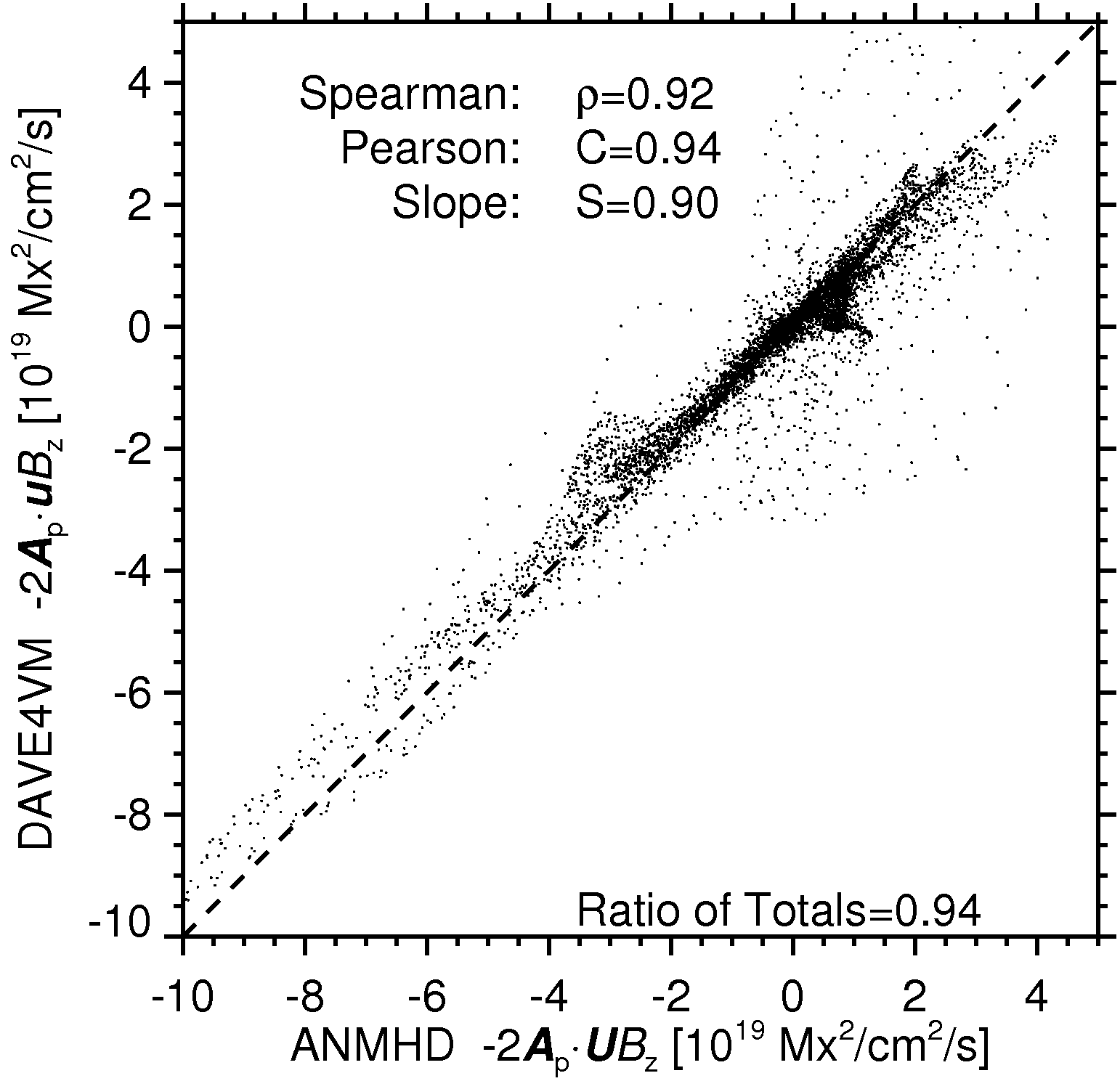}}
\caption{Scatter plots of the estimated helicity flux DAVE assuming
$\u=\uF$ (\textit{left}) and DAVE4VM (\textit{right}) versus ANMHD's helicity flux. The
nonparametric Spearman rank-order correlation coefficients ($\RO$),
Pearson correlation coefficients ($\PCC$), and slopes ($\Slopes$)
estimated by the least absolute deviation method are shown, as is the
ratio of the integrated estimated helicity flux to the integrated
ANMHD helicity flux.\label{fig:helicity}}
\end{figure}
\cite{Demoulin2003} show that the gauge-invariant helicity
flux \cite[]{Berger1984} can be expressed concisely in terms of the
flux transport vectors $\uF\,B_z$
\begin{equation}
\gA\left(\x\right)=-2\,\A_p\cdot\left(B_z\,\v_h-v_z\,\B_h\right)=-2\,\A_p\cdot\left(\uF\,B_z\right)
\end{equation}
where $\A_p=\zhat\cross\grad\Phi_p$ is the potential reference field (with
zero helicity) which satisfies
\begin{equation}
\zhat\cdot\left(\grad\cross\A_p\right)=\nabla_h^2\Phi_p=B_z,
\label{eqn:poisson}
\end{equation}
and $\grad\cdot\A_p=\zhat\cdot\A_p=0$.  To estimate the 
helicity flux density, $\Phi_p$ was computed on a $257\times257$
square centered on the region of interest with Dirichlet boundary
conditions using MUDPACK \cite[]{Adams1993}.
While interpretation of maps of helicity flux $\gA\left(\x\right)$
through the photosphere is problematic \cite[]{Pariat2005,Pariat2007},
a comparison of $\gA\left(\x\right)$ estimated from the DAVE or
DAVE4VM verses $\GA\left(\x\right)$ calculated from ANMHD indicates
the \textit{accuracy} of the estimated flux transport vectors in the
direction of the vector potential. Figure~\ref{fig:helicity} shows
scatter plots of the estimated helicity flux from DAVE assuming
$\u=\uF$ (\textit{left}) and DAVE4VM (\textit{right}) versus ANMHD's helicity flux. The
nonparametric Spearman rank-order correlation coefficients ($\RO$) and
Pearson correlation coefficients ($\PCC$) are shown, as is the ratio
of the integrated estimated helicity flux to the integrated ANMHD
helicity flux $\mathcal{R}_{\gA}=\sum{\gA}/\sum{\GA}$.  The DAVE4VM's
estimates represent a significant improvement over the DAVE's,
improving the correlation coefficients by roughly $0.1$ and the slope
by $0.2$. Furthermore, the ratio of totals has improved by roughly a
factor of 4 from $0.22$ for the DAVE to $0.94$ for the DAVE4VM. The
``ground truth''\footnote{This estimate differs by about 10\% from the
helicity estimate in \cite{Welsch2007}. The discrepancy is caused by
the different methodologies and boundaries used for computing the
vector potential $\A_p$. MUDPACK was used in this study
with~(\ref{eqn:poisson}) whereas \cite{Welsch2007} used a Green's
function scheme to compute $\A_p$.} helicity injected through the
surface is
$dH_A/dt=\sum{\GA}=-2.8\times10^{37}\,\mathrm{Mx}^2/\mathrm{s}$.
\section{Discussion and Conclusions \label{sec:conclusions}}
\begin{deluxetable}{cccccccc}
\tablecaption{Comparison between the DAVE and
DAVE4VM over the 3815 pixels that satisfy $\left|B_z\right|>370$~G in
Figure~\ref{fig:ftv}. This corresponds roughly to the mask used in
\cite{Welsch2007}.\label{tab:strong}}
\tablehead{&&\multicolumn{3}{c}{DAVE (Assuming $\u=\uF$) }&\multicolumn{3}{c}{DAVE4VM}\\
\multicolumn{2}{c}{Quantities}&\colhead{Spearman}&\colhead{Pearson}&\colhead{Slope}&\colhead{Spearman}&\colhead{Pearson}&\colhead{Slope}}
\startdata
                                             $u_x\,B_z$&                                             $U_x\,B_z$& 0.43& 0.60& 0.33& 0.87& 0.88& 0.81\\
                                             $u_y\,B_z$&                                             $U_y\,B_z$& 0.73& 0.86& 1.07& 0.90& 0.90& 0.95\\
                                         $v_{\perp{x}}$&                                         $V_{\perp{x}}$& 0.88& 0.91& 0.86& 0.90& 0.92& 0.93\\
                                         $v_{\perp{y}}$&                                         $V_{\perp{y}}$& 0.93& 0.93& 1.17& 0.93& 0.93& 0.99\\
                                         $v_{\perp{z}}$&                                         $V_{\perp{z}}$& 0.27& 0.39& 0.29& 0.76& 0.77& 0.72\\
                    $\grad_h\cdot\left(\uF\,B_z\right)$&                                  $\Delta B_z/\Delta t$&-0.92&-0.95&-0.99&-0.84&-0.95&-0.97\\
                                         $e_{\perp{x}}$&                                         $E_{\perp{x}}$& 0.73& 0.86& 1.07& 0.90& 0.90& 0.95\\
                                         $e_{\perp{y}}$&                                         $E_{\perp{y}}$& 0.43& 0.60& 0.33& 0.87& 0.88& 0.81\\
                                         $e_{\perp{z}}$&                                         $E_{\perp{z}}$& 0.97& 0.96& 1.17& 0.96& 0.97& 0.98\\
                                                  $s_z$&                                                  $S_z$& 0.09& 0.05& 0.02& 0.83& 0.79& 0.68\\

\enddata
\end{deluxetable}
\begin{deluxetable}{lcccccccccc}
\tablecaption{Comparison of accuracy of the velocity estimates between
the DAVE and DAVE4VM over the 3815 pixels that satisfy
$\left|B_z\right|>370$~G in Figure~\ref{fig:ftv}.  This corresponds
roughly to the mask used in \cite{Welsch2007}.\label{tab:vstrong}}
\tablehead{&\multicolumn{4}{c}{DAVE (Assuming $\u=\uF$)}&\multicolumn{4}{c}{DAVE4VM}\\
\colhead{\f}&\colhead{$\left\langle\left|\delta\smash{\widetilde{\f}}\right|\right\rangle$}&\colhead{$\left\langle\delta\left|\smash{\widetilde{\f}}\right|\right\rangle$}&\colhead{$C_{\mathrm{vec}}$}&\colhead{$C_{\mathrm{CS}}$}&\colhead{$\left\langle\cos\theta\right\rangle_W$}&
\colhead{$\left\langle\left|\delta\smash{\widetilde{\f}}\right|\right\rangle$}&\colhead{$\left\langle\delta\left|\smash{\widetilde{\f}}\right|\right\rangle$}&\colhead{$C_{\mathrm{vec}}$}&\colhead{$C_{\mathrm{CS}}$}&\colhead{$\left\langle\cos\theta\right\rangle_W$}}
\startdata
  $\uF\,B_z$& 0.75$\pm$0.37&-0.26$\pm$0.40&0.68&0.69&0.74& 0.42$\pm$0.30&-0.08$\pm$0.24&0.91&0.91&0.93\\
  $\v_\perp$& 0.57$\pm$0.30& 0.06$\pm$0.27&0.87&0.87&0.88& 0.39$\pm$0.28&-0.01$\pm$0.17&0.93&0.90&0.92\\

\enddata
\end{deluxetable}
For completeness, Tables~\ref{tab:strong} and~\ref{tab:vstrong}
provide a summary of metrics and correlation coefficients for the DAVE
and DAVE4VM on the original mask $\left|B_z\right|>370$~G used by
\cite{Welsch2007} in the same format as in Tables~\ref{tab:weak}
and~\ref{tab:vweak}.  MEF performed the best overall in the original
study by \cite{Welsch2007} although there were some metrics where the
DAVE outperformed MEF such as in the accuracy of the plasma velocities
listed in Table~\ref{tab:vstrong} (compare with Figure~8 in
\cite{Welsch2007}).  The DAVE4VM's estimates are a substantial
improvement over the results of the DAVE assuming $\u=\uF$ on this
mask.  Comparing the rank-order Spearman correlation coefficients for
the flux transport vectors, perpendicular plasma velocity, and
electric field in Table~\ref{tab:strong}, the DAVE4VM equals or
out-performs MEF. Particularly, the DAVE4VM's estimate of the vertical
perpendicular plasma velocity is substantially better than MEF with a
rank-order of 0.76 in the former and 0.61 in the latter case. Accurate
vertical flows are necessary to diagnose flux emergence and accurately
estimate the helicity flux. The one area where MEF exhibits
superiority is in the estimate of the Poynting flux where the DAVE4VM
captures 76\% and MEF captures 100\% \cite[]{Welsch2007}.  The
fractional errors
$\left\langle\left|\delta\smash{\widetilde{\f}}\right|\right\rangle$
and
$\left\langle\delta\left|\smash{\widetilde{\f}}\right|\right\rangle$
are substantially lower than the DAVE for both the flux transport
vectors and plasma velocities. The DAVE had the largest vector
correlation $C_{\mathrm{vec}}$ and the direction correlation
$C_{\mathrm{CS}}$ in the original study and the DAVE4VM improves over
this performance exhibiting correlation coefficients of roughly
$0.9$. The improvement for the flux transport vectors is particularly
dramatic.  The plasma velocities are more accurate and exhibit
considerably less bias than those reported for MEF in
\cite{Welsch2007}.\par
The DAVE4VM offers some minor advantages over MEF.  The DAVE4VM is
somewhat faster than MEF; the DAVE4VM(DAVE) requires 30(10) seconds to
process\footnote{The routines were all coded in Interactive Data
  Language \cite[]{IDL_5.6} and the computations were performed on a
  dual processor AMD Opteron 240 running at 1.4~GHz with a one
  megabyte memory cache and ten gigabytes of Random Access Memory.}
the full $288\times288$ pixel frame from ANMHD whereas MEF requires
roughly 10~minutes to converge on a reduced mask of the ANMHD data
(private communication with Belur Ravindra). The DAVEVM is local and
directly estimates velocities across neutral lines and across broader
weak field regions whereas MEF is a iterative global method that
requires judicious choice of boundaries to ensure convergence.  In
concert, the DAVE4VM's velocity estimate might be used with MEF either
as an initial guess for the electrostatic potential $\psi$
via~(\ref{eqn:psi}) or as ancillary inaccurate velocity measurements
in the MEF variational term that constrains the photospheric plasma
velocities \cite[]{Longcope2004}. Since the DAVE4VM is fast and does
not require supervision beyond choosing a window size (and even this
could be automated according to the criteria discussed in
\S~\ref{sec:ANMHD}), this approach is appropriate for real-time
monitoring of helicity and energy fluxes through the photosphere from
observatories such as \textit{Solar Dynamics Observatory}.\par
What is responsible for the DAVE4VM's improved performance?  The only
differences between the DAVE and DAVE4VM are the terms $\sss_{ij}$ in
the structure tensor~(\ref{eqn:structure}) that describe the local
structure of the horizontal magnetic fields necessary for the
description of vertical flows. There are two circumstances when
line-of-sight methods such as the DAVE, LMSAL's LCT, and FLCT will
produce accurate estimates of the flux transport velocity:
\begin{mathletters}
\begin{enumerate}
\item $\B_h=0$: \textit{The magnetic field is purely vertical} and
$\u=\v_h=\v_{\perp{h}}$. If the horizontal magnetic fields and their
associated derivatives are zeroed, the DAVE and DAVE4VM produce
\textit{identical} flux transport and perpendicular plasma velocity estimates.
The DAVE is consistent with the assumption that the magnetic field is
purely vertical:\\
\begin{equation}
\lim_{\B_h\rightarrow0}\partial_t B_z+\gradh\cdot\left(B_z\,\vt-v_z\,\B_h\right)=\partial_t B_z+\gradh\cdot\left(B_z\,\vt\right).\label{iso:A}
\end{equation}
\item $v_z=0$: \textit{There are no net upflow/downflows} and
  $\u=\v_h\neq\v_{\perp{h}}$. In this situation, there must be
  projected vertical flows along the magnetic field to cancel any
  projected vertical flow perpendicular to the magnetic field with
  $v_{\perp{z}}=-v_{\parallel{z}}=-B_z\,\left(\v_h\cdot\B_h\right)/B^2$. Consequently,
  $\v_\parallel\neq0$. The DAVE is consistent with the assumption that
  there are no vertical flows $v_z=0$:\\
\begin{equation}
\lim_{v_z\rightarrow0}\partial_t B_z+\gradh\cdot\left(B_z\,\vt-v_z\,\B_h\right)=\partial_t B_z+\gradh\cdot\left(B_z\,\vt\right).\label{iso:B}
\end{equation}
\end{enumerate}
\end{mathletters}
Both limits~(\ref{iso:A}) and (\ref{iso:B}) are isomorphic
with~(\ref{eqn:DAVE}). By induction, ~(\ref{eqn:DAVE}) is consistent
with the assumptions leading to~(\ref{iso:A}) and (\ref{iso:B}). Since
DAVE does not consider corrections $\sss_{ij}$ due to the horizontal
magnetic field, \textit{$\u$ should generally be considered a biased
estimate of the horizontal plasma velocity $\u=\v_h$ and \textit{not}
the flux transport velocity!} Formally the alternative hypothesis
$H_1:\u=\v_h$ may be tested against the null hypothesis $H_0:\u=\uF$
of \cite{Demoulin2003}.
\begin{figure}
\centerline{\includegraphics[width=\size]{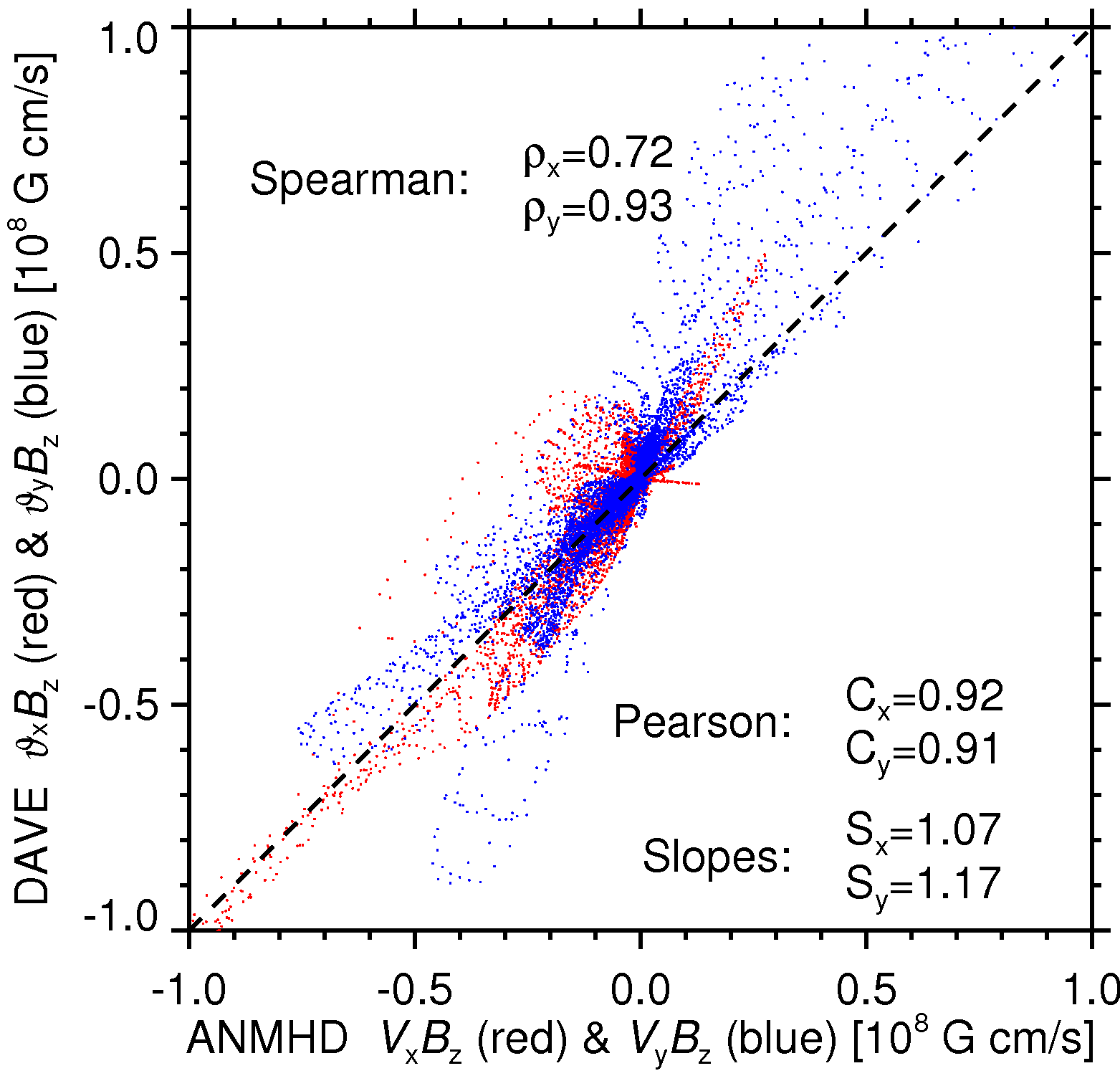}}
\caption{Scatter plots of the estimated velocities $\u\,B_z$ from DAVE
  versus the  horizontal plasma velocities $\V_h\,B_z$ from ANMHD.
 The nonparametric Spearman rank-order correlation
  coefficients ($\RO$), Pearson correlation coefficients ($\PCC$), and
  slopes ($\Slopes$) estimated by the least absolute deviation method
  are shown.\label{fig:DAVE}}
\end{figure}
The null hypothesis $H_0:\u=\uF$ is represented by the left panel in
Figure~\ref{fig:ftv_corr}. The alternative hypothesis, represented by
$H_1:\u=\v_h$, is characterized by the scatter plot of the estimated
velocities $\u\,B_z$ for DAVE versus the horizontal plasma velocities
$\V_{h}\,B_z$ from ANMHD in Figure~\ref{fig:DAVE}. The nonparametric
Spearman rank-order correlation coefficients ($\RO$), Pearson
correlation coefficients ($\PCC$), and slopes ($\Slopes$)
estimated by the least absolute deviation method are all significantly
better for the alternative hypothesis than for the null
hypothesis. The null hypothesis that the velocities inferred by DAVE
represent the flux transport velocities may be 
rejected in favor\footnote{This does not imply that the alternative
hypothesis is correct.} of the alternative hypothesis
$\u=\v_h$.\par
\begin{figure}
\centerline{\includegraphics[width=\size]{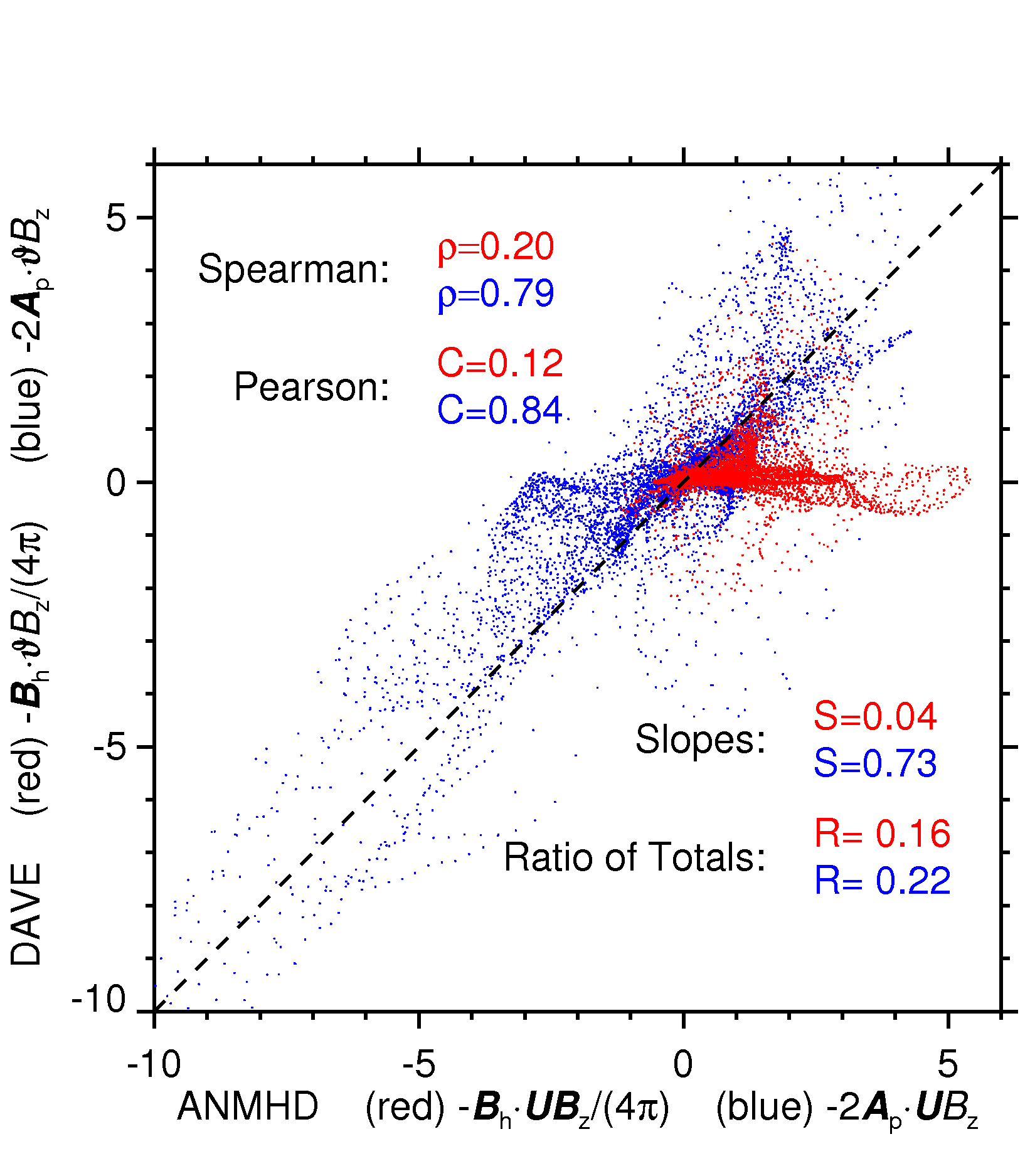}\hspc\hspc\hspc\includegraphics[width=\size]{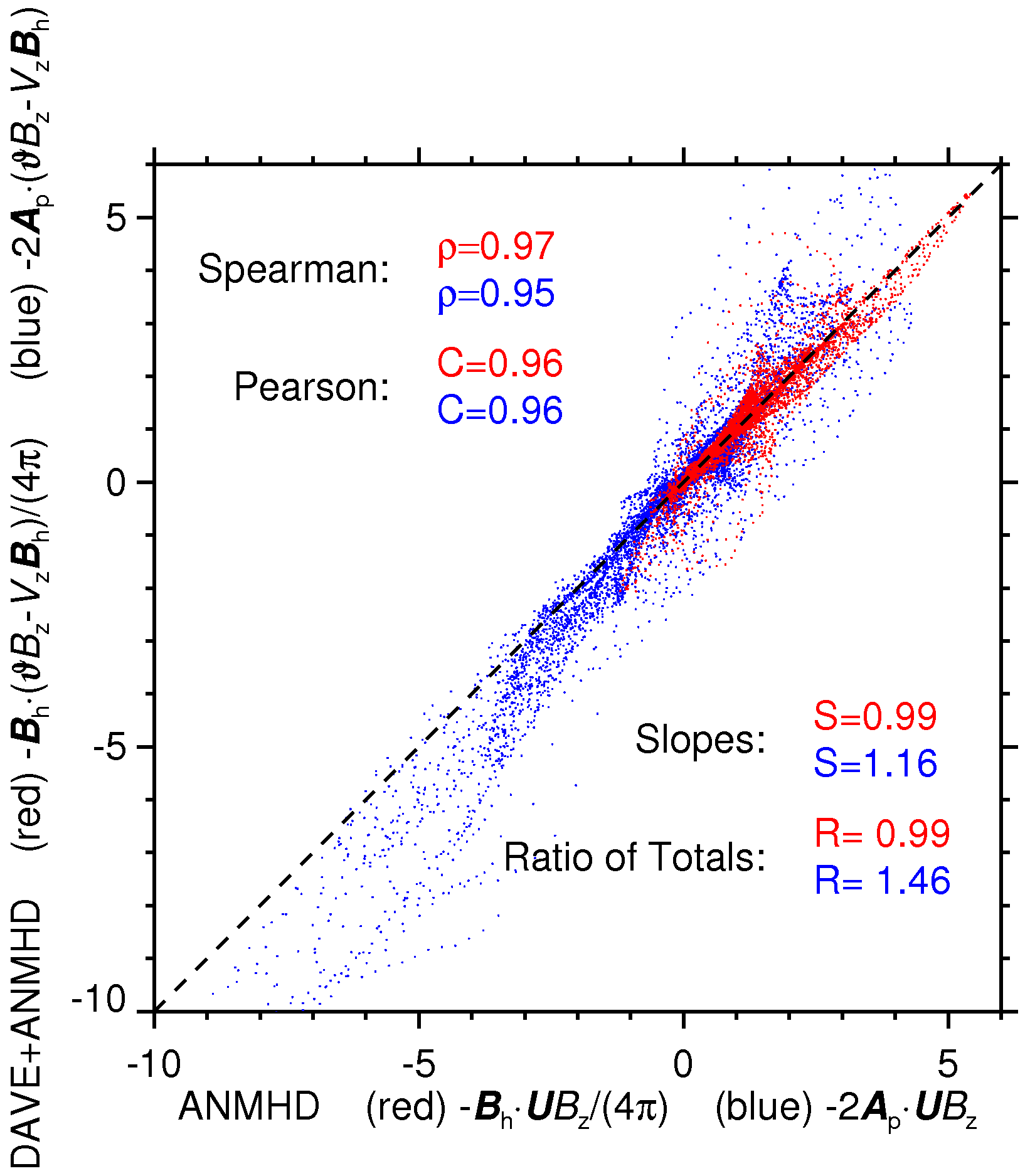}}\vskip0.2in
\caption{(\textit{left}) Scatter plots of the estimated Poynting flux (red) and
  helicity flux (blue) from DAVE assuming $\u=\uF$ versus the Poynting
  and helicity flux from ANMHD. (\textit{right}) Scatter plots of the estimated
  Poynting flux (red) and helicity flux (blue) combining DAVE assuming
  $\u=\v_h$ with the emergence term from ANMHD versus the Poynting and
  helicity flux from ANMHD. The nonparametric Spearman rank-order
  correlation coefficients ($\RO$), Pearson correlation coefficients
  ($\PCC$), and slopes ($\Slopes$) estimated by the least absolute
  deviation method are shown.\label{fig:NAIL}}
\end{figure}
These results explain why $V_{\perp{z}}$ and $v_{\perp{z}}$ are poorly
correlated for the DAVE in Figure~\ref{fig:vp_corr} and the slope
between them is nearly zero \---- the DAVE is consistent with the
\textit{assumption} $v_z=0$ when $\B_h\neq0$. Generally, in regions of
flux emergence, the accuracy $v_{\perp{z}}$ is critical for estimating
the flux transport vectors which in turn is critical for estimating
the helicity and Poynting fluxes. When horizontal magnetic fields and
vertical flows are present, the flux transport vectors estimated from
methods that rely exclusively on the line-of-sight or vertical
component (DAVE, LMSAL's LCT, FLCT) cannot be trusted to provide the
total fluxes. This is particularly true along neutral lines where flux
is emerging or submerging! Under the \textit{best case scenarios},
only the shearing or ``horizontal
fluxes''\footnote{\cite{Demoulin2003} terms these fluxes the
``tangential fluxes'' but ``horizontal'' is more appropriate in the
context of Welsch's terminology used in this paper.} across the
photosphere.
\begin{mathletters}
\begin{equation}
\left.\frac{dp}{dt}\right|_h=-\frac{1}{4\,\pi}\,\int_S{dx^2}\,\B_h\cdot\left(\v_h\,B_z\right),
\end{equation}
and
\begin{equation}
\left.\frac{dh_A}{dt}\right|_h=-2\,\int_S{dx^2}\,\A_p\cdot\left(\v_h\,B_z\right),
\end{equation}
\end{mathletters}
may be estimated from the line-of-sight tracking methods.  In the best
case scenarios only the shearing fluxes are captured by line-of-sight
tracking methods in \textit{partial} agreement with the
\textit{Ansatz} of \cite{Chae2001b} and in disagreement with the
geometrical arguments of \cite{Demoulin2003} who argue that
line-of-sight tracking methods capture both the shearing and
emergence. Again, for the energy and helicity, the alternative
hypothesis $H_1:\u=\v_h$ (shearing) may be tested against the null
hypothesis $H_0:\u=\uF$ (shearing and emergence). The left panel of
Figure~\ref{fig:NAIL}, representing the null hypothesis is a
combination of the left-hand panels of Figure~\ref{fig:poynting} and
Figure~\ref{fig:helicity}. The right panel of Figure~\ref{fig:NAIL},
representing the alternative hypothesis, combines the shearing term
estimated from DAVE with the emergence term from ANMHD.  Generally, in
the presence of vertical flows and horizontal magnetic fields,
line-of-sight tracking methods do not accurately capture the complete
footpoint dynamics and the null hypothesis that the velocities inferred
by DAVE represent the flux transport velocities may be rejected in
favor of the alternative hypothesis $\u=\v_h$. \par
The implementation of vector magnetograms in optical flow methods
presents practical challenges.  First, the transverse magnetic field
components are known to be noisier than the line-of-sight component
and the noise variance will likely change from pixel to pixel due to
variable photon statistics (heteroscedastic errors). Second the
line-of-sight component and transverse components are determined from
different polarizations and require inter-calibration. Third, the
orientation of the transverse component is ambiguous by
$180^\circ$. The first issue may be addressed within the total least
squares framework discussed by \cite{Schuck2006a}, \cite{Branham1999}
and others \cite[See references in][]{Schuck2006a}.  The main obstacle
to resolving the first issue is estimating a covariance matrix for
the structure tensor $\SSS$.\par
The second and third issues both may be interpreted as
inter-calibration bias where the estimated horizontal magnetic field
$\widehat{\B}_h=\alpha\,\B_h$ is proportional to the true horizontal
magnetic field $\B_h$; these errors are not random.  The flux
transport velocities estimated from DAVE4VM are robust to
\textit{overall} inter-calibration errors which include the
$180^\circ$ ambiguity resolution errors. Changing the \textit{overall}
magnitude or sign of $\B_h$ has no effect on the flux transport
velocities because~(\ref{eqn:DAVE4VM}) is invariant with respect to
the transformation $v_z\rightarrow{v_z}/\alpha$ and
$\B_h\rightarrow\alpha\B_h$ (private communication with Pascal
Demoulin). However, the estimated vertical perpendicular plasma
velocity and vertical perpendicular electric field will be
anti-correlated with the ground truth when $\alpha<0$.  Nonetheless,
an \textit{overall} rescaling of the horizontal magnetic field will
have no effect on the helicity flux. However, the Poynting flux will
be incorrect by a factor of $\alpha$ including perhaps a sign error
because of the rescaling horizontal magnetic field which is inherent
in the energy estimate~(\ref{eqn:poynting}). More troublesome are the
effects of \textit{spatially varying} bias errors in inter-calibration
or ambiguity resolution. The consequences of these errors,
particularly along the boundaries between proper and improper
ambiguity resolution, are presently unknown and should be investigated
with future end-to-end analysis of synthetic magnetograms. However,
local methods such as DAVE4VM are probably more robust than global
methods to spatially dependent errors in inter-calibration or
ambiguity resolution because local methods inherently localize the
effect of bias errors by isolating subregions with the window aperture
whereas global methods couple the entire solution region together
permitting bias errors in one subregion to influence the solution in
other subregions. \par
In light of the DAVE4VM's dramatic improvement in performance by
simply including horizontal magnetic fields, speculation that the
ANMHD simulation data are not appropriate for testing tracking methods
cannot be correct. Rather, aside from issues of image structure, the
ANMHD simulation data represent an \textit{ideal} case for the
line-of-sight methods because the vertical magnetic field is known
(not simply the line-of-sight component). The results of this study
suggest that horizontal magnetic fields and vertical flows will render
velocity estimates from ``pure'' tracking methods inaccurate if they
are treated as the flux transport velocities $\uF$.  This conclusion
holds equally true for velocity estimates near disk center as the
ANMHD simulations represents disk-center data!  The good agreement
between the performance of MEF and the DAVE4VM on the ANMHD data
implies that incorporating the right physics is more important for
producing accurate velocity estimates than is the particular method
used the solve the equations.\par
Presently, the \textit{only} way to
explore the ``image'' physics is by testing the ``optical flow''
methods on synthetic data from well-designed MHD simulations that
attempt to reproduce the physics of the Sun. Naive ``moving paint''
experiments \cite[]{Schuck2006a} cannot critically test optical flow
methods for magnetograms because the test data are consistent with the
two circumstances when pure tracking methods will certainly perform
well: $\B_h=0$ and $v_z=0$. Consequently, good performance of an
optical flow method in naive ``moving paint'' experiments should not
be considered evidence that a method will produce accurate estimates
of plasma physics quantities.\par
In the interest of reproducibility \cite[]{Joyner2007}, all of the
software used to perform the calculations, create the figures, and
draw the conclusions for this paper are archived with the
\textit{Astrophysical Journal} as a tgz file. Updates to the 
DAVE/DAVE4VM software are also available.\footnote{\url{http://wwwppd.nrl.navy.mil/whatsnew/}.}
\acknowledgments 
I thank the referee for constructive criticism, Graham Barnes for
encouraging the publication of this work, Bill Abbett for providing
the ANMHD data that formed the core of this research, and Pascal
Demoulin and Brian Welsch for encouraging the clarification of several
issues discussed in this manuscript.  I also gratefully acknowledge
useful conversations with George Fisher, Bill Amatucci, and Etienne
Pariat. I thank Julie Schuck for editing the manuscript.  This work
was supported by NASA LWS TR\&T grant NNH06AD87I, LWS TR\&T Strategic
Capability grant NNH07AG26I, and ONR.
\appendix
\section{Matrix Elements of $\SS$\label{app:matrix}}
\newcommand{\spc}{\,}
\newcommand{\lp}{\left(}
\newcommand{\rp}{\right)}
\begin{mathletters}
\begin{eqnarray*}
\GGG_{00}&=&{\lp\partial_x{B_z}\rp}^2\\ 
\GGG_{10}&=&\lp\partial_x{B_z}\rp{\spc}\lp\partial_y{B_z}\rp\\ 
\GGG_{11}&=&{\lp\partial_y{B_z}\rp}^2\\ 
\GGG_{20}&=&{B_z}{\spc}\lp\partial_x{B_z}\rp + {\lp\partial_x{B_z}\rp}^2{\spc}x'\\ 
\GGG_{21}&=&{B_z}{\spc}\lp\partial_y{B_z}\rp + \lp\partial_x{B_z}\rp{\spc}\lp\partial_y{B_z}\rp{\spc}x'\\ 
\GGG_{22}&=&B_z^2 + 2{\spc}{B_z}{\spc}\lp\partial_x{B_z}\rp{\spc}x' + {\lp\partial_x{B_z}\rp}^2{\spc}x'^2\\ 
\GGG_{30}&=&{B_z}{\spc}\lp\partial_x{B_z}\rp + \lp\partial_x{B_z}\rp{\spc}\lp\partial_y{B_z}\rp{\spc}y'\\ 
\GGG_{31}&=&{B_z}{\spc}\lp\partial_y{B_z}\rp + {\lp\partial_y{B_z}\rp}^2{\spc}y'\\ 
\GGG_{32}&=&B_z^2 + {B_z}{\spc}\lp\partial_x{B_z}\rp{\spc}x' + {B_z}{\spc}\lp\partial_y{B_z}\rp{\spc}y' + \lp\partial_x{B_z}\rp{\spc}\lp\partial_y{B_z}\rp{\spc}x'{\spc}y'\\ 
\GGG_{33}&=&B_z^2 + 2{\spc}{B_z}{\spc}\lp\partial_y{B_z}\rp{\spc}y' + {\lp\partial_y{B_z}\rp}^2{\spc}y'^2\\ 
\GGG_{40}&=&{\lp\partial_x{B_z}\rp}^2{\spc}y'\\ 
\GGG_{41}&=&\lp\partial_x{B_z}\rp{\spc}\lp\partial_y{B_z}\rp{\spc}y'\\ 
\GGG_{42}&=&{B_z}{\spc}\lp\partial_x{B_z}\rp{\spc}y' + {\lp\partial_x{B_z}\rp}^2{\spc}x'{\spc}y'\\ 
\GGG_{43}&=&{B_z}{\spc}\lp\partial_x{B_z}\rp{\spc}y' + \lp\partial_x{B_z}\rp{\spc}\lp\partial_y{B_z}\rp{\spc}y'^2\\ 
\GGG_{44}&=&{\lp\partial_x{B_z}\rp}^2{\spc}y'^2\\ 
\GGG_{50}&=&\lp\partial_x{B_z}\rp{\spc}\lp\partial_y{B_z}\rp{\spc}x'\\ 
\GGG_{51}&=&{\lp\partial_y{B_z}\rp}^2{\spc}x'\\ 
\GGG_{52}&=&{B_z}{\spc}\lp\partial_y{B_z}\rp{\spc}x' + \lp\partial_x{B_z}\rp{\spc}\lp\partial_y{B_z}\rp{\spc}x'^2\\ 
\GGG_{53}&=&{B_z}{\spc}\lp\partial_y{B_z}\rp{\spc}x' + {\lp\partial_y{B_z}\rp}^2{\spc}x'{\spc}y'\\ 
\GGG_{54}&=&\lp\partial_x{B_z}\rp{\spc}\lp\partial_y{B_z}\rp{\spc}x'{\spc}y'\\ 
\GGG_{55}&=&{\lp\partial_y{B_z}\rp}^2{\spc}x'^2\\ 
\sss_{60}&=&- \lp\partial_x{B_x}\rp{\spc}\lp\partial_x{B_z}\rp   - \lp\partial_y{B_z}\rp{\spc}\lp\partial_x{B_z}\rp\\ 
\sss_{61}&=&- \lp\partial_x{B_x}\rp{\spc}\lp\partial_y{B_z}\rp   - \lp\partial_y{B_z}\rp{\spc}\lp\partial_y{B_z}\rp\\ 
\sss_{62}&=&- \lp\partial_x{B_x}\rp{\spc}{B_z}   - \lp\partial_y{B_z}\rp{\spc}{B_z} - \lp\partial_x{B_x}\rp{\spc}\lp\partial_x{B_z}\rp{\spc}x' - \lp\partial_y{B_z}\rp{\spc}\lp\partial_x{B_z}\rp{\spc}x'\\ 
\sss_{63}&=&- \lp\partial_x{B_x}\rp{\spc}{B_z}   - \lp\partial_y{B_z}\rp{\spc}{B_z} - \lp\partial_x{B_x}\rp{\spc}\lp\partial_y{B_z}\rp{\spc}y' - \lp\partial_y{B_z}\rp{\spc}\lp\partial_y{B_z}\rp{\spc}y'\\ 
\sss_{64}&=&- \lp\partial_x{B_x}\rp{\spc}\lp\partial_x{B_z}\rp{\spc}y'   - \lp\partial_y{B_z}\rp{\spc}\lp\partial_x{B_z}\rp{\spc}y'\\ 
\sss_{65}&=&- \lp\partial_x{B_x}\rp{\spc}\lp\partial_y{B_z}\rp{\spc}x'   - \lp\partial_y{B_z}\rp{\spc}\lp\partial_y{B_z}\rp{\spc}x'\\ 
\sss_{66}&=&{\lp\partial_x{B_x}\rp}^2 + 2{\spc}\lp\partial_x{B_x}\rp{\spc}\lp\partial_y{B_z}\rp + {\lp\partial_y{B_z}\rp}^2\\ 
\sss_{70}&=&- {B_x}{\spc}\lp\partial_x{B_z}\rp   - \lp\partial_x{B_x}\rp{\spc}\lp\partial_x{B_z}\rp{\spc}x' - \lp\partial_y{B_z}\rp{\spc}\lp\partial_x{B_z}\rp{\spc}x'\\ 
\sss_{71}&=&- {B_x}{\spc}\lp\partial_y{B_z}\rp   - \lp\partial_x{B_x}\rp{\spc}\lp\partial_y{B_z}\rp{\spc}x' - \lp\partial_y{B_z}\rp{\spc}\lp\partial_y{B_z}\rp{\spc}x'\\ 
\sss_{72}&=&- {B_x}{\spc}{B_z}   - \lp\partial_x{B_x}\rp{\spc}{B_z}{\spc}x' - \lp\partial_y{B_z}\rp{\spc}{B_z}{\spc}x' - {B_x}{\spc}\lp\partial_x{B_z}\rp{\spc}x' - \lp\partial_x{B_x}\rp{\spc}\lp\partial_x{B_z}\rp{\spc}x'^2 - \lp\partial_y{B_z}\rp{\spc}\lp\partial_x{B_z}\rp{\spc}x'^2\\ 
\sss_{73}&=&- {B_x}{\spc}{B_z}   - \lp\partial_x{B_x}\rp{\spc}{B_z}{\spc}x' - \lp\partial_y{B_z}\rp{\spc}{B_z}{\spc}x' - {B_x}{\spc}\lp\partial_y{B_z}\rp{\spc}y' - \lp\partial_x{B_x}\rp{\spc}\lp\partial_y{B_z}\rp{\spc}x'{\spc}y' - \lp\partial_y{B_z}\rp{\spc}\lp\partial_y{B_z}\rp{\spc}x'{\spc}y'\\ 
\sss_{74}&=&- {B_x}{\spc}\lp\partial_x{B_z}\rp{\spc}y'   - \lp\partial_x{B_x}\rp{\spc}\lp\partial_x{B_z}\rp{\spc}x'{\spc}y' - \lp\partial_y{B_z}\rp{\spc}\lp\partial_x{B_z}\rp{\spc}x'{\spc}y'\\ 
\sss_{75}&=&- {B_x}{\spc}\lp\partial_y{B_z}\rp{\spc}x'   - \lp\partial_x{B_x}\rp{\spc}\lp\partial_y{B_z}\rp{\spc}x'^2 - \lp\partial_y{B_z}\rp{\spc}\lp\partial_y{B_z}\rp{\spc}x'^2\\ 
\sss_{76}&=&{B_x}{\spc}\lp\partial_x{B_x}\rp + {B_x}{\spc}\lp\partial_y{B_z}\rp + {\lp\partial_x{B_x}\rp}^2{\spc}x' + 2{\spc}\lp\partial_x{B_x}\rp{\spc}\lp\partial_y{B_z}\rp{\spc}x' + {\lp\partial_y{B_z}\rp}^2{\spc}x'\\ 
\sss_{77}&=&B_x^2 + 2{\spc}{B_x}{\spc}\lp\partial_x{B_x}\rp{\spc}x' + 2{\spc}{B_x}{\spc}\lp\partial_y{B_z}\rp{\spc}x' + {\lp\partial_x{B_x}\rp}^2{\spc}x'^2 + 2{\spc}\lp\partial_x{B_x}\rp{\spc}\lp\partial_y{B_z}\rp{\spc}x'^2 + {\lp\partial_y{B_z}\rp}^2{\spc}x'^2\\ 
\sss_{80}&=&- {B_y}{\spc}\lp\partial_x{B_z}\rp   - \lp\partial_x{B_x}\rp{\spc}\lp\partial_x{B_z}\rp{\spc}y' - \lp\partial_y{B_z}\rp{\spc}\lp\partial_x{B_z}\rp{\spc}y'\\ 
\sss_{81}&=&- {B_y}{\spc}\lp\partial_y{B_z}\rp   - \lp\partial_x{B_x}\rp{\spc}\lp\partial_y{B_z}\rp{\spc}y' - \lp\partial_y{B_z}\rp{\spc}\lp\partial_y{B_z}\rp{\spc}y'\\ 
\sss_{82}&=&- {B_y}{\spc}{B_z}   - {B_y}{\spc}\lp\partial_x{B_z}\rp{\spc}x' - \lp\partial_x{B_x}\rp{\spc}{B_z}{\spc}y' - \lp\partial_y{B_z}\rp{\spc}{B_z}{\spc}y' - \lp\partial_x{B_x}\rp{\spc}\lp\partial_x{B_z}\rp{\spc}x'{\spc}y' - \lp\partial_y{B_z}\rp{\spc}\lp\partial_x{B_z}\rp{\spc}x'{\spc}y'\\ 
\sss_{83}&=&- {B_y}{\spc}{B_z}   - \lp\partial_x{B_x}\rp{\spc}{B_z}{\spc}y' - \lp\partial_y{B_z}\rp{\spc}{B_z}{\spc}y' - {B_y}{\spc}\lp\partial_y{B_z}\rp{\spc}y' - \lp\partial_x{B_x}\rp{\spc}\lp\partial_y{B_z}\rp{\spc}y'^2 - \lp\partial_y{B_z}\rp{\spc}\lp\partial_y{B_z}\rp{\spc}y'^2\\ 
\sss_{84}&=&- {B_y}{\spc}\lp\partial_x{B_z}\rp{\spc}y'   - \lp\partial_x{B_x}\rp{\spc}\lp\partial_x{B_z}\rp{\spc}y'^2 - \lp\partial_y{B_z}\rp{\spc}\lp\partial_x{B_z}\rp{\spc}y'^2\\ 
\sss_{85}&=&- {B_y}{\spc}\lp\partial_y{B_z}\rp{\spc}x'   - \lp\partial_x{B_x}\rp{\spc}\lp\partial_y{B_z}\rp{\spc}x'{\spc}y' - \lp\partial_y{B_z}\rp{\spc}\lp\partial_y{B_z}\rp{\spc}x'{\spc}y'\\ 
\sss_{86}&=&\lp\partial_x{B_x}\rp{\spc}{B_y} + {B_y}{\spc}\lp\partial_y{B_z}\rp + {\lp\partial_x{B_x}\rp}^2{\spc}y' + 2{\spc}\lp\partial_x{B_x}\rp{\spc}\lp\partial_y{B_z}\rp{\spc}y' + {\lp\partial_y{B_z}\rp}^2{\spc}y'\\ 
\sss_{87}&=&{B_x}{\spc}{B_y} + \lp\partial_x{B_x}\rp{\spc}{B_y}{\spc}x' + {B_y}{\spc}\lp\partial_y{B_z}\rp{\spc}x' + {B_x}{\spc}\lp\partial_x{B_x}\rp{\spc}y' + {B_x}{\spc}\lp\partial_y{B_z}\rp{\spc}y' \\
&&+ {\lp\partial_x{B_x}\rp}^2{\spc}x'{\spc}y' + 2{\spc}\lp\partial_x{B_x}\rp{\spc}\lp\partial_y{B_z}\rp{\spc}x'{\spc}y' + {\lp\partial_y{B_z}\rp}^2{\spc}x'{\spc}y'\\ 
\sss_{88}&=&B_y^2 + 2{\spc}\lp\partial_x{B_x}\rp{\spc}{B_y}{\spc}y' + 2{\spc}{B_y}{\spc}\lp\partial_y{B_z}\rp{\spc}y' + {\lp\partial_x{B_x}\rp}^2{\spc}y'^2 + 2{\spc}\lp\partial_x{B_x}\rp{\spc}\lp\partial_y{B_z}\rp{\spc}y'^2 + {\lp\partial_y{B_z}\rp}^2{\spc}y'^2\\ 
\GGG_{90}&=&\lp\partial_t{B_z}\rp{\spc}\lp\partial_x{B_z}\rp\\ 
\GGG_{91}&=&\lp\partial_t{B_z}\rp{\spc}\lp\partial_y{B_z}\rp\\ 
\GGG_{92}&=&{B_z}{\spc}\lp\partial_t{B_z}\rp + \lp\partial_t{B_z}\rp{\spc}\lp\partial_x{B_z}\rp{\spc}x'\\ 
\GGG_{93}&=&{B_z}{\spc}\lp\partial_t{B_z}\rp + \lp\partial_t{B_z}\rp{\spc}\lp\partial_y{B_z}\rp{\spc}y'\\ 
\GGG_{94}&=&\lp\partial_t{B_z}\rp{\spc}\lp\partial_x{B_z}\rp{\spc}y'\\ 
\GGG_{95}&=&\lp\partial_t{B_z}\rp{\spc}\lp\partial_y{B_z}\rp{\spc}x'\\ 
\sss_{96}&=&- \lp\partial_x{B_x}\rp{\spc}\lp\partial_t{B_z}\rp   - \lp\partial_y{B_z}\rp{\spc}\lp\partial_t{B_z}\rp\\ 
\sss_{97}&=&- {B_x}{\spc}\lp\partial_t{B_z}\rp   - \lp\partial_x{B_x}\rp{\spc}\lp\partial_t{B_z}\rp{\spc}x' - \lp\partial_y{B_z}\rp{\spc}\lp\partial_t{B_z}\rp{\spc}x'\\ 
\sss_{98}&=&- {B_y}{\spc}\lp\partial_t{B_z}\rp   - \lp\partial_x{B_x}\rp{\spc}\lp\partial_t{B_z}\rp{\spc}y' - \lp\partial_y{B_z}\rp{\spc}\lp\partial_t{B_z}\rp{\spc}y'\\ 
\GGG_{99}&=&{\lp\partial_t{B_z}\rp}^2\\ 
\end{eqnarray*}
The primed coordinates are defined relative to the center of the aperture $\x'=\x-\X$.
\end{mathletters}
\bibliographystyle{apj}
\singlespace
\bibliography{bibliography}

\end{document}